\documentclass[]{aa} 

\usepackage{txfonts}
\usepackage{natbib,graphicx,amssymb}
\usepackage[english]{babel}
\usepackage{xspace}
\usepackage{lscape}
\usepackage{breqn}
\usepackage{url}

\newcommand{\kms}{km\,s$^{-1}$\xspace}
\newcommand{\mum}{$\mu$m\xspace}
\newcommand{\av}{$A_{\rm V}$\xspace}
\newcommand{\cii}{[C\,{\sc ii}]\xspace}
\newcommand{\nii}{[N\,{\sc ii}]\xspace}
\newcommand{\niii}{[N\,{\sc iii}]\xspace}
\newcommand{\oi}{[O\,{\sc i}]\xspace}
\newcommand{\oiii}{[O\,{\sc iii}]\xspace}
\newcommand{\ciil}{[C\,{\sc ii}]$_{157}$\xspace}
\newcommand{\niila}{[N\,{\sc ii}]$_{122}$\xspace}
\newcommand{\niilb}{[N\,{\sc ii}]$_{205}$\xspace}
\newcommand{\niiil}{[N\,{\sc iii}]$_{57}$\xspace}
\newcommand{\oila}{[O\,{\sc i}]$_{63}$\xspace}
\newcommand{\oilb}{[O\,{\sc i}]$_{145}$\xspace}
\newcommand{\oiiil}{[O\,{\sc iii}]$_{88}$\xspace}
\newcommand{\hers}{\textit{Herschel}\xspace}
\newcommand{\spit}{\textit{Spitzer}\xspace}
\newcommand{\siii}{[S\,{\sc iii}]\xspace}
\newcommand{\siiila}{[S\,{\sc iii}]$_{18}$\xspace}
\newcommand{\siiilb}{[S\,{\sc iii}]$_{33}$\xspace}
\newcommand{\sivl}{[S\,{\sc iv}]$_{10}$\xspace}
\newcommand{\neiil}{[Ne\,{\sc ii}]$_{12}$\xspace}
\newcommand{\neiiil}{[Ne\,{\sc iii}]$_{15}$\xspace}
\newcommand{\siv}{[S\,{\sc iv}]\xspace}
\newcommand{\neii}{[Ne\,{\sc ii}]\xspace}
\newcommand{\neiii}{[Ne\,{\sc iii}]\xspace}
\newcommand{\nev}{[Ne\,{\sc v}]\xspace}
\newcommand{\si}{[Si\,{\sc ii}]\xspace}
\newcommand{\htwo}{H$_{\rm 2}$\xspace}
\newcommand{\hii}{H\,{\sc ii}\xspace}
\newcommand{\zsun}{$Z_{\odot}$\xspace}
\newcommand{\msun}{$M_{\odot}$\xspace}
\newcommand{\lsun}{$L_{\odot}$\xspace}
\newcommand{\lb}{${L_{\rm B}}$\xspace}
\newcommand{\ltir}{${L_{\rm TIR}}$\xspace}

\newcommand{\arcm}{$^{\prime}$\xspace}
\newcommand{\arcs}{$^{\prime\prime}$\xspace}
 \newcommand{\cm}{cm$^{-3}$\xspace}

\begin{document}

\title{The \hers Dwarf Galaxy Survey: \\
I.~Properties of the low-metallicity ISM from PACS spectroscopy
\thanks{{\it Herschel} is an ESA space observatory with science instruments 
provided by European-led Principal Investigator consortia and with important 
participation from NASA.}}

\author{
  D.~Cormier\inst{1}
  \and S.~C.~Madden\inst{2}
  \and V.~Lebouteiller\inst{2}
  \and N.~Abel\inst{3}
  \and S.~Hony\inst{1}
  \and F.~Galliano\inst{2}
  \and A.~R{\'e}my-Ruyer\inst{2}
  \and F.~Bigiel\inst{1}
 \and M.~Baes\inst{4}
 \and A.~Boselli\inst{5}
 \and M.~Chevance\inst{2}
 \and A.~Cooray\inst{6}
 \and I.~De~Looze\inst{4,7}
 \and V.~Doublier\inst{8}
 \and M.~Galametz\inst{9}
 \and T.~Hughes\inst{4}
 \and O.~\L.~Karczewski\inst{10}
 \and M.-Y.~Lee\inst{2}
 \and N.~Lu\inst{11}
 \and L.~Spinoglio\inst{12}
}

\institute{ 
Institut f\"ur theoretische Astrophysik, 
Zentrum f\"ur Astronomie der Universit\"at Heidelberg, 
Albert-Ueberle Str. 2, 69120 Heidelberg, Germany. 
\email{diane.cormier@zah.uni-heidelberg.de} 
\and
Laboratoire AIM, CEA/DSM - CNRS - Universit\'e Paris
  Diderot, Irfu/Service d'Astrophysique, CEA Saclay, 91191
  Gif-sur-Yvette, France 
\and
 University of Cincinnati, Clermont College, Batavia, OH, 45103, USA
\and
 Sterrenkundig Observatorium, Universiteit Gent, 
 Krijgslaan 281 S9, B-9000 Gent, Belgium
 \and 
 Laboratoire d'Astrophysique de Marseille - 
 LAM, Universit{\'e} d'Aix-Marseille \& CNRS, UMR 7326, 
 38 rue F. Joliot-Curie, 13388, Marseille Cedex 13, France
 \and 
 Department of Physics \& Astronomy,
 University of California, 
 Irvine, CA, 92697, USA
 \and 
Institute of Astronomy, University of Cambridge, 
Madingley Road, Cambridge, CB3 0HA, UK
\and
 Max-Planck-Institut f\"ur extraterrestrische Physik, 
 Giessenbachstra{\ss}e, 85748 Garching, Germany
\and
 ESO, Karl-Schwarzschild-Stra{\ss}e 2, 85748 Garching, Germany
\and
Department of Physics and Astronomy, University of Sussex, Brighton, BN1 9QH, UK 
\and
 Infrared Processing and Analysis Center, California Institute of Technology, 
 MS 100-22, Pasadena, CA 91125, USA
\and
Istituto di Fisica dello Spazio Interplanetario, INAF, 
Via Fosso del Cavaliere 100, I-00133 Roma, Italy
}

\date{Received 24 October 2014; accepted 5 February 2015.}

\abstract
{The far-infrared (FIR) lines are important tracers of the cooling and 
physical conditions of the interstellar medium (ISM), and are rapidly 
becoming workhorse diagnostics for galaxies throughout the universe. 
There are clear indications of a different behavior of these lines 
at low metallicity that needs to be explored.
}
{Our goal is to explain the main differences and trends observed in the 
FIR line emission of dwarf galaxies compared to more metal-rich galaxies, 
and how this translates in ISM properties. 
}
{We present \hers PACS spectroscopic observations of the \cii157\,\mum, 
\oi63 and 145\,\mum, \oiii88\,\mum, \nii122 and 205\,\mum, and \niii57\,\mum 
fine-structure cooling lines in a sample of 48 low-metallicity star-forming 
galaxies of the guaranteed time key program Dwarf Galaxy Survey.   
We correlate PACS line ratios and line-to-\ltir ratios with \ltir, \ltir/\lb, metallicity, 
and FIR color, and interpret the observed trends in terms of ISM conditions and 
phase filling factors with Cloudy radiative transfer models. 
}
{We find that the FIR lines together account for up to 3 percent 
of \ltir and that star-forming regions dominate the overall emission in dwarf galaxies. 
Compared to metal-rich galaxies, the ratios of \oiiil/\niila and \niiil/\niila 
are high, indicative of hard radiation fields. 
In the photodissociation region (PDR), the \ciil/\oila ratio is slightly higher than 
in metal-rich galaxies, with a small increase with metallicity, and the \oilb/\oila ratio 
is generally lower than $0.1$, demonstrating that optical depth effects should 
be small on the scales probed. 
The \oiiil/\oila ratio can be used as an indicator of the ionized gas/PDR filling factor, 
and is found to be $\sim$4 times higher in the dwarfs than in metal-rich galaxies. 
The high \cii/\ltir, \oi/\ltir, and \oiii/\ltir ratios, which decrease with increasing \ltir 
and \ltir/\lb, are interpreted as a combination of moderate FUV fields and a low 
PDR covering factor. 
Harboring compact phases of a low filling factor and a large volume filling factor 
of diffuse gas, the ISM of low-metallicity dwarf galaxies has a more porous 
structure than that of metal-rich galaxies. 
}
{}

\titlerunning{PACS spectroscopy of the \hers Dwarf Galaxy Survey}
\authorrunning{Cormier et al.}
\maketitle

\section{Introduction}
\label{sect:intro}
Low-metallicity environments, such as nearby star-forming dwarf galaxies, 
are ideal laboratories for examining conditions for star formation in objects 
as chemically unevolved as high-redshift galaxies.
They are good candidates with which to study and address fundamental 
questions such as: 
How do the physical conditions of star-forming regions vary as a function 
of the elemental enrichment? 
What controls the energy balance, heating and cooling, in the different 
interstellar medium (ISM) phases and shapes the morphology of dwarf galaxies?
What are the roles of the different ISM phases in the integrated view of galaxies?  

Spectroscopy in the far-infrared (FIR) provides a unique and 
unobscured view of the ISM properties and conditions for star formation. 
The FIR cooling lines, less affected by dust attenuation than optical lines, 
are powerful probes of the star formation activity, linking them directly to the 
surrounding medium from which these stars are born. 
The fine-structure lines \cii157\,\mum, \oi63\,\mum, and \oi145\,\mum 
are important coolants of moderately dense 
($n_{\rm H} > 10$\,\cm) warm ($T\sim100-300$\,K) neutral ISM 
and are among the brightest cooling lines in star-forming galaxies. 
The \nii122 and 205\,\mum and \oiii52 and 88\,\mum lines are 
tracers of ionized gas. 
Together, they can be used as diagnostics of the FUV flux, gas density, 
temperature, and filling factor of the photodissociation regions (PDR) 
and ionized regions \citep[e.g.,][]{tielens-1985,wolfire-1990,kaufman-2006}.
In combination with MIR features (\siii18.71 and 33.48\,\mum, 
\neiii15.56\,\mum, \nev14.32\,\mum, \si34.82\,\mum, \htwo, 
polycyclic aromatic hydrocarbons or PAHs, etc.), they provide a comprehensive 
view on conditions and excitation processes in star-forming regions. 

The \cii157\,\mum line has been studied in a variety of objects, 
such as Galactic PDRs \citep[e.g.,][]{bennett-1994,pineda-2013}, 
normal spiral galaxies \citep{stacey-1991,malhotra-1997,negishi-2001,brauher-2008}, 
low-metallicity galaxies \citep{poglitsch-1995,madden-1997,hunter-2001,israel-2011,cigan-2015}, 
ultra-luminous IR galaxies \citep[ULIRG,][]{luhman-1998,farrah-2013}, 
and high-redshift galaxies \citep{maiolino-2009,stacey-2010,hailey-dunsheath-2010,swinbank-2012,de-breuck-2014}. 
From those studies \citep[see][for a review]{genzel-2000}, 
the \cii line emerges as the brightest cooling line of the ISM 
in star-forming galaxies and as a good tracer of the star formation activity, 
accounting for $\sim$0.1-1\% of the FIR luminosity 
\citep[e.g.,][]{stacey-2010,pierini-2003,boselli-2002,delooze-2011,delooze-2014,sargsyan-2014}. 
A deficit in the FIR line intensities is, however, found in luminous IR galaxies 
\citep[e.g.,][]{luhman-2003,gracia-carpio-2011,croxall-2012,diaz-santos-2013}. 
The origin of the \cii line has been debated in extragalactic studies 
as it can arise from molecular, neutral atomic, and ionized gas phases. 
Characterizing unambiguously the ISM conditions thus requires the use 
of additional FIR cooling lines (such as \nii or \oi). 
In dwarf galaxies, the FIR lines were only detected by the KAO and 
ISO LWS in bright or Local Group dwarf galaxies 
\citep[e.g.,][]{poglitsch-1995,israel-1996,madden-1997,hunter-2001}. 
These studies found that the FIR lines are exceptionally bright 
compared to the dust emission. 
The brightness of the FIR lines contrasts with the faint CO emission, 
resulting in high \cii-to-CO(1-0) ratios in low-metallicity galaxies 
\citep{madden-2000,cormier-2014,madden-2015}. 
While CO suffers from photodissociation, large amounts of \htwo can 
potentially be present in low-metallicity PDRs owing to self-shielding. 
This \htwo gas not traced by CO is known as the CO-dark gas \citep[e.g.,][]{wolfire-2010}. 
Interpreting the FIR observations in terms of excitation led to the first 
evidence of such a reservoir of dark gas, not seen by CO but by C$^+$ 
\citep{poglitsch-1995,israel-1996,madden-1997}. 
This demonstrates the potential of the FIR lines as calibrators of the dark gas 
at low metallicity and star formation tracers \citep{delooze-2014} that 
could be used for high-redshift studies.

The advent of the \hers Space Observatory \citep{pilbratt-2010} has provided 
a unique opportunity to study the dust and gas properties in larger samples 
of chemically unevolved systems. 
The \hers Guaranteed Time Key Program ``The Dwarf Galaxy Survey'' 
(DGS; \citealt{madden-2013}) is a 230~hours project which 
observed 50 dwarf galaxies with the PACS \citep{poglitsch-2010} 
and SPIRE \citep{griffin-2010} instruments. 
Details about the program are presented in \cite{madden-2013}, which 
also contains a list of the dwarf galaxies and their general properties. 
We summarize the main characteristics here. 
The galaxies of the sample have metallicities ranging from $\sim$1/40\,\zsun 
(I\,Zw\,18) to near solar (He\,2-10), and star formation rates ranging from 
$\sim$5$\times$10$^{-4}$\,\msun yr$^{-1}$ (UGC\,4483), to 
$\sim$25\,\msun yr$^{-1}$ (Haro\,11). 
They have different shapes (irregulars, spirals, etc.), and all of them are 
nearby galaxies, no further than 200\,Mpc. 
The PACS (70, 100, and 160\,\mum) and SPIRE (250, 350, and 500\,\mum) 
photometry probe the dust emission in the FIR/submillimeter (submm). 
These data are essential to constrain spectral energy distribution (SED) 
models and investigate the dust properties. The photometry of the DGS is 
presented in \cite{remy-2013} and has revealed, among other results, 
the presence of warmer dust than in metal-rich galaxies, the presence of an excess 
emission in the submm in several dwarf galaxies, and lower dust-to-gas mass 
ratios than expected at the lowest metallicities \citep{remy-2014}. 
The PACS (55-210\,\mum) and SPIRE (195-670\,\mum) spectrometers 
give access to the FIR fine-structure cooling lines, the CO ladder, 
from $J=4-3$ to $J=13-12$, and fainter lines of other molecules 
such as H$_{\rm 2}$O, HCN, OH$^+$, etc. 
By probing different ISM phases, they provide information on the filling 
factors and physical conditions of these phases, as well as on the ISM 
morphology. The dust and gas observations together reveal a coherent 
multiphase picture of the ISM. 

In this paper, we present the observations and results of the FIR lines 
obtained with the PACS instrument onboard \hers in 48 galaxies of the DGS. 
Section~\ref{sect:data} describes the data reduction and flux extraction. 
The observations are interpreted empirically with correlation diagrams 
and compared to the ISO-LWS work of \cite{brauher-2008} on 
mostly solar metallicity objects in Sect.~\ref{sect:corr}. 
The data are further analyzed with \hii region/PDR models in Sect.~\ref{sect:models} 
to contrast the physical conditions characteristic of the ISM 
of dwarf galaxies to those of more metal-rich galaxies. 
We discuss our results in terms of ISM properties and filling factors 
of the ionized and neutral phases in Sect.~\ref{sect:discuss}. 
More specific modeling of the PDR conditions in the dwarf galaxies will be 
performed in a subsequent paper that we refer to as {\sc Paper~II}.

\section{Data}
\label{sect:data}
\subsection{\hers PACS spectroscopy}
\subsubsection{Observing details}
The PACS spectrometer observed 48 galaxies as part of the DGS and 
SHINING programs, in a total time of 152.7\,h. 
Only the two, faint galaxies Tol\,0618-402 and UGCA\,20, which are part 
of the DGS sample (50 galaxies), were dropped from the PACS spectroscopy 
program because of time constraints. 
The PACS array consists or 5$\times$5 spatial pixels (or spaxels) 
covering a total field of view of 47\arcs$\times$47\arcs. 
The beam size is $\sim$9\arcs and 12\arcs and the spectral resolution 
90\,\kms and 240\,\kms at 60\,\mum and 160\,\mum, respectively \citep{poglitsch-2010}. 
We obtained pointed observations for the most compact objects, 
small mappings for more extended galaxies, and partial mappings of specific 
star-forming regions for the most extended galaxies ($d_{25}>$6\arcm typically), 
such as the Magellanic Clouds. 
We define an $extended$ source sample for the galaxies partially mapped 
or for which the size of the maps do not match for all observed lines. 
This sample comprises 4 Local Group galaxies, IC\,10, NGC\,6822, 
LMC (8 regions targeted), and SMC, and the nearby galaxy NGC\,4449. 
The other 43 galaxies of the DGS are classified as $compact$ 
(see also Sect.~\ref{sect:incomplete}). 

The dwarf galaxies were observed between November 2009 and August 2012. 
46 sources were done in chop-nod mode and 6 sources in unchopped mode. 
The number of targeted spectral line varies from just \cii157\,\mum, 
in the faintest cases, up to 7 lines, \cii157\,\mum, \oiii88\,\mum, \oi63\,\mum, 
\oi145\,\mum, \niii57\,\mum, \nii122\,\mum, and \nii205\,\mum, in the brightest galaxies. 
We provide details on the observations, such as OBSIDs, map sizes, 
lines observed, etc. in Appendix~\ref{app:append-a}.

\subsubsection{Properties of the FIR lines}
\label{sect:firintro}
General characteristics of the FIR fine-structure cooling lines 
observed with PACS are given in Table~\ref{table:general}. 

\vspace{2pt}\textit{\cii157\,\mum:}\\
The \cii line is one of the most important coolants of the ISM, 
as carbon is the fourth most abundant element. 
\cii is excited by collisions with $e^-$, hydrogen atoms, or molecules. 
The ionization potential of C$^0$ being 11.26\,eV, below that of hydrogen, 
C$^+$ can be found outside of \hii regions, in the neutral phase.  
It requires only 91.3\,K to be excited hence it can cool any warm neutral phase. 
Thus C$^+$ can originate from diffuse ionized gas as well as diffuse neutral gas 
or the surface layers of PDRs (up to \av$\sim$5\,mag). 
The relative contribution of each medium to its observed intensity 
is a function of density and ionization degree \citep[e.g.,][]{kaufman-2006}. 

\vspace{2pt}\textit{\nii122\,\mum and \nii205\,\mum:}\\
\nii is only found in the ionized gas. The critical densities with $e^-$ are 
310\,\cm and 50\,\cm for \nii122\,\mum and \nii205\,\mum, respectively. 
Being in the same ionization stage, their ratio is a good electron density 
tracer of the low-density diffuse ionized gas. 
Because \cii can be found in the low-density ionized gas, part of its emission can 
correlate with the \nii emission. Hence the \nii lines can, in principle, be used 
to disentangle the fraction of \cii from the ionized gas to that from the PDR 
alone \citep{oberst-2006}. 

\vspace{2pt}\textit{\niii57\,\mum:}\\
The \niii57\,\mum emission only arises in ionized gas and is usually associated 
with \hii regions. Being at two different ionization stages, the ratio of 
\niiil/\niila is a measure of the effective temperature of the ionizing stars 
\citep{rubin-1994}. 

\vspace{2pt}\textit{\oi63\,\mum and \oi145\,\mum:}\\
$\rm{O^0}$ has an ionization potential of 13.62\,eV, just above that of hydrogen. 
The first two fine-structure transitions require excitation energies of 228 and 325\,K 
above the ground state, corresponding to the \oi63\,\mum and 145\,\mum lines. 
\oi is only found in neutral gas and usually arises from warm, dense regions. 
Along with \cii, \oi63\,\mum is one of the brightest PDR cooling lines \citep[e.g.,][]{bernard-salas-2012}. 
\oi can also exist deeper in the PDR than \cii since the formation of CO occurs 
at larger \av ($\sim$10\,mag) than the transition of $\rm{C^+}$ into $\rm{C^0}$ ($\sim$3\,mag). 
The ratio of the two \oi lines is an indicator of the gas temperature 
for temperatures in the vicinity of 300\,K. 
It is a density tracer for high temperatures and high densities. 
The \oi emission can be affected by optical depth effects 
(self-absorption, optical thickness), particularly for the lower level transition 
at 63\,\mum \citep{liseau-2006}. In the optically thin limit, the \oi63\,\mum, 
just above the ground state, is brighter than the 145\,\mum line. 

\vspace{2pt}\textit{\oiii88\,\mum:}\\
The \oiii88\,\mum line is only found in the ionized gas, and because 
it requires energetic photons (35\,eV), it is commonly accepted that 
it comes from \hii regions rather than inter-clump diffuse media 
where radiation fields are expected to be softer. 
The \oiii transition at 52\,\mum ($^3P_2-\,^3P_1$) is at the edge of 
the PACS wavelength coverage and was not observed in our survey.

\begin{center}
\begin{table}[!t]\small
  \caption{Characteristics of the PACS FIR fine-structure cooling lines.} 
  \hfill{}
\begin{tabular}{l c c c c c}
    \hline\hline
     \vspace{-8pt}\\
    \multicolumn{1}{l}{Species} & 
    \multicolumn{1}{c}{$\lambda$} &
    \multicolumn{1}{c}{Transition} &
    \multicolumn{1}{c}{IP} &
    \multicolumn{1}{c}{$\Delta E/k^{(a)}$} &
    \multicolumn{1}{c}{$n_{\rm crit}$} \\ 
    \multicolumn{1}{l}{ } & 
    \multicolumn{1}{c}{[$\mu$m]} &
    \multicolumn{1}{c}{ } &
    \multicolumn{1}{c}{[eV]} &
    \multicolumn{1}{c}{[K]} &
    \multicolumn{1}{c}{[$\rm{cm^{-3}}$]} \\ 
    \hline
    \vspace{-8pt}\\
	{\cii}		& $157.7$ 	& $^2P_{3/2}-\,^2P_{1/2}$	& 11.3 	& 91		& $50^{(b)}$, $2.8\times10^3$ \\
	{\nii}		& $121.9$ 	& $^3P_2-\,^3P_1$		& 14.5 	& 188	& $310$ \\
	{\nii}		& $205.2$ 	& $^3P_1-\,^3P_0$		& 14.5 	& 70		& $48$ \\
	{\niii}		& $57.3$ 		& $^3P_{3/2}-\,^3P_{1/2}$	& 29.6 	& 251	& $3.0\times10^3$ \\
	{\oi}		& $63.2$ 		& $^3P_1-\,^3P_2$		& -- 		& 228	& $4.7\times10^5$ \\
	{\oi}		& $145.5$ 	& $^3P_0-\,^3P_1$		& -- 		& 327	& $9.5\times10^4$ \\
	{\oiii}		& $88.4$		& $^3P_1-\,^3P_0$		& 35.1 	& 163	& $510$ \\
    \hline \hline
    \vspace{-5pt}\\
  \end{tabular}
  \hfill{}
  \newline
Values taken from \cite{madden-2013}. The IP for [O\,{\sc ii}] is 13.62\,eV. 
$(a)$~Excitation temperature $\Delta E/k$ required 
to populate the transition level from the ground state. 
$(b)$~Critical density for collisions with electrons. 
  \label{table:general}
\end{table}
\end{center}

\subsubsection{Data reduction}
The data were downloaded from the \hers Science Archive 
and reduced with the \hers Interactive Processing Environment 
\citep[HIPE,][]{ott-2010} user release 12 and the PACS calibration tree v65. 
We reprocessed the observations from the raw data (level~0) to spectral 
data cubes (level~2) using the standard pipeline scripts available within HIPE. 
We opted for the telescope background normalization calibration method 
for all chop-nod observations, except for the \niilb observations which 
were processed with the de-leaked RSRF provided in HIPE\footnote{more 
information available in the PACS Observer's Manual (v.2.5.1)}.
In the first part of the pipeline (level~0 to level~0.5), the astrometry and 
information from the instrument are read. Masks for bad and noisy pixels, 
and for when the chopper and grating are moving are also applied. 
From level~0.5 to level~1, glitches are flagged, and the calibration is applied. 
In the chop-nod case, signal from the chopper position is subtracted. 
From level~1 to level~2, a spectral flat-field correction is applied, 
and outlier data points are masked by a sigma-clipping. 
The flat-field scaling is both multiplicative and additive for chop-nod 
observations, and additive only for unchopped observations. 
In the unchopped case, a transient correction and background 
subtraction are also executed. 
For the transient correction, we applied a multi-resolution method as 
described in \cite{lebouteiller-2012}. 
For the background subtraction, we used a procedure that smoothes 
and averages the OFF-spectra taken before and after an ON-spectrum 
in time (when several OFF-frames are available), and then subtracts it 
from the ON-frame.

\subsubsection{Spectral maps}
\label{sect:maps}
The data are saved as spectral cubes before they are rebinned and projected 
onto a final spatial grid in the pipeline script. Once the spectral cubes 
(one cube per raster position) are exported from HIPE, we used the software 
\textsc{PACSman} version 3.57 for the analysis \citep{lebouteiller-2012}. 

The first step is to fit the spectral lines and measure the line flux in all the 
spatial positions. A polynomial curve of order 1 or 2 is first fit to the continuum, 
and then a simultaneous fit to the continuum and line is performed with the 
IDL procedure \texttt{mpfitfun} \citep{markwardt-2009}. 
The line profile is assumed to be Gaussian, although skewed profiles 
can be seen if the observed object is a point-source not centered on the array. 
Only UM\,448 shows clear asymmetric and wider profiles in several spaxels, 
but since we have made mini-maps for this galaxy, the better centered spaxels 
are well-enough fitted by a Gaussian profile, and we can reliably recover the 
total flux of this galaxy. 
By default, the continuum is defined as the part of the spectrum within $2-6$ 
FWHM of the line center. 
The width of the Gaussian is taken as the instrumental width, 
plus a broadening when necessary. 
The PACS line profiles are sometimes found larger than the instrumental profile, 
necessitating broadened line fits. We report on the broadening parameters 
of the FIR lines in Appendix~\ref{app:append-e}. 

To estimate the uncertainty on the fit parameters, we use a Monte Carlo approach. 
We randomly perturb the data within their error bars, fit the line, iterate the process 
$100$ times, and take as final fit parameters the median of the resulting values, 
and as error on those parameters the standard deviation. 
An additional 15\% absolute flux calibration uncertainty is to be considered\,$^1$.

For the irregular galaxy IC\,10, the \cii spectra are contaminated 
by emission from the Milky Way. This emission appears at $+320$\,\kms relative 
to IC\,10, with a peak of $\sim$1.8\,Jy, and is fairly uniform across our \cii map. 
The separation between the two emission peaks is large enough 
to clearly identify the respective contributions. 
Thus to extract the flux only from IC\,10, we fit simultaneously two Gaussians. 
We also note that OFF-spectra are not available for several observations in 
LMC-30\,Doradus ({\sc OBSIDs 1342231280, 1342231282, 1342231283, 1342231285}), 
LMC-N11\,B ({\sc OBSIDs 1342188940, 1342188941}), and NGC\,4449 
({\sc OBSIDs 1342197813, 1342197814}), resulting in variations across wavelength 
of a few Jy typically in the baselines of the ON-spectra. 
For NGC\,4449 and LMC-N11\,B, we model those variations with a sinusoidal 
component of low frequency to improve the baseline fit. 
For LMC-30\,Dor, no sinusoid is required as those variations are not significant 
compared to the line strength and thus do not affect our line fits. 

Once flux values are calculated for each spatial position, the data are projected 
on a new grid of sub-pixel resolution of 3.1\arcs (roughly 3 times oversampled) 
with a drizzling scheme. 
When more than one spaxel falls on the same sub-pixel, a fraction of the flux 
from the corresponding spaxel of each raster is considered, and the uncertainties 
are combined quadratically. 
The final flux maps, with spectra and line fits are shown for each galaxy 
in Appendix~\ref{app:append-b}.

\subsubsection{Flux extraction}
\label{sect:fluxextract}
We have explored several methods to extract accurate line fluxes. 
For compact objects, we consider 4 methods to measure the total line flux, 
as it depends on how well-centered the source is on the PACS array 
and on the spatial extent of the source:\\
--~{\it central}: we apply a point-source correction only to the central (or brightest) spaxel. 
This method works only if the source is well centered on the spaxel. \\
--~{\it psf}: we fit a point spread function (PSF) to the array. \\
--~{\it 3$\times$3}: we combine the inner 3$\times$3 spaxels before fitting 
the line and apply a 3$\times$3 point-source correction. \\
--~{\it 5$\times$5}: we combine the 5$\times$5 spaxels of the PACS array 
before fitting the line and apply a 5$\times$5 point-source correction. \\
For final flux, we select the method that gives the highest response with a 
good signal-to-noise ratio. \\
In the case of extended or elongated objects we simply perform an 
aperture extraction with sky subtraction when possible or consider 
the total flux in the area mapped. The errors are added quadratically 
from the error maps and multiplied by a factor of $\sqrt{9}$ to account 
for the number of sub-pixels per independent beam. 
We consider a spectral line detected if the S/N is $\ge$3, and upper limits 
are given as the $3\sigma$ level, from the method used for the detected 
lines or from the {\it central} method otherwise. 
Table~\ref{table:totfluxes} lists line fluxes as well as the method used 
for the flux extraction (aperture size when applicable). To summarize, 
these values are the most reliable measurements of the total line fluxes 
for the DGS galaxies. For the $extended$ sample, we report peak surface 
brightnesses instead of total fluxes.

\subsubsection{Summary of detections}
\begin{figure}[!t]
\centering
\includegraphics[clip,trim=0 1cm 0 0,width=7cm]{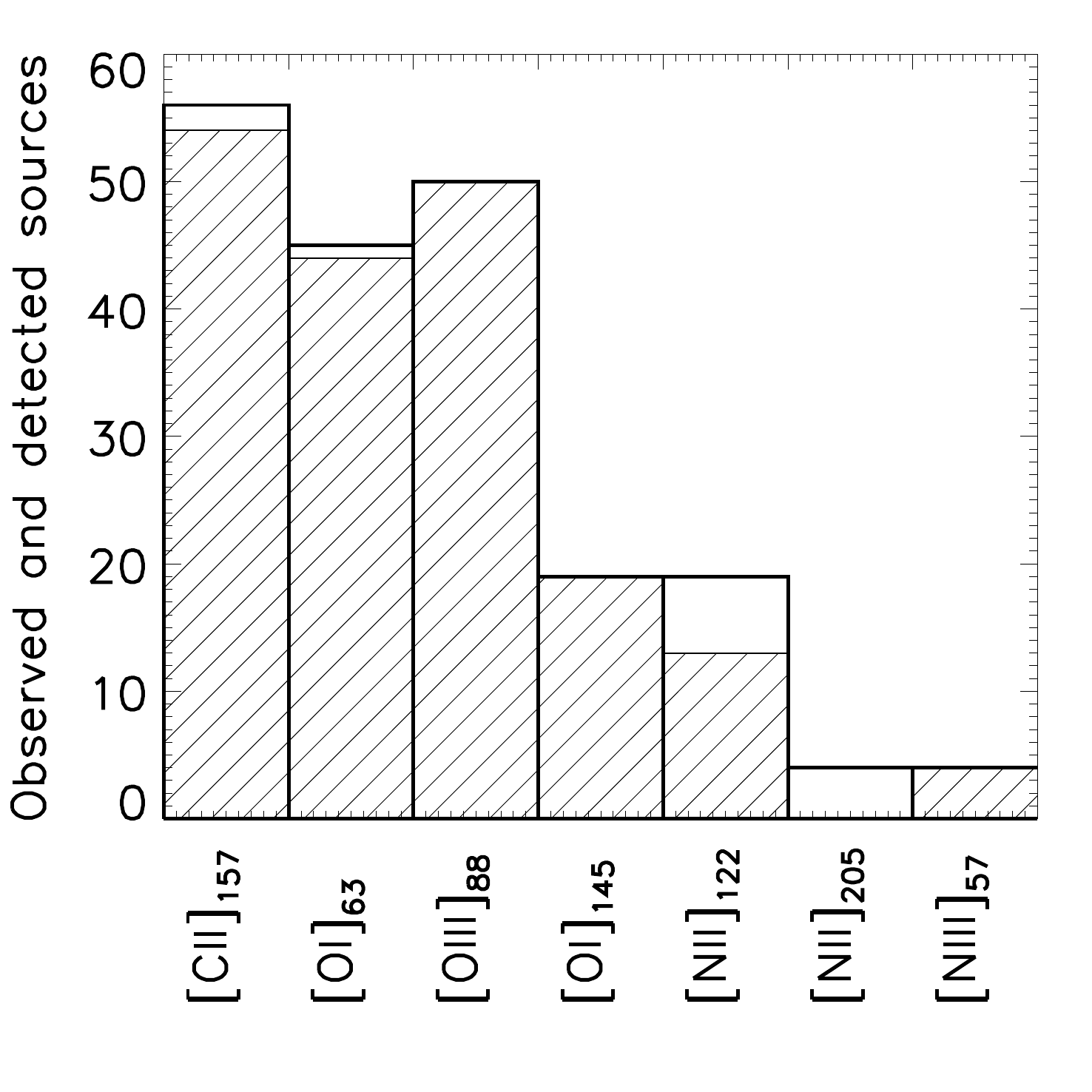}
\caption{For each PACS spectral line, the number of sources 
observed and detected (hashed) in the \hers Dwarf Galaxy Survey.}
\label{fig:obshisto}
\end{figure}

The \ciil line was targeted in all objects, along with the \oiiil and 
\oila lines for almost all objects ($\sim$85\%). 
For the brightest galaxies (30\%), the \oilb and \niila 
lines were also observed, and in a very few cases (7\%) the \niilb 
and \niiil lines as well. 
Figure~\ref{fig:obshisto} shows a histogram of how many galaxies 
were observed and detected ($>$3$\sigma$ level) in each spectral line. 
The \ciil, \oiiil, and \oila lines are the brightest, detected in more than 90\% of the 
cases, and on average are at least 10 times brighter than the other lines. 
The \nii\,205\,\mum line has only been attempted a few times but not detected 
as it is located at the edge of the spectral band, suffering from leakage.

\subsubsection{Additional sources of uncertainty}
\label{sect:incomplete}
In addition to the measurement errors and uncertainties from the PACS 
calibration, there are other sources of uncertainty to consider when 
interpreting the data. We discuss two main factors. 

\vspace{2pt}\textit{Inhomogeneous line mapping:}\\
Although we have measured the total flux within each line map, the spatial 
coverages of those maps, in a given galaxy, may not match exactly. 
To estimate whether the total line fluxes can be directly compared with each other, 
we consider the 100\,\mum PACS photometry maps as a reference for the distribution 
of the FIR emission. For each spectral line, we measure the fraction of the 100\,\mum 
emission seen inside the map coverage. We find that this fraction does not vary by 
more than 15\% for all galaxies, except for NGC\,1140 for which we consider a correction 
factor of 20\% because the \ciil map covers a larger area than the other lines. 
Hence we can confidently use the total fluxes reported in Table~\ref{table:totfluxes} 
to analyze line ratios.

\vspace{2pt}\textit{Incomplete mapping with PACS:}\\
For 11 galaxies of the $compact$ sample, the \ciil line is detected at 
a $>$5$\sigma$ level at the edges of the PACS map. The \ciil emission 
is extended but not fully mapped, hence we may be missing signal 
when reporting total fluxes. 
We coarsely estimate the flux loss by measuring the dust 
emission from the PACS 100\,\mum maps inside and outside of the area 
covered by the \ciil maps. This way, we find a missing flux fraction of 
15\% for Mrk\,153, Mrk\,1089, Mrk\,930, NGC\,1569, and of 
55\%, 35\%, 35\%, 20\%, and 60\% for NGC\,2366, NGC\,4214, 
NGC\,4861, NGC\,625, and UM\,311, respectively, and a fraction that 
we consider negligible (lower than 15\%) for the other galaxies. 
Table~\ref{table:totfluxes} reports the uncorrected fluxes, and the correction 
factor to apply if one wants to compare the line fluxes to integrated quantities.

\subsubsection{Comparison with ISO}
Three galaxies of the DGS (Haro\,11, NGC\,1569, and NGC\,625) 
were detected by ISO and reported as compact in \cite{brauher-2008}. 
The LWS and PACS fluxes in Haro\,11 agree within 20-40\%, the \oila and \oiiil 
fluxes in NGC\,1569 agree within 10\%, while the \ciil fluxes are found higher with 
PACS by a factor of 2 and 3 in NGC\,625 and NGC\,1569, respectively. 
In those galaxies (especially NGC\,1569), the PACS maps show clear 
extended structure, beyond the LWS beam. Since they were considered 
as point-sources in the LWS flux extraction and the LWS beam profile 
is more peaked at 160\,\mum than at the shorter wavelengths, this may 
explain why the \ciil fluxes recovered by PACS are higher. 
Given that mini maps were performed with PACS for those three galaxies, 
we are confident that we have better recovered their total fluxes.  
More generally, no systematic differences between the LWS and PACS 
fluxes have been established by the PACS instrument team. Therefore 
in Sect.~\ref{sect:corr}, we compare our PACS data with the LWS data 
of other galaxies as they are reported in the literature (without correction factors).

\subsection{Ancillary data}
\label{sect:prepare}
\subsubsection{\spit IRS spectroscopy}
\label{sect:irsdata}
For the modeling carried out in Sect.~\ref{sect:models}, we compiled 
mid-IR line fluxes for the \textit{compact} sources of the DGS, measured 
both with the low-resolution modules ($R\sim60-127$) and with the 
high-resolution modules ($R\approx600$) of the IRS instrument \citep{houck-2004}. 
The line fits were performed assuming a Gaussian profile and 
constraining the line width to be the instrumental broadening since 
there is no evidence of lines being resolved. 
The IRS line fluxes are provided in Appendix~\ref{app:append-c}. 
We provide in the following some details on the spectral extraction. 

The low-resolution data were retrieved from the CASSIS spectral database 
v6 \citep{lebouteiller-2011}. 
For point-sources, we used the optimal extraction spectra, which provide 
the best S/N. For compact sources, we used the tapered column extraction, 
in which the flux is integrated within a spatial window whose width scales 
with the source extent and with wavelength. While scaling the width to account 
for all the emission along the long (cross-dispersion) axis allows a better 
flux calibration for extended sources, it does not correct for the emission outside 
the slit along the short (dispersion) axis. For this reason, we have applied 
a custom flux calibration correction in which we have assumed that sources 
can be modeled as 2D Gaussians with cropped wings (representing 
the source function) convolved with the instrument PSF. The fraction of light 
falling outside the slit is then calculated for all wavelengths and for any source extent. 

High resolution spectra were extracted with two methods used by the 
CASSIS high-resolution pipeline \citep{lebouteiller-2015}. 
For point-like sources we used an optimal extraction similar to the 
low-resolution algorithm while for extended sources we used a full 
aperture extraction (integrating the flux falling inside the aperture). 
For the full-aperture extraction, we decided to use a point-source 
calibration because in all cases the source is dominated by emission 
that is not uniform across the slit and that is instead characterized 
either by a PSF profile or by a broad Gaussian-like profile. The full 
aperture extraction ensures that all the light entering the aperture 
is accounted for. While we should in principle also correct for the 
emission outside the aperture, like for the low-resolution data, the 
custom flux calibration for the high-resolution spectra of partially-extended 
sources is not available yet. We chose therefore a correction factor 
similar to what is used for point-sources in the lack of a better estimate. 

Fluxes measured in the low- or high-resolution spectra can differ. 
While the high-resolution allows in general a more reliable determination 
of the line fluxes, the low-resolution module slits are better adapted for 
extended sources. More light enters the low-resolution long slits (SL, LL) 
than the high-resolution apertures (SH, LH), so when using a 
point-like source calibration, the high-resolution line fluxes will be smaller 
than the low-resolution line fluxes. The more extended the source is, 
the larger this effect. Since the size ratio SH/SL (0.24) is larger than 
LH/LL (0.14), the effect is also larger for the long-wavelength modules. 
In Sect.~\ref{sect:models}, we use line ratios from high-resolution 
data when available and from low-resolution data otherwise. 
Line ratios of \siv10.5\,\mum/\siii18.7\,\mum, \neiii15.6\,\mum/\neii12.8\,\mum, 
and \siii18.7\,\mum/\siii33.5\,\mum agree within $\sim$30\% between 
the low-resolution and high-resolution data.

\subsubsection{Photometry and reference sample data}
In this paper, the dwarf galaxy sample is primarily compared to the 
ISO LWS extragalactic sample of FIR fine-structure lines from \cite{brauher-2008} 
(hereafter {B08}), which is one of the most complete in the literature. 
The {B08} sample contains a larger variety of galaxies than the DGS, 
with normal star-forming galaxies, starbursts, and active galactic nuclei (AGN). 
We only use their data for unresolved sources, and with clear detections ($>$3$\sigma$). 
The DGS sample extends the analysis of \cite{brauher-2008} to lower 
metallicities and less luminous galaxies, as the {B08} sample is mostly 
composed of bright objects. 
Additionally, we contrast our objects with the ULIRG samples of the \hers 
SHINING \citep{gracia-carpio-2011} and HERUS \citep{farrah-2013} surveys.

We study the behavior of the FIR cooling lines of the dwarf galaxies as 
a function of four important parameters: metallicity, F60/F100, \ltir, and \ltir/\lb. 
The broadband ratio of IRAS flux densities at 60\,\mum and 100\,\mum, 
denoted F60/F100, probes the peak of the SED 
(which is often at wavelengths shorter than 70\,\mum in our dwarf galaxies; 
\citealt{remy-2013}), and thus the dust temperature. 
\ltir is the total infrared luminosity, measured from 3 to 1\,100\,\mum. 
It reflects a large part of a galaxy's energy budget and is an indirect probe of the 
star formation activity. 
\ltir is commonly used as the proxy for the gas heating \citep[e.g.,][]{rubin-2009} 
since \ltir includes all contributions from the dust emission, and in particular 
the small grains and PAHs which are the main agents of the gas photoelectric 
heating. The ratio of the \cii and \oi lines to \ltir is a measure of the 
photoelectric efficiency, which we discuss in more detail in Sect.~\ref{sect:pe}. 
Finally, \lb is the blue band luminosity, corrected for Galactic extinction 
using the \cite{schlegel-1998} values provided by the NASA/IPAC 
Infrared Science Archive but not for internal dust attenuation, and 
converted from the magnitudes reported in \cite{madden-2013}. 
\ltir/\lb relates the stellar light processed by dust to that escaping, 
i.e. the mean extinction towards the star-forming region. 
The mean extinction is sensitive to the compactness of the cold 
surrounding medium and thus to the covering factor of PDRs. 
We note that we opt for the B-band luminosity rather than FUV in order to 
have a sufficient number of galaxies with reliable data. 

For the normal galaxies ({B08}, SHINING, and HERUS samples), 
the IRAS photometry is used to compute F60/F100 and to calculate \ltir 
using the \cite{dale-2002} formula. 
For the $compact$ sample of the DGS, we consider SED integrated 
values from \cite{remy-2015} for \ltir and observed IRAS fluxes from 
\cite{engelbracht-2008} for F60/F100 (when available). 
For the galaxies undetected or not observed by IRAS, we create synthetic 
IRAS fluxes based on the \hers photometry and SED modeling of those galaxies, 
by convolving the SED models with the IRAS response curves. 
For the $extended$ sample, we create \ltir and synthetic IRAS maps based on 
the model of \cite{galliano-2011} that we convolve to the 
largest beam (12\arcs, corresponding to the PACS 160\,\mum beam size). 
For the comparison to the spectral lines, we integrate those quantities 
over the same area mapped for each PACS line. 

Although less sensitive, IRAS being an all-sky survey, we prefer to use 
the IRAS photometry for our analysis with FIR colors rather than the \hers 
photometry to ease the comparison with the literature. We list the IRAS 
and \hers/PACS band ratios and \ltir in Appendix~\ref{app:append-d}. 
We find median ratios of IRAS60/PACS70\,$=0.86$ and IRAS100/PACS100\,$=1.05$.  
Because the method used to derive \ltir for the {B08} galaxies 
and for the dwarf galaxies is not the same (IRAS formula versus SED modeling), 
we assess its influence by comparing the results for those galaxies 
where we can apply both methods (i.e. the dwarf galaxy sample). 
We find that the difference in the resulting \ltir is small, with a median 
difference of 15\%.

\section{Observed ISM properties}
\label{sect:corr}
In this section, we present and analyze the correlations between several 
FIR tracers to identify observational trends in the sample, and link them to 
general parameters (e.g., metallicity, dust temperature). 
We focus our analysis on the integrated emission of galaxies. 
To quantify possible correlations, we compute the Spearman's rank 
correlation with the IDL procedure \texttt{r_\,correlate} only for the 
$compact$ sample, when it contains more than $8$ data pairs. 
We identify correlations at a significance level of 5\%. 
The interpretation of identified trends is supported by radiative transfer 
modeling of the photoionized and PDR gas in Sect.~\ref{sect:models}.

\subsection{Luminosity of the FIR lines}
 \begin{figure*}[thp]
\centering
\includegraphics[clip,trim=0 2.1cm 0.5cm 0,width=6cm]{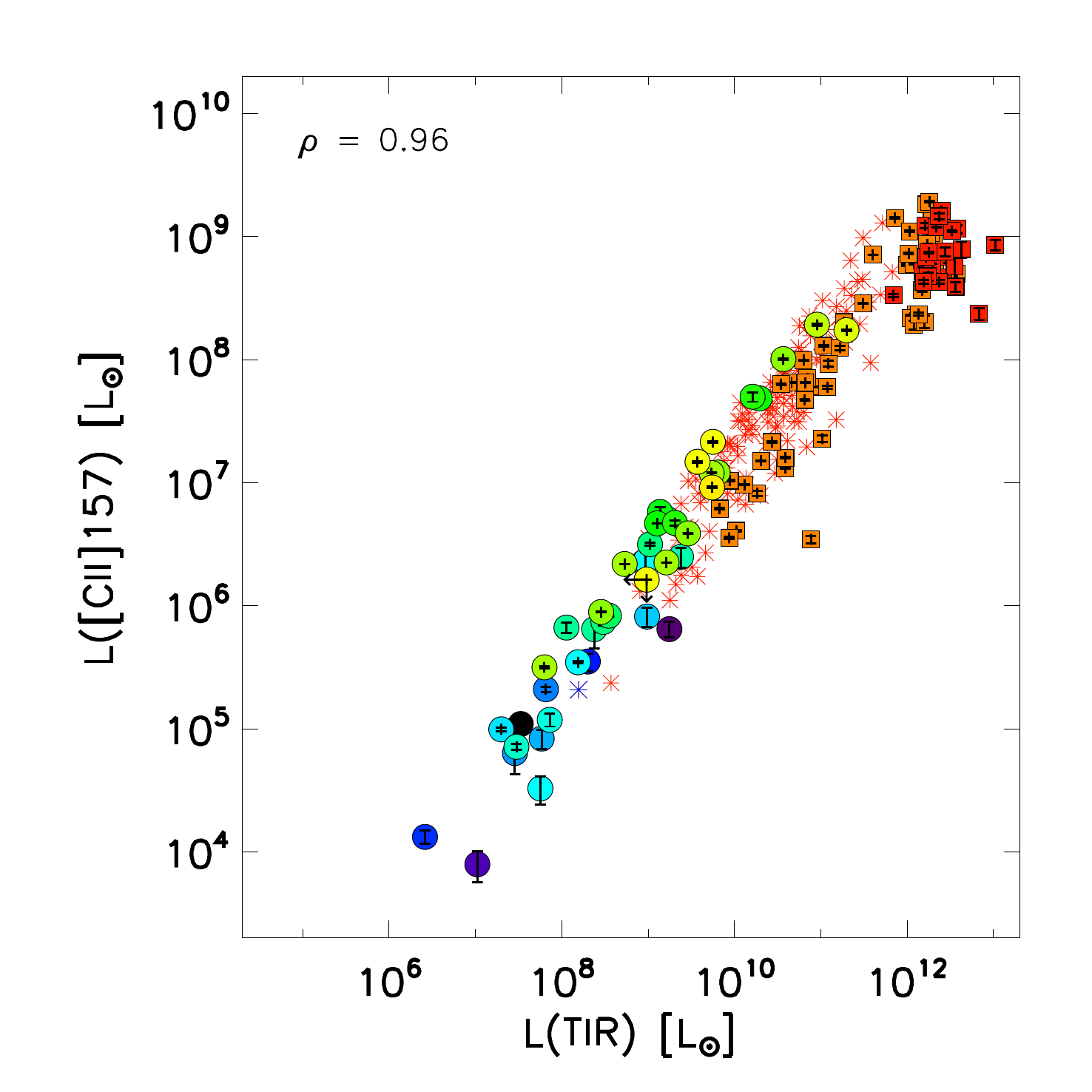} \vspace{-0.5mm}
\includegraphics[clip,trim=0 2.1cm 0.5cm 0,width=6cm]{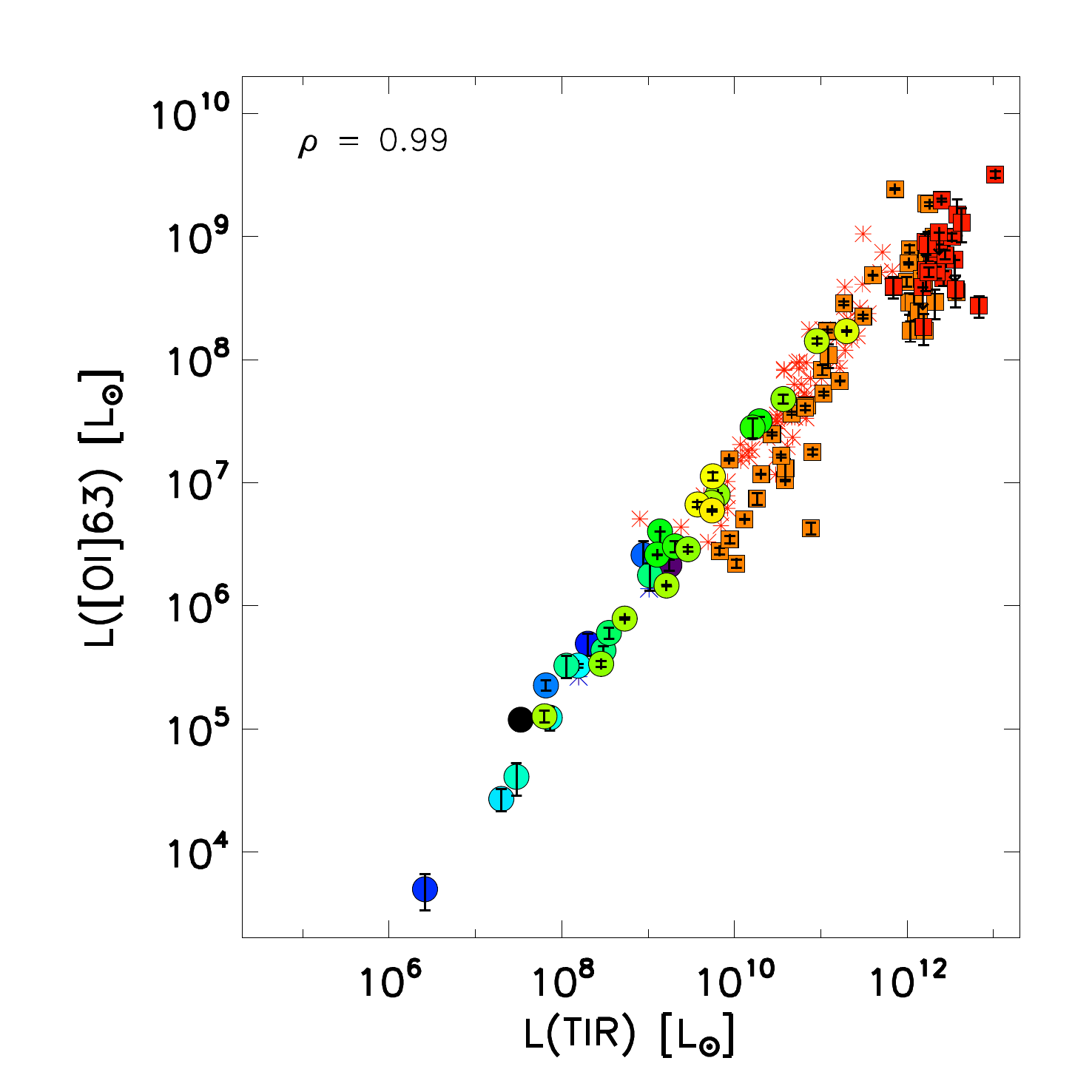}
\includegraphics[clip,trim=0 2.1cm 0.5cm 0,width=6cm]{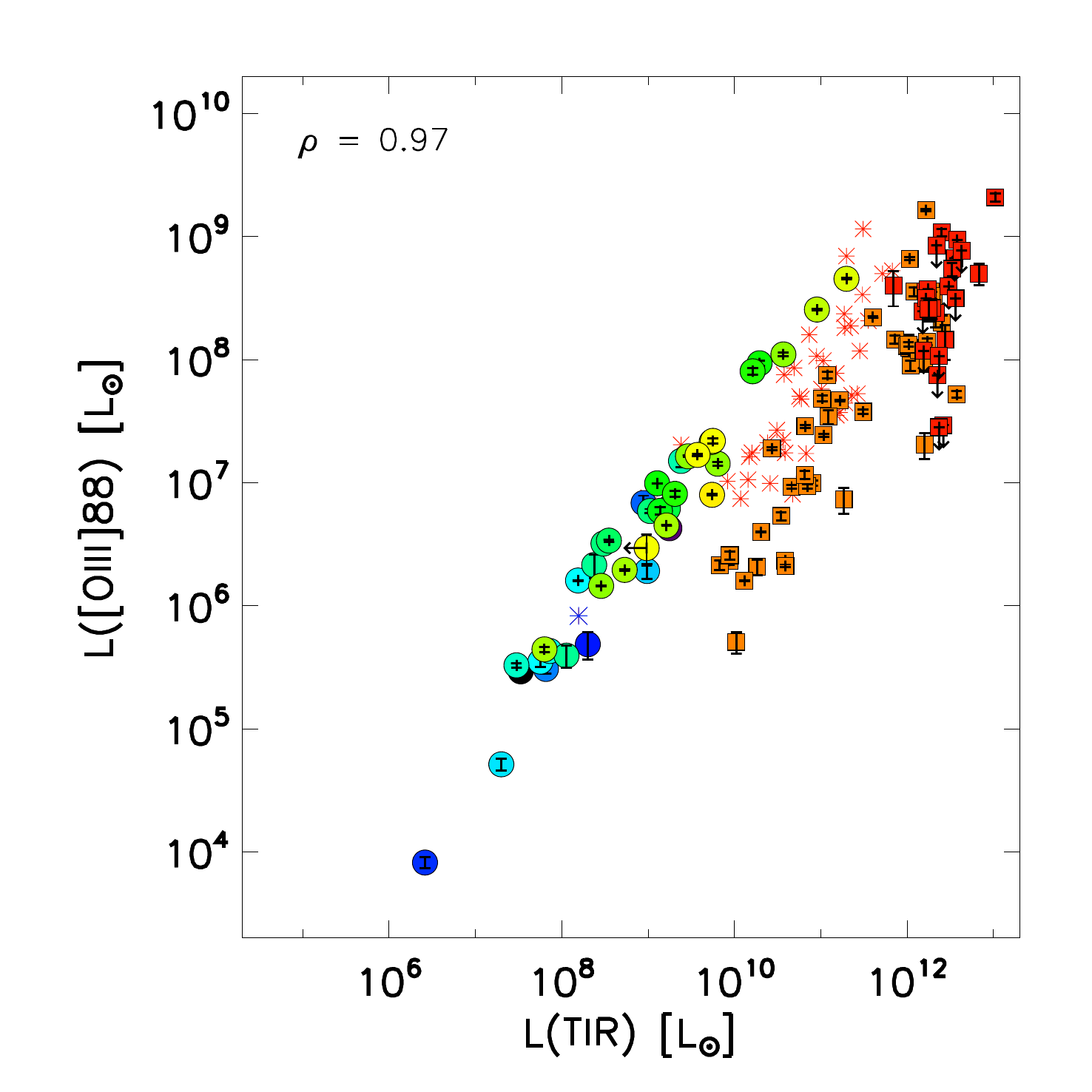}
\includegraphics[clip,trim=0 0 0.5cm 1.03cm,width=6cm]{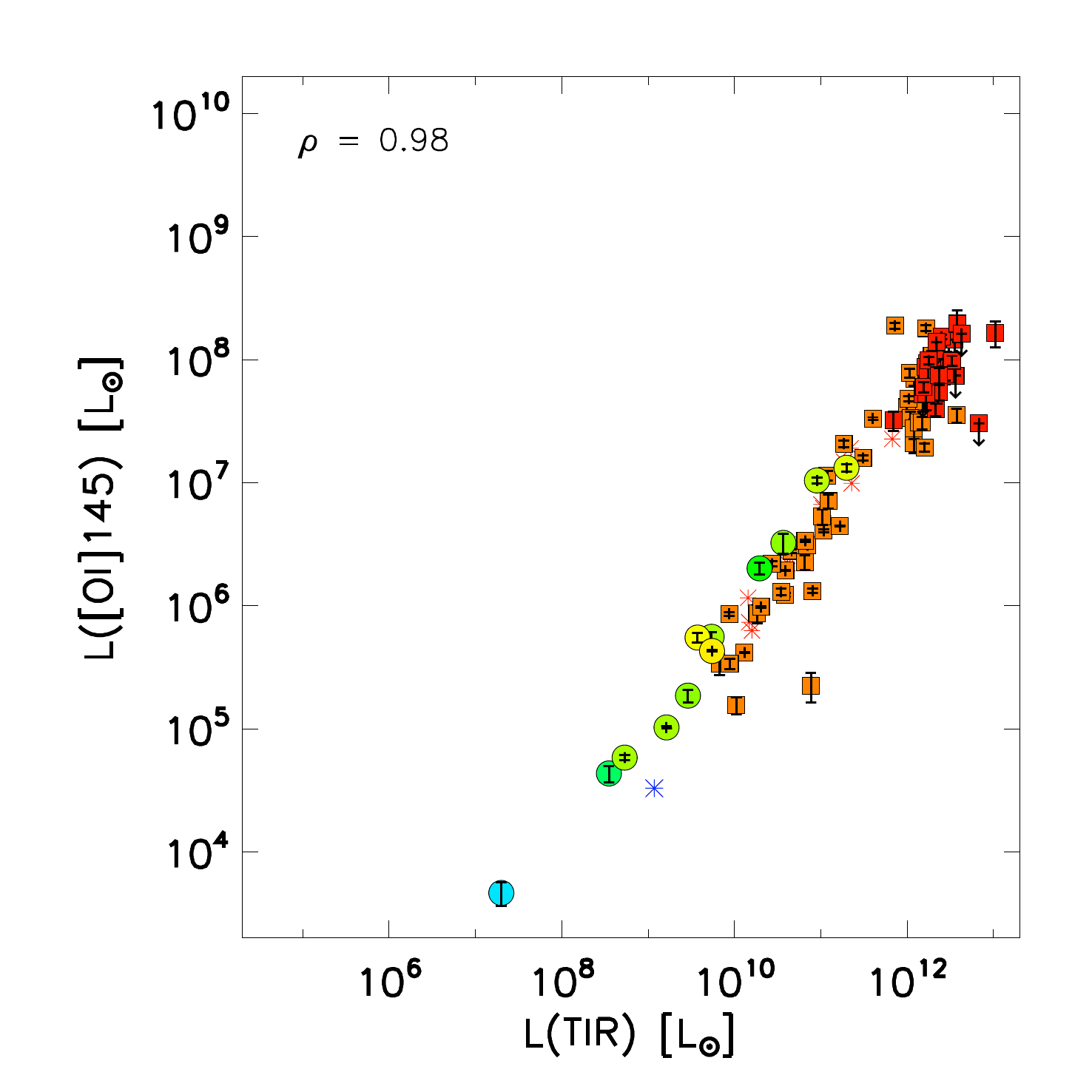}
\includegraphics[clip,trim=0 0 0.5cm 1.03cm,width=6cm]{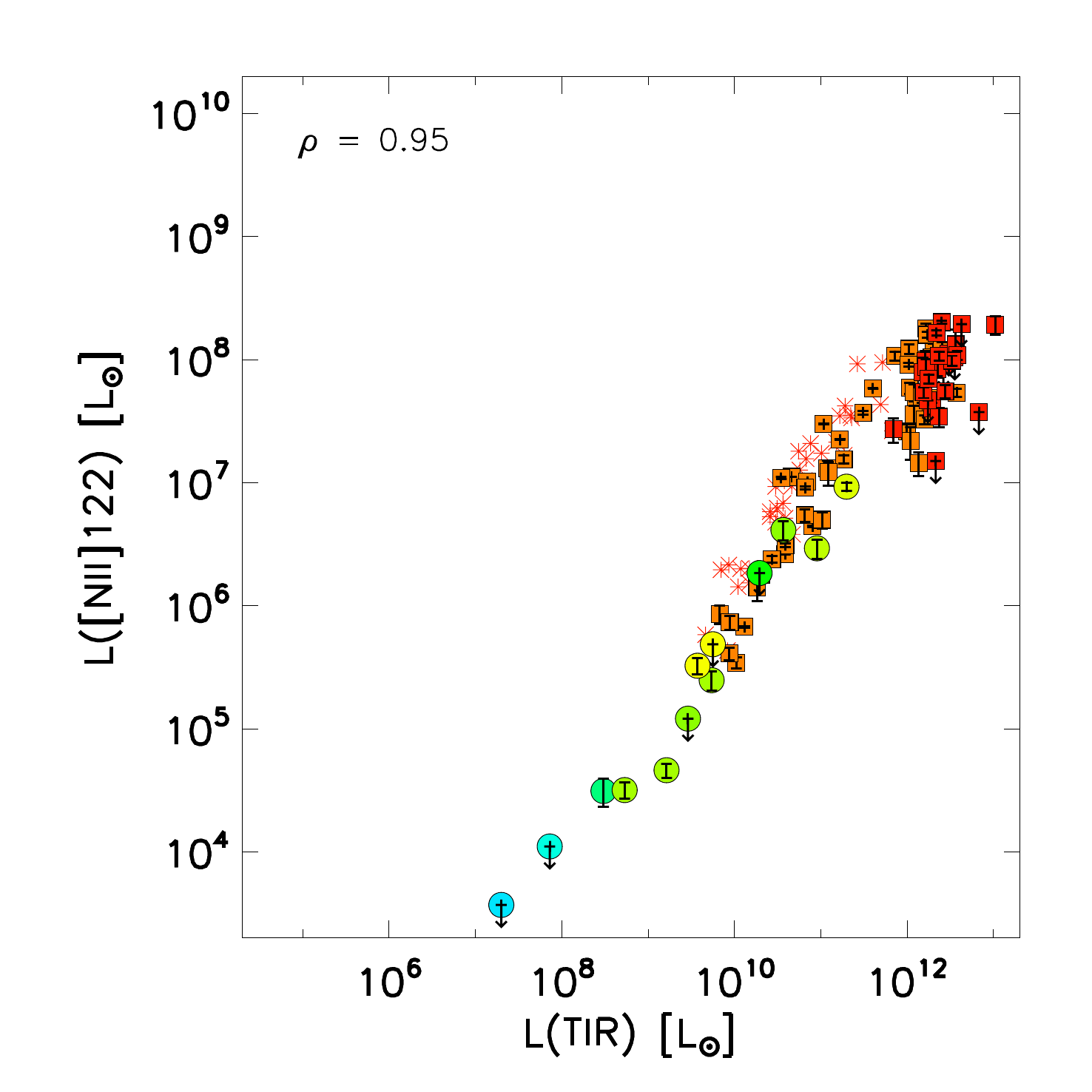}
\includegraphics[clip,trim=0 0 0.5cm 1.03cm,width=6cm]{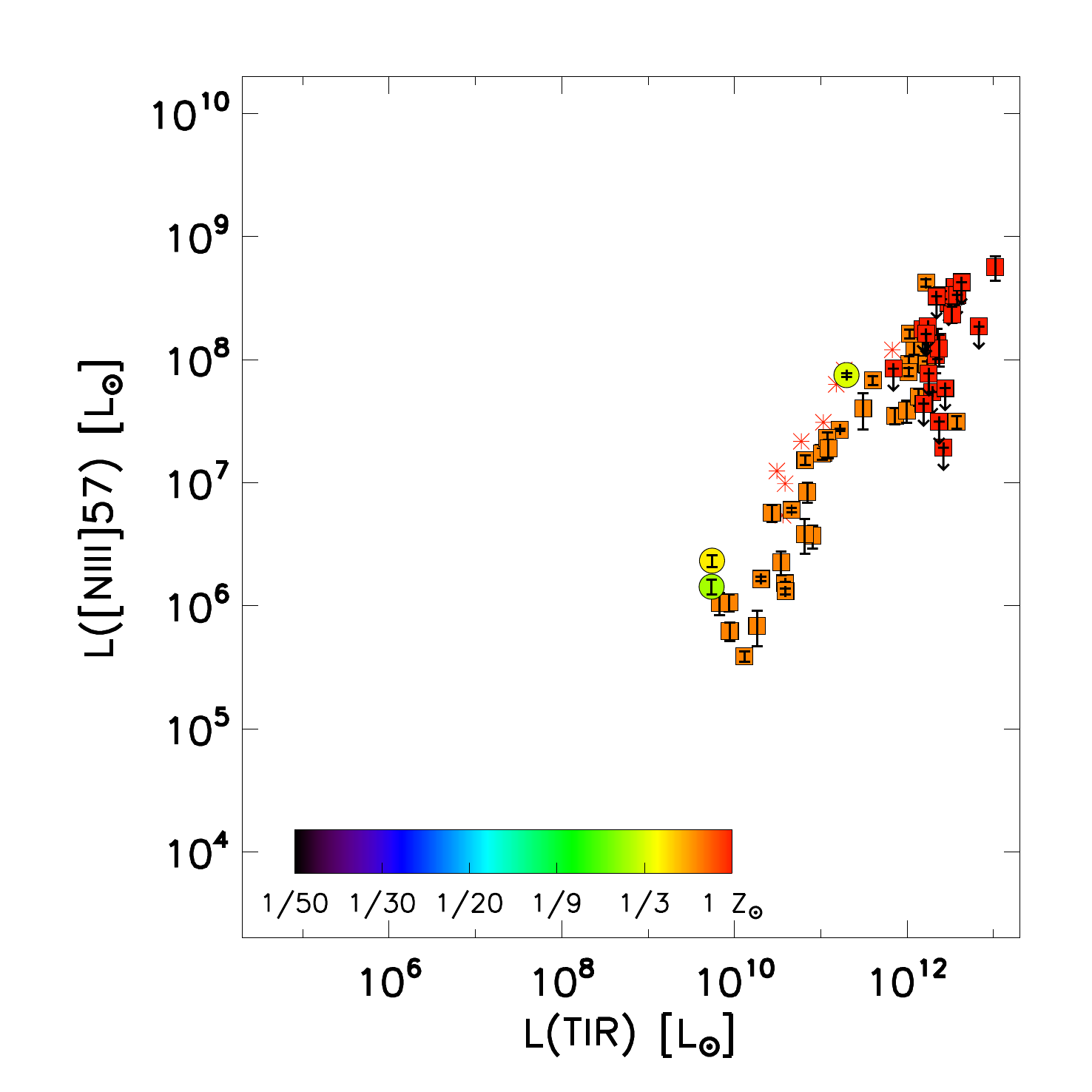}
\caption{
PACS line fluxes versus \ltir. 
The DGS galaxies are represented by filled circles ($compact$ sample) 
color-coded by metallicity. 
The red asterisks are the galaxies from \cite{brauher-2008} 
and the blue asterisks are the low-metallicity 
sources originally published by \cite{hunter-2001}. 
The orange rectangles are the HERUS data from \cite{farrah-2013} 
and the red rectangles are the SHINING data from \cite{gracia-carpio-2011}. 
Spearman correlation coefficients applied to the DGS $compact$ sample 
are indicated in the top left corner (when found above the chosen 
significance level and with more than 8 data points). 
}
\label{fig:dgs_panel1}
\end{figure*}

The \oiii88\,\mum line is found to be very bright in the DGS galaxies 
and is the brightest line in most cases. This is true, locally in extended 
sources where the peaks of emission are $\sim$10 times brighter than 
the \cii157\,\mum line (e.g., LMC-30\,Dor), and even on galaxy-wide scales. 
This is surprising, at first, as it requires 35\,eV photons to create a $\rm{O^{++}}$ 
ion while the PDR lines are more easily excited. Bright \oiii is likely related to 
the starbursting nature of the dwarfs, which host stars with higher effective 
temperatures than normal galaxies, as observed by \cite{hunter-2001}. 
The \ciil line is the second brightest FIR line and \oila is the third brightest line. 
\oila usually dominates the cooling in resolved PDRs 
\citep[e.g., the Orion Bar,][]{bernard-salas-2012}, and we find that it is often 
brighter than \ciil on peaks of emission within extended sources (e.g., SMC-N\,66), 
but fainter than \ciil on galaxy-wide scales. 

We compare the FIR line luminosities with \ltir in Fig.~\ref{fig:dgs_panel1}. 
The emission from the FIR lines is strongly correlated with \ltir 
($\rho\simeq0.97$ in the dwarfs), a result expected since all 
those quantities scale with the size, i.e. luminosity, of the galaxy. 
Those correlations being tight further indicates that the bulk of 
the FIR emission is related to star formation. Calibrating the 
FIR lines as star formation rate tracers is done in \cite{delooze-2014}. 
The dwarfs extend the relation observed, at high luminosities 
with the ULIRGs from HERUS and SHINING and moderate luminosities 
with the {B08} galaxies, to much lower luminosities, thanks to 
the sensitivity of \hers. The dispersion is smaller in the star-forming 
galaxies than in the ULIRGs, which are known to present a line deficit, 
especially visible in the HERUS ionic line fluxes \citep{gracia-carpio-2011,farrah-2013}. 
What is striking in the figure is the offset ($\sim1$\,dex) between 
the dwarf galaxies and the ULIRGs from HERUS and SHINING regarding 
their \oiiil emission. The {B08} galaxies fill the gap between the two samples. 
Moreover, we notice an offset of the DGS sample relative to the {B08} sample, 
with fainter \niila emission in the dwarfs.

\subsection{Effects of metallicity}
 \begin{figure*}[thp]
\centering
\includegraphics[clip,trim=0 2.1cm 0.5cm 0,width=6cm]{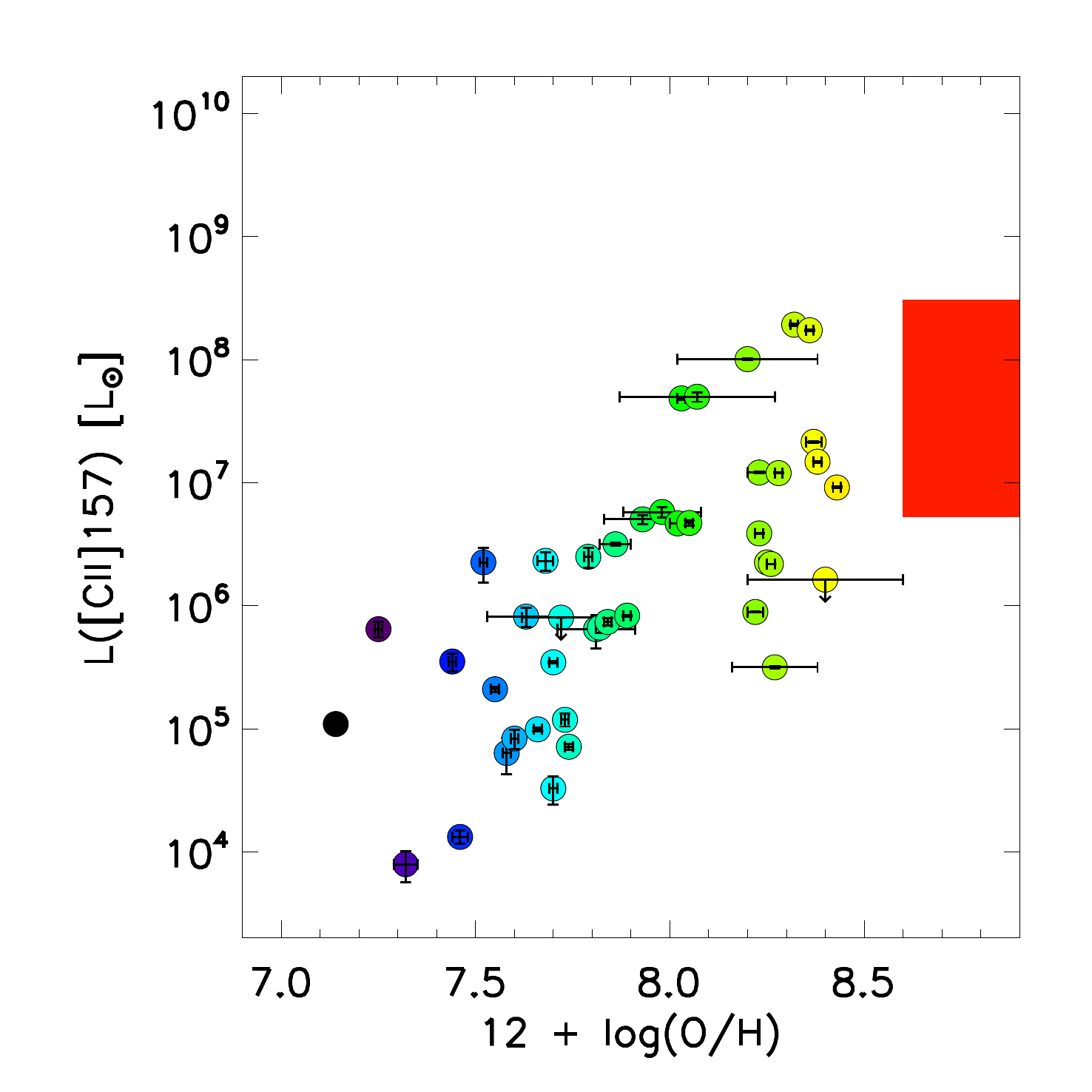} \vspace{-0.5mm}
\includegraphics[clip,trim=0 2.1cm 0.5cm 0,width=6cm]{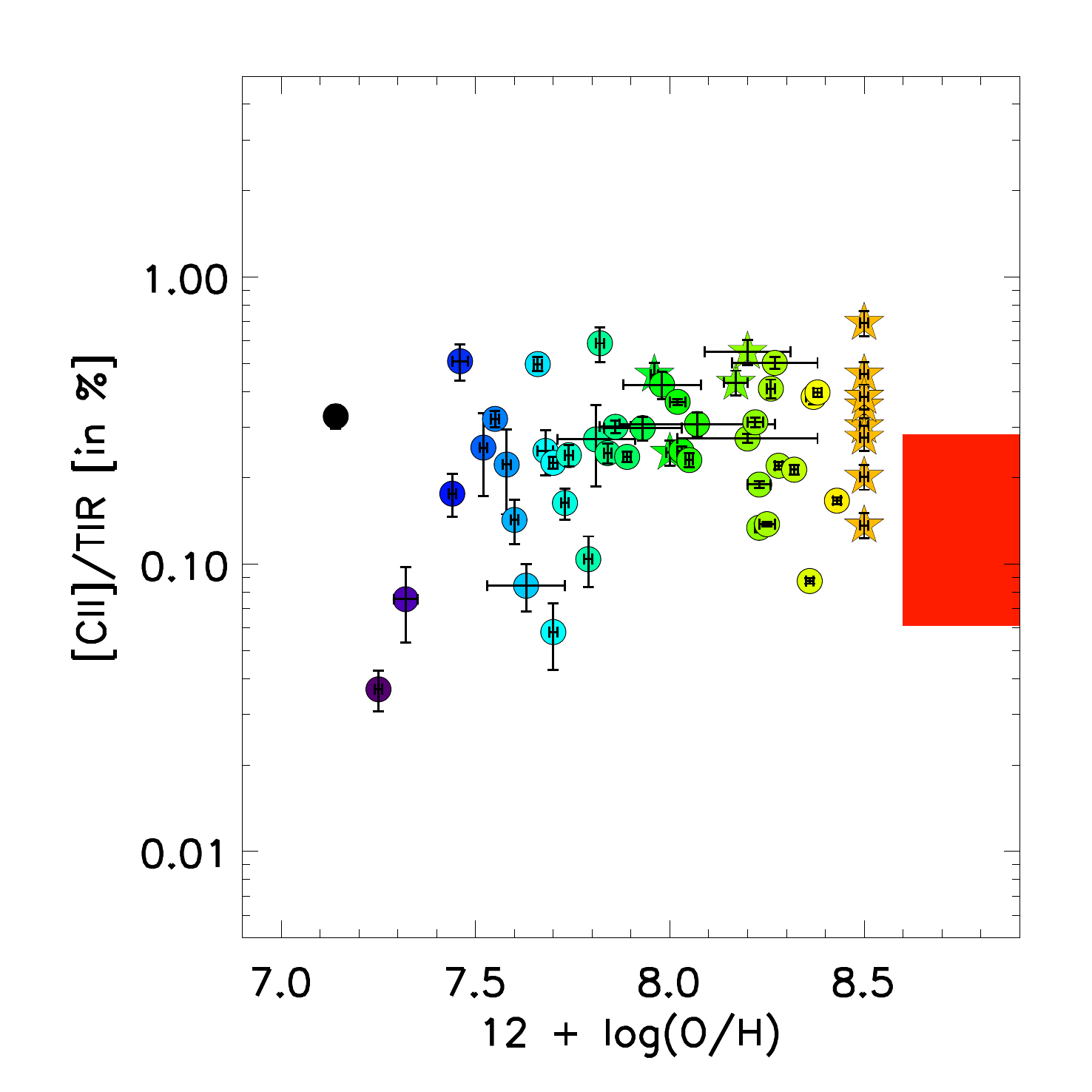}
\includegraphics[clip,trim=0 2.1cm 0.5cm 0,width=6cm]{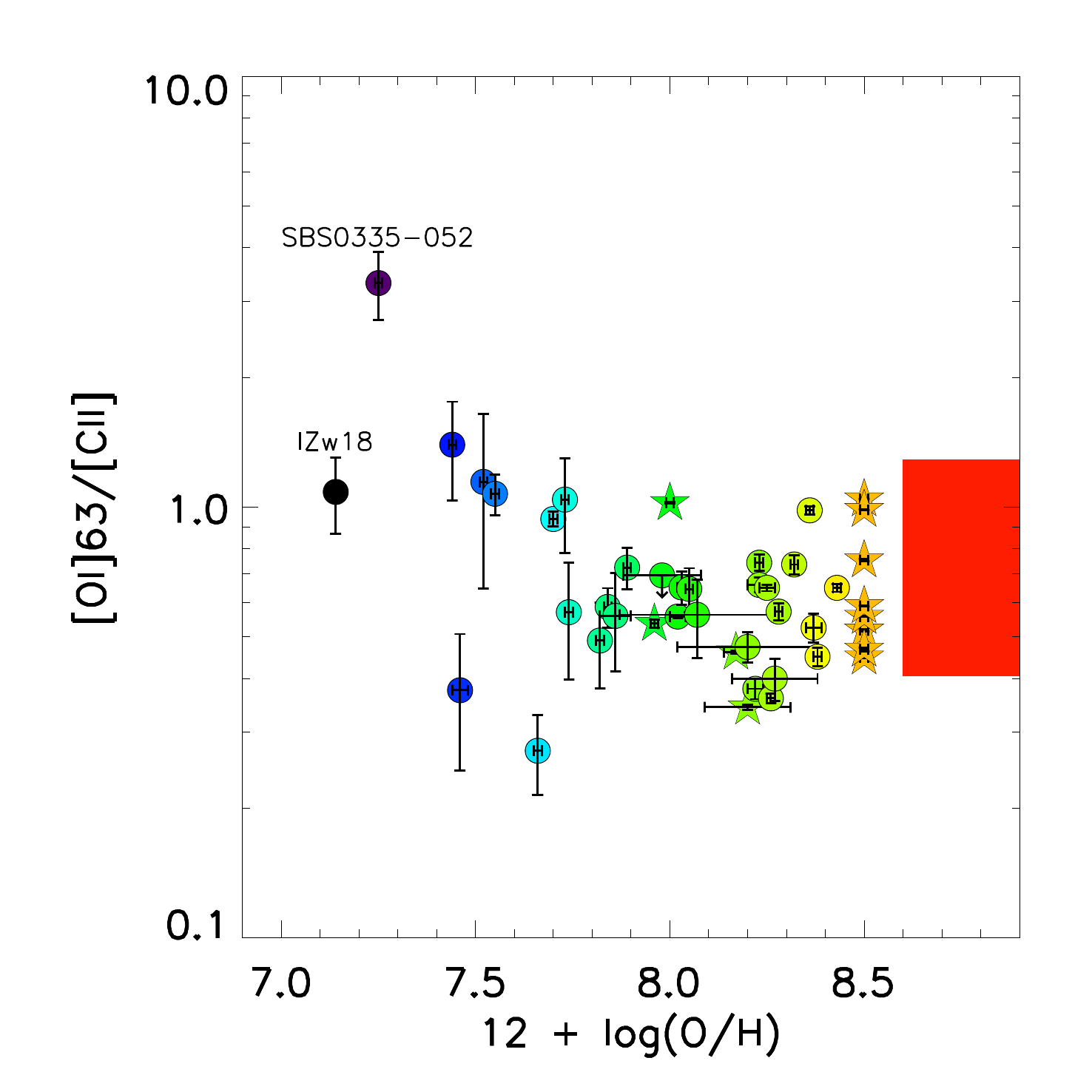}
\includegraphics[clip,trim=0 2.1cm 0.5cm 1.03cm,width=6cm]{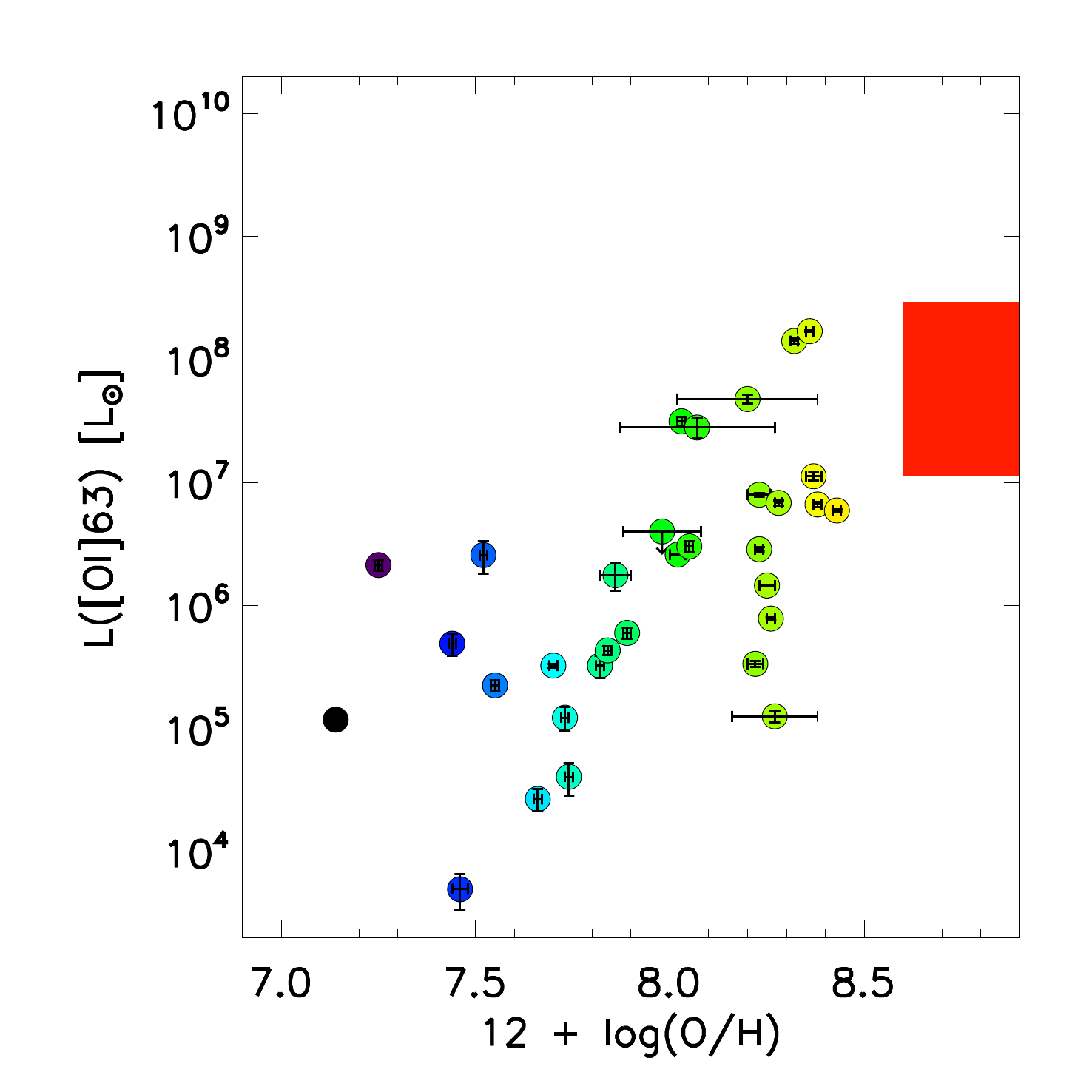} \vspace{-0.5mm}
\includegraphics[clip,trim=0 2.1cm 0.5cm 1.03cm,width=6cm]{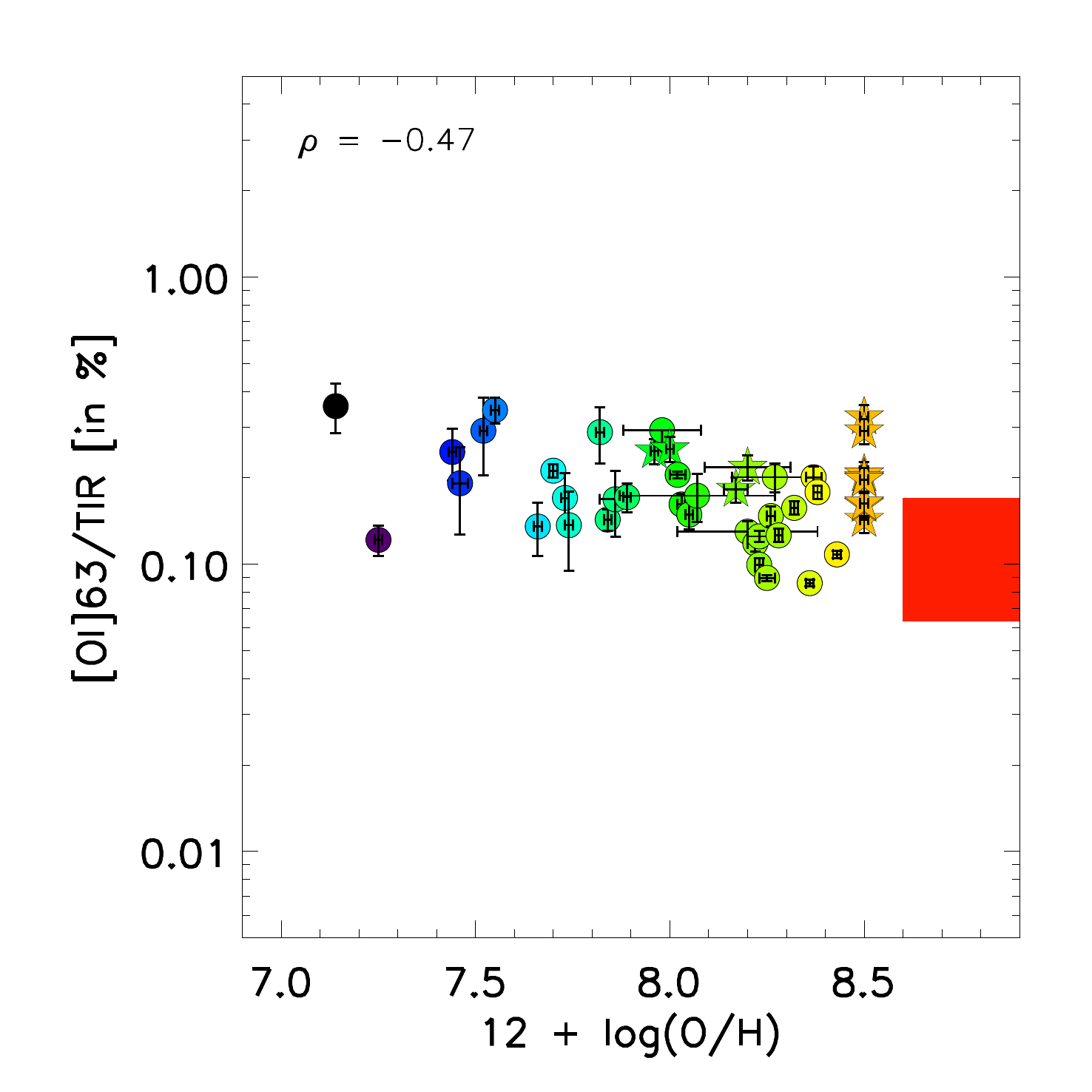}
\includegraphics[clip,trim=0 2.1cm 0.5cm 1.03cm,width=6cm]{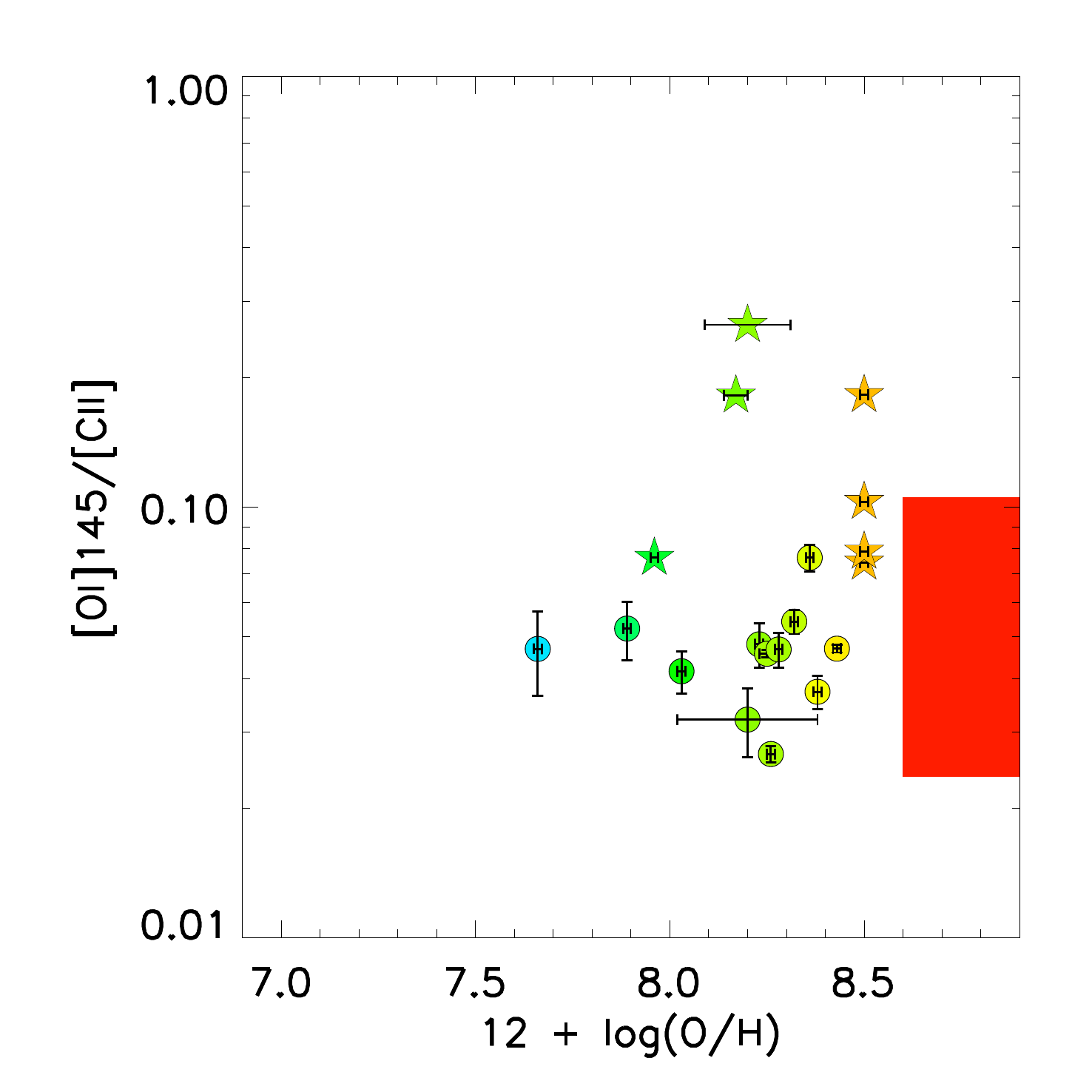}
\includegraphics[clip,trim=0 0 0.5cm 1.03cm,width=6cm]{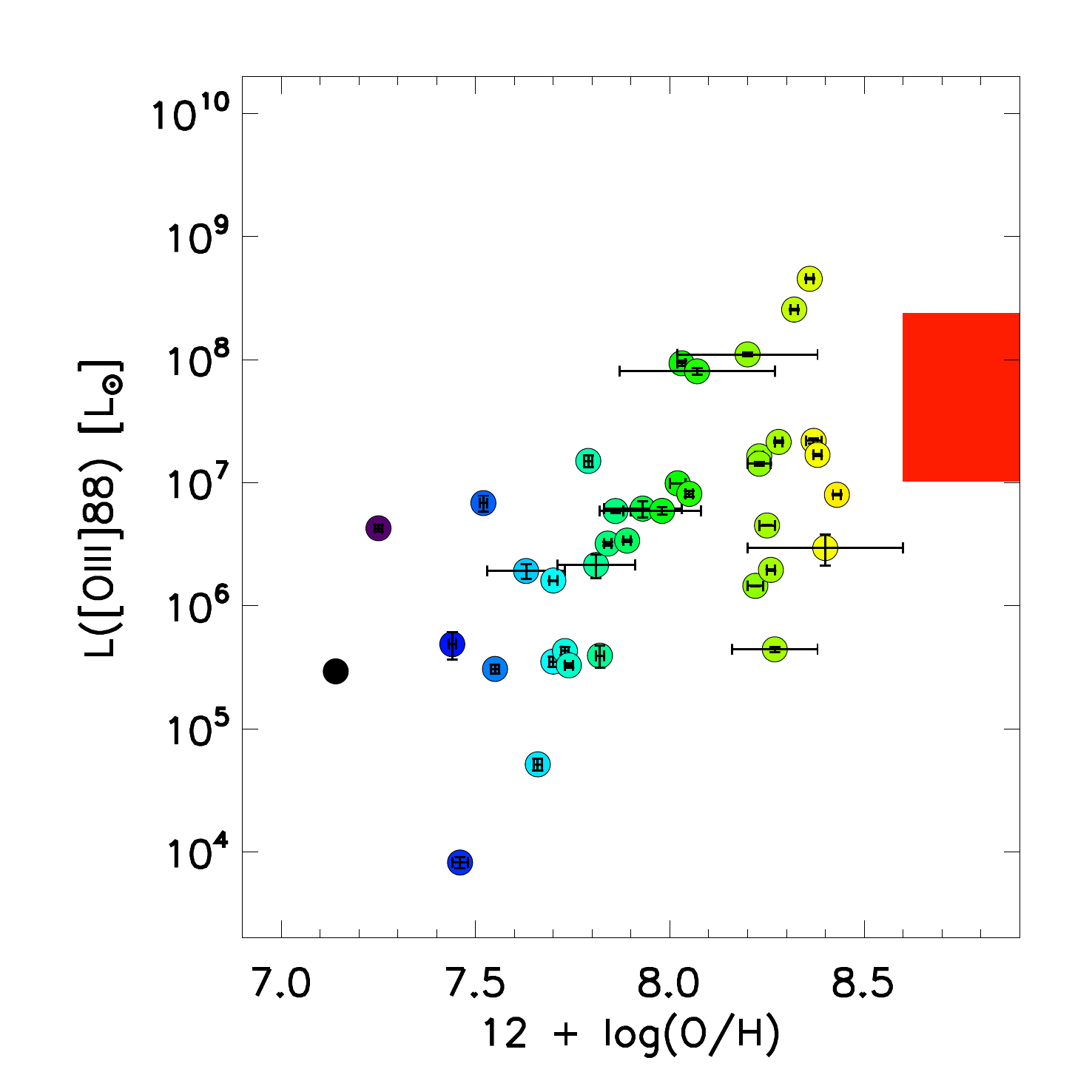}
\includegraphics[clip,trim=0 0 0.5cm 1.03cm,width=6cm]{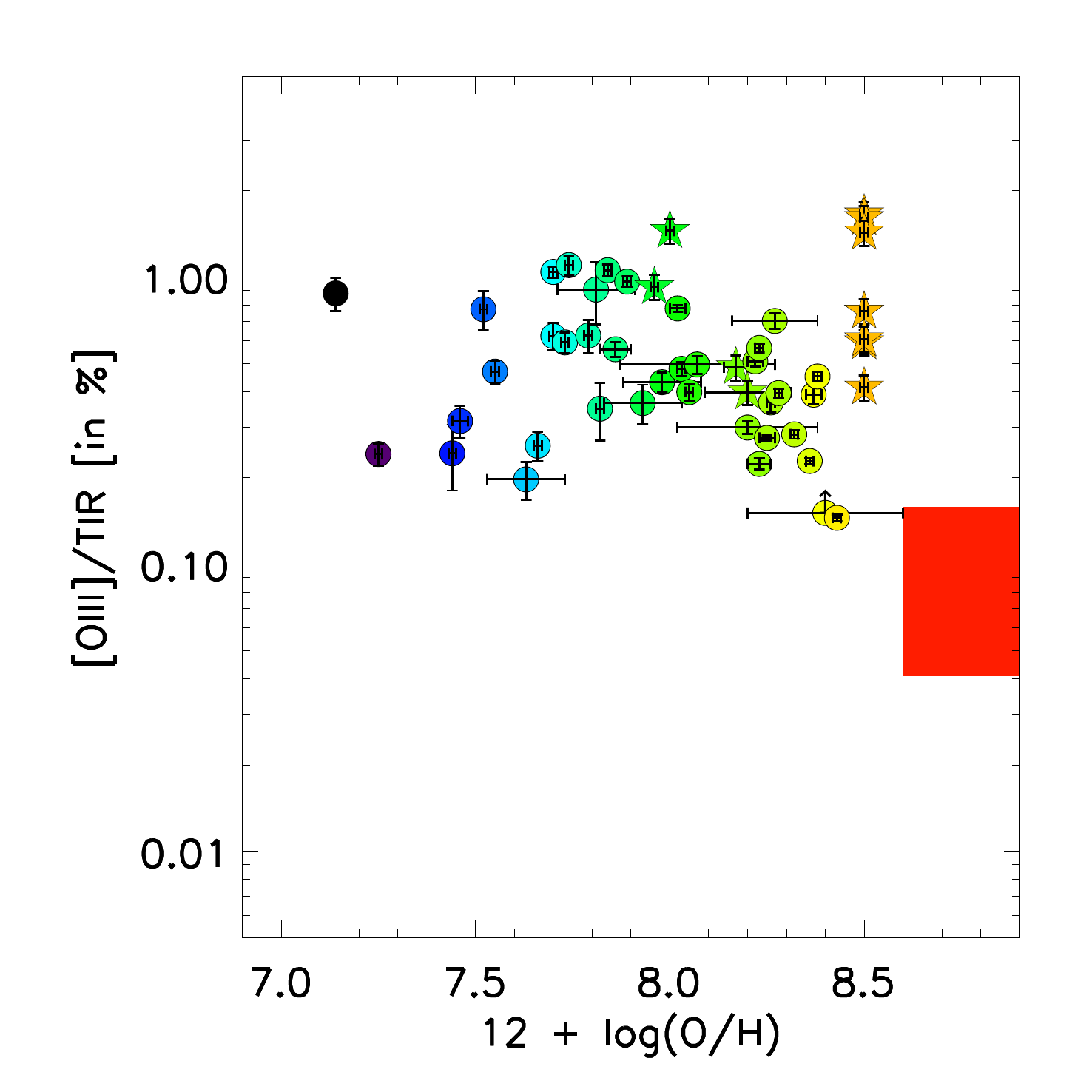}
\includegraphics[clip,trim=0 0 0.5cm 1.03cm,width=6cm]{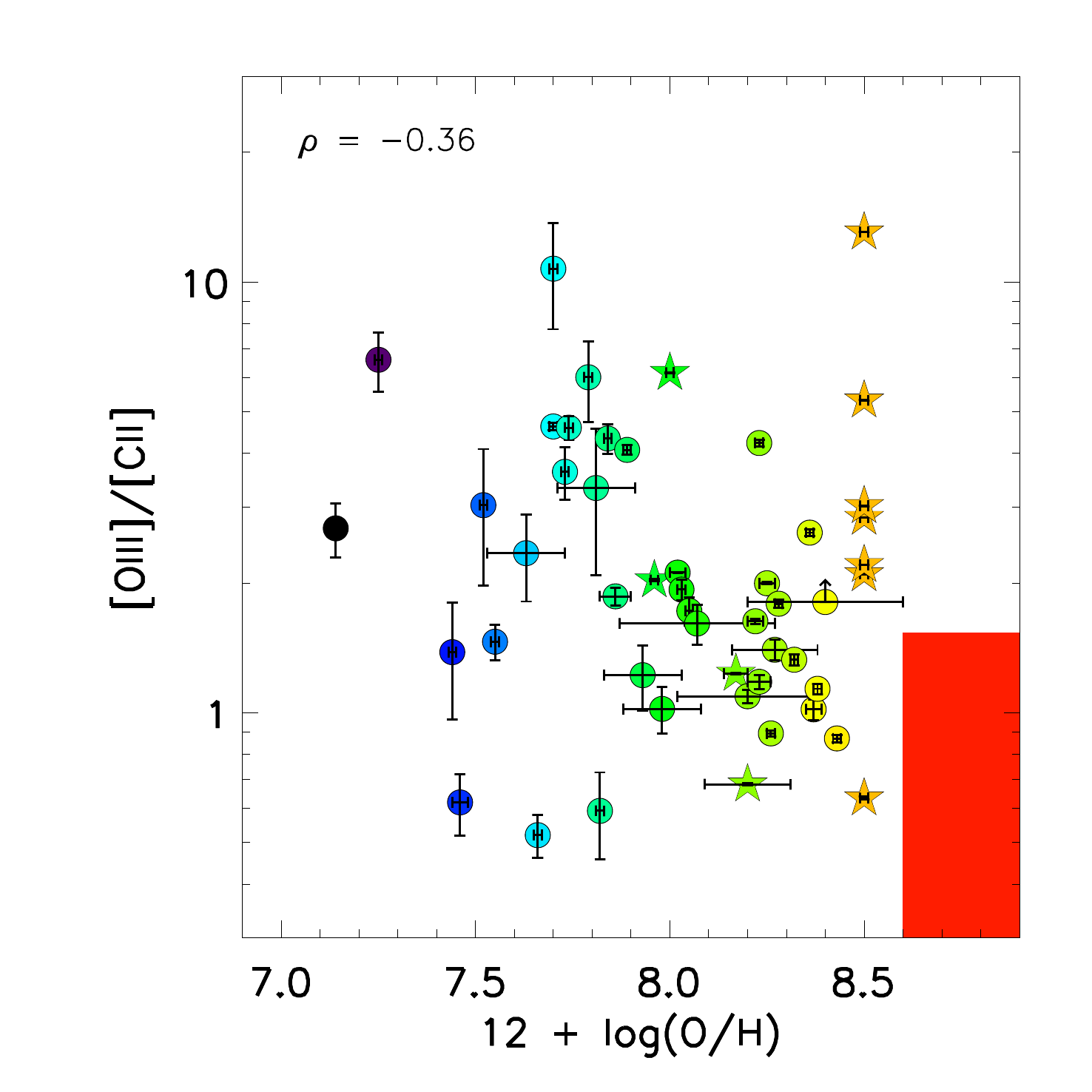}
\caption{
PACS line fluxes, ratios, and line-to-\ltir ratios as a function of metallicity. 
The DGS galaxies are represented by filled circles ($compact$ sample) 
and stars ($extended$ sample), color-coded by metallicity. 
Spearman correlation coefficients applied to the DGS $compact$ sample 
are indicated in the top left corner (when found above the chosen 
significance level and with more than 8 data points).
The red, horizontal lines show the mean values of the \cite{brauher-2008} 
sample, at metallicity around solar. 
}
\label{fig:dgs_panel2}
\end{figure*}

We correlate the FIR measurements with metallicity in Fig.~\ref{fig:dgs_panel2}. 
As expected from the metallicity-luminosity relationship, the luminosity of 
the FIR lines increases with metallicity ($\rho\sim0.65$), with large scatter 
(dispersion $\sim$0.8\,dex). 
For reference, we measure the following slope: 
$\displaystyle L{\rm (line)} \propto (12+\log{O/H})^2$. 
No clear trends are observed between the FIR line ratios or line-to-\ltir ratios and 
metallicity, except a slight decrease of \oila/\ltir ($\rho\simeq-0.5$) and \oiiil/\ciil 
($\rho\simeq-0.4$) with increasing metallicity. 
No trend is detected if we include the regions of the LMC (orange stars). 
The \oila/\ciil ratio is also found higher on average at lower metallicity 
(c.f. I\,Zw\,18, SBS\,0335-052), but the scatter is important. 
To first order, variations in metal abundances (caused for example 
by delayed injection of carbon in the gas phase compared to oxygen) are not 
responsible for the trends with \oila as they would affect the other lines, and 
in particular the \oilb/\ciil ratio, similarly. 
Optical depths effects on the \oila line (see Sect.~\ref{sect:firintro}) could reduce 
the amount of observed \oila emission and account for the trends observed with 
\oila but not with \oilb. 
However, they should also translate into a decrease of \oila/\ciil with \ltir, which 
we do not observe, and we do not have enough observations of the \oilb line 
at the lower metallicities to quantify those effects. 
Therefore, the weak trends that we observe with metallicity are likely due to 
a change of cloud conditions, i.e. increased \oila emission in denser/warmer clumps 
at lower metallicities (where cosmic rays or soft X-rays may 
become important heating sources; e.g., \citealt{pequignot-2008}). 
This also agrees with the fact that the \oila/\ciil and \oila/\ltir ratios are higher 
in the extended galaxies, particularly in the Magellanic Clouds observations  
that provide a zoomed in view on the star-forming regions and for which \oila 
is the main coolant of the dense PDR.  
We further test those hypotheses with radiative transfer models in {\sc Paper~II}. 
Overall, this indicates that metallicity is not the only regulator of the FIR line 
emission.

\subsection{PACS line ratios}
\label{sect:pacsratios}
\begin{figure*}[thp]
\centering
\includegraphics[clip,width=6cm]{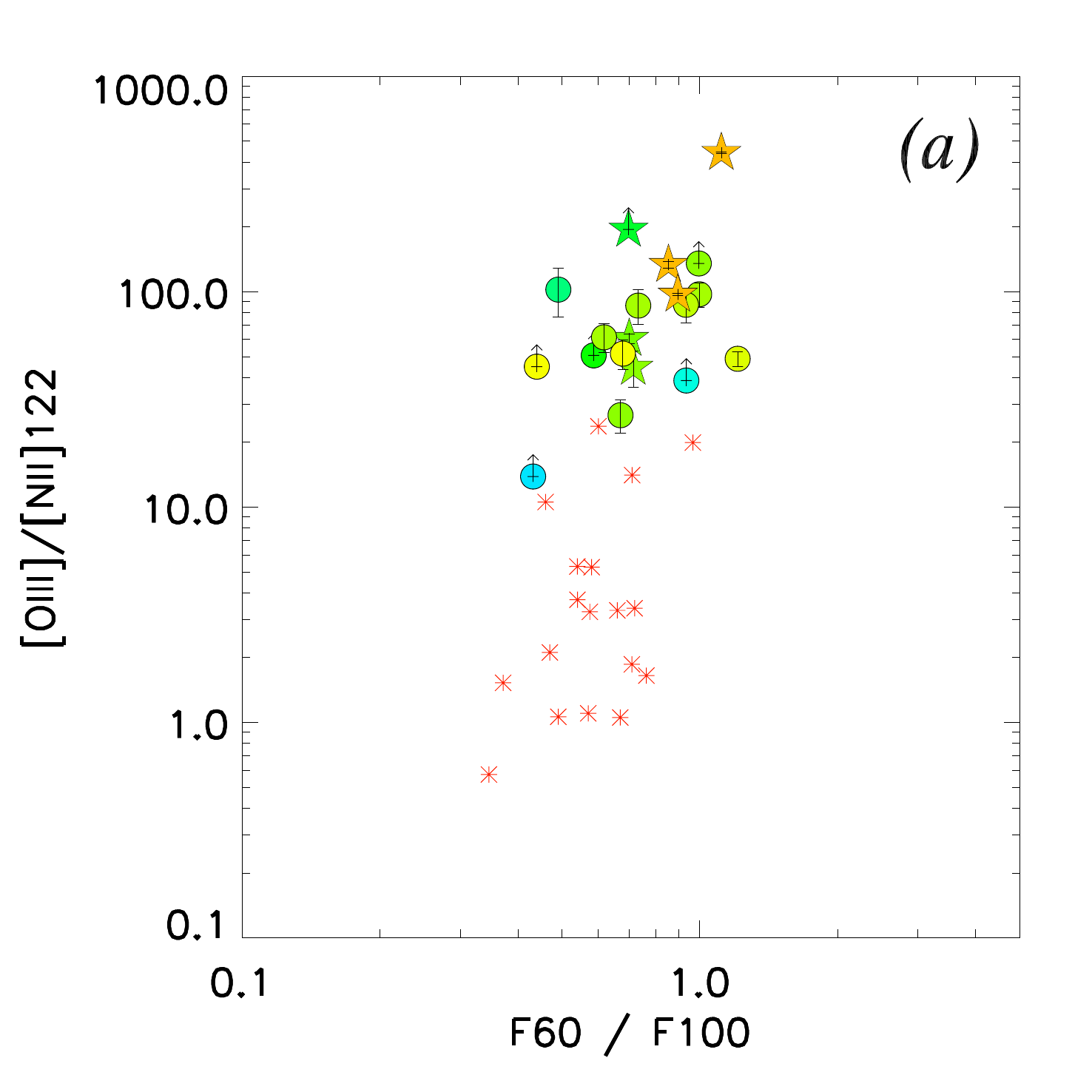}
\includegraphics[clip,width=6cm]{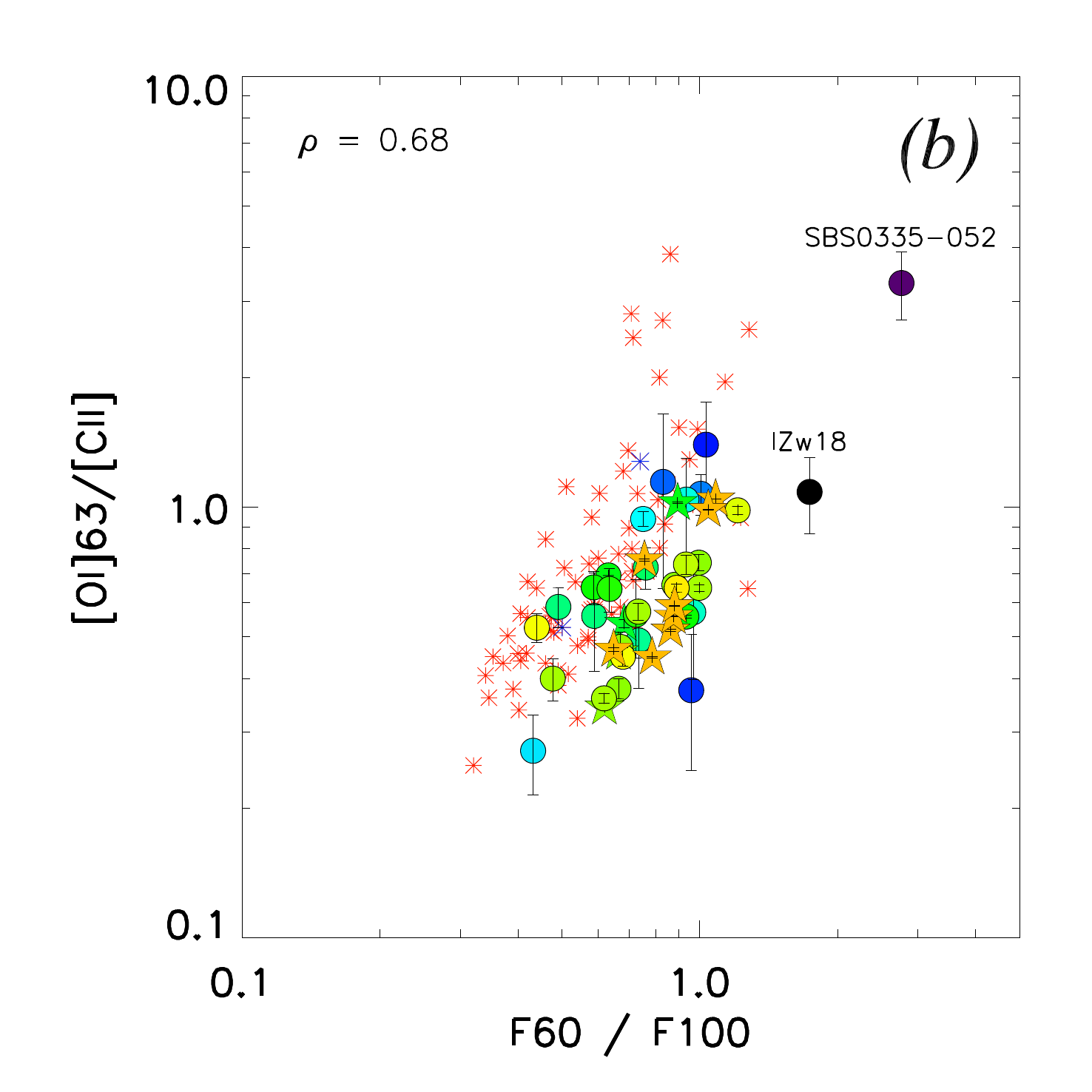}
\includegraphics[clip,width=6cm]{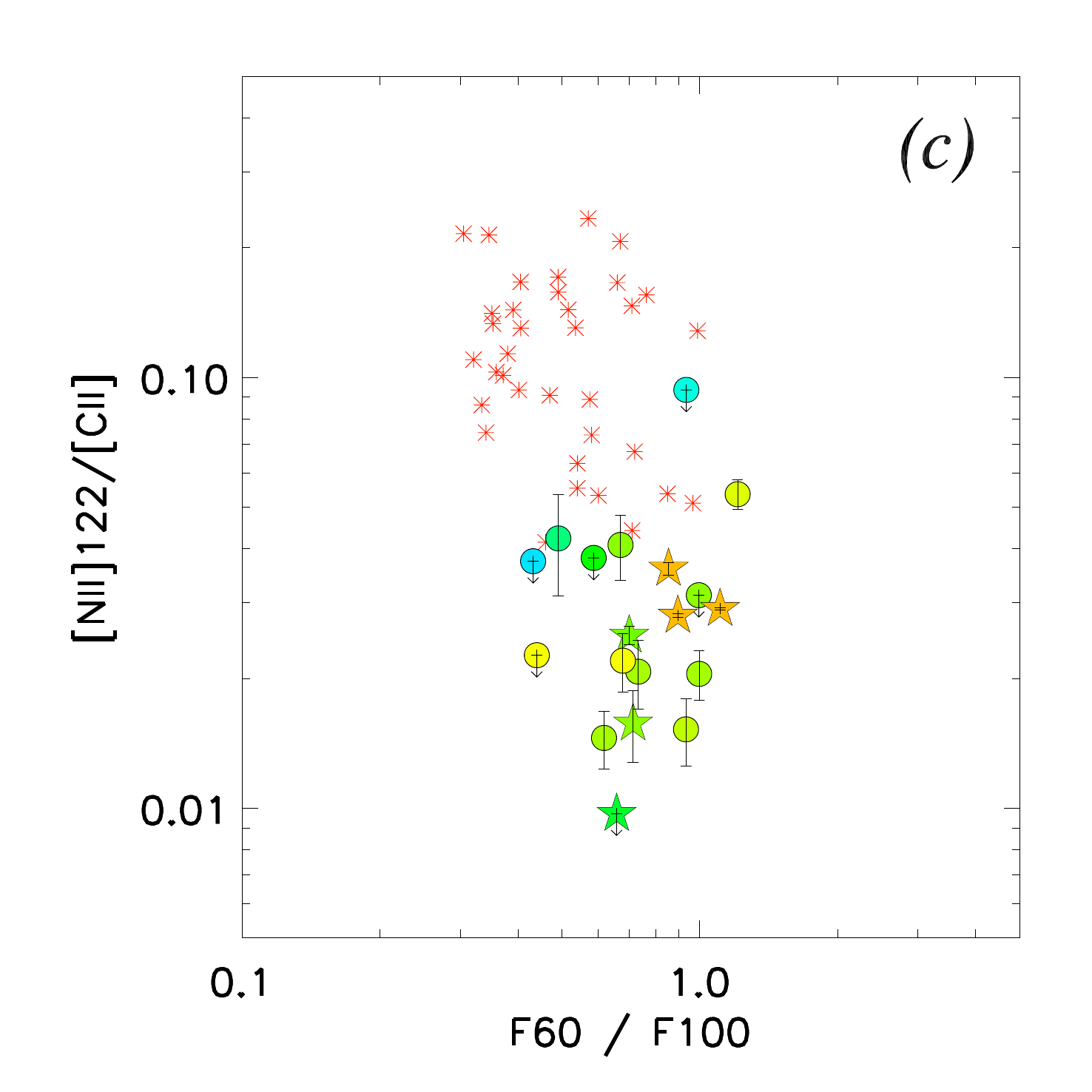}
\includegraphics[clip,width=6cm]{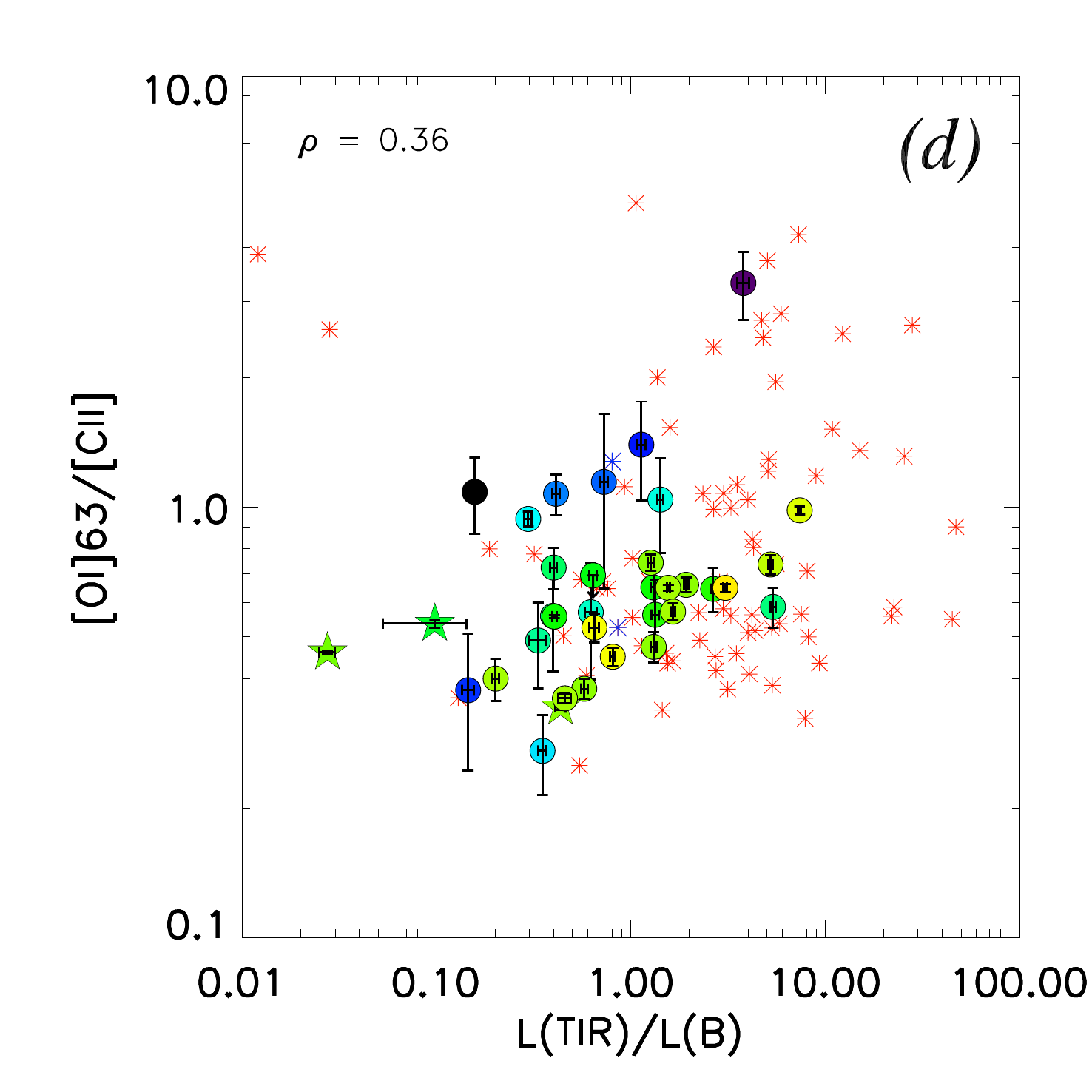}
\includegraphics[clip,width=6cm]{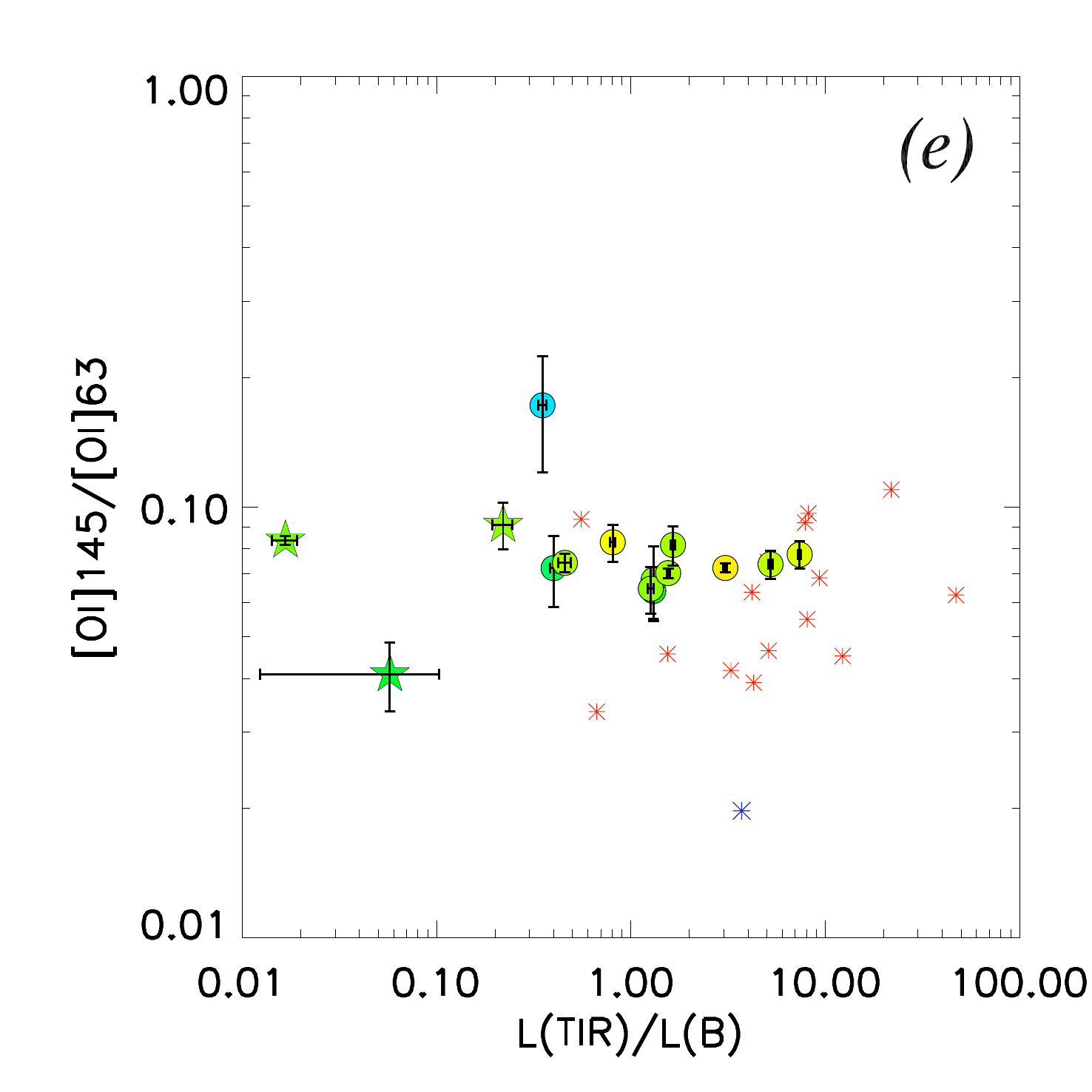}
\includegraphics[clip,width=6cm]{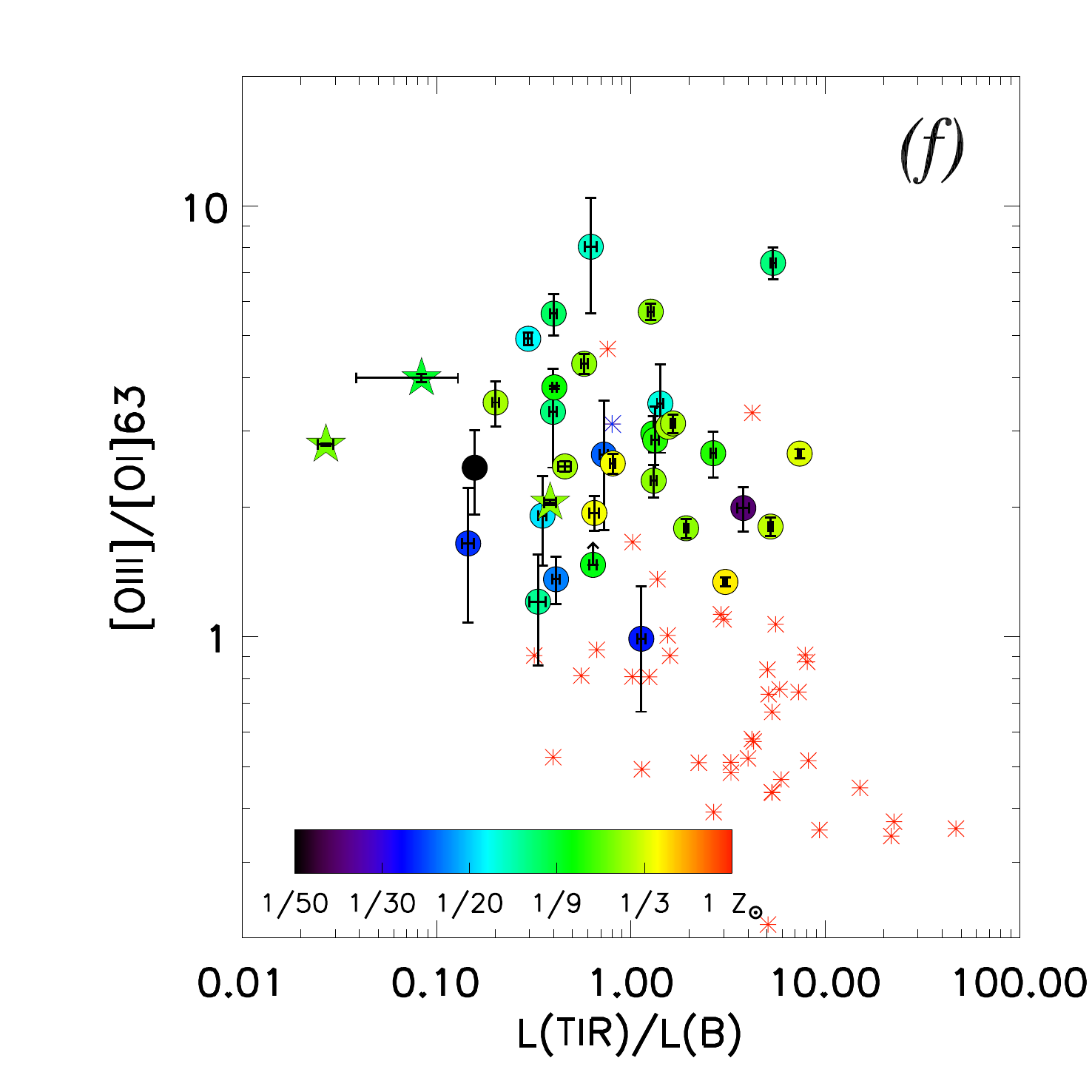}
\caption{
PACS line ratios against F60/F100 and \ltir/\lb. 
The DGS galaxies are represented by filled circles ($compact$ sample) 
and stars ($extended$ sample), color-coded by metallicity. 
See also the caption of Fig.~\ref{fig:dgs_panel1}. 
}
\label{fig:dgs_panel3}
\end{figure*}

We compare the PACS line ratios with FIR color, \ltir, 
and \ltir/\lb for the DGS and {B08} samples (Fig.~\ref{fig:dgs_panel3}). 
Median, minimum, and maximum values of the PACS line ratios are 
summarized in Table~\ref{table:dgsratios}. We indicate values for the 
complete sample (compact and extended objects), $compact$ sample 
alone, and for the {B08} dataset. 
Overall, the data clearly indicate large real variations in the line 
ratios as well as systematic offsets in the ionized gas tracers. Our dwarf galaxies 
show weak trends within the sample, but behave differently than the more 
metal-rich galaxies {\it as a group} (not with respect to metallicity/abundances).

\begin{center}
\begin{table*}[thp]\small
  \caption{PACS line and line-to-\ltir median ratios of the DGS galaxies.} 
  \hfill{}
\begin{tabular}{l c c c}
    \hline\hline
     \vspace{-8pt}\\
    \multicolumn{1}{l}{} & 
    \multicolumn{1}{c}{Full DGS sample} & 
    \multicolumn{1}{c}{$Compact$ sample$^{(a)}$} &
    \multicolumn{1}{c}{{B08} sample} \\ 
    \hline
    \vspace{-8pt}\\
    \multicolumn{1}{l}{{\it PACS line ratio}} & 
    \multicolumn{3}{c}{} \\
    \hline
    \vspace{-6pt}\\
	{\oila/\ciil \dotfill}			& $0.59^{3.31}_{0.27}$~~(0.15~dex)		& $0.65$~~(0.20~dex) 	& $0.72$~~(0.25~dex) \\
	{\oiiil/\ciil \dotfill}				& $2.00^{13.0}_{0.52}$~~(0.34~dex) 	& $1.79$~~(0.34~dex)	& $0.54$~~(0.45~dex) \\
	{\oilb/\oila \dotfill}			& $0.074^{0.17}_{0.041}$~~(0.06~dex)	& $0.074$~~(0.05~dex)	& $0.063$~~(0.25~dex) \\
	{\oiiil/\oila \dotfill}			& $2.96^{11.8}_{0.99}$~~(0.24~dex)		& $2.66$~~(0.22~dex)	& $0.74$~~(0.25~dex) \\
	{\oiiil/\niila \dotfill}			& $86.3^{442}_{26.7}$~~(0.23~dex)		& $61.5$~~(0.22~dex)	& $3.27$~~(0.37~dex) \\
	{\niiil/\niila \dotfill}			& $8.06^{9.86}_{5.75}$~~(0.13~dex)		& $8.06$~~(-)			& $1.91$~~(0.57~dex) \\
	{\niila/\ciil \dotfill}			& $0.025^{0.054}_{0.015}$~~(0.23~dex)	& $0.022$~~(0.27~dex)	& $0.12$~~(0.21~dex) \\
	{\niiil/\oiiil \dotfill}			& $0.17^{0.29}_{0.067}$~~(0.52~dex)	& $0.17$~~(0.37~dex)	& $1.08$~~(0.44~dex) \\
	{\niilb/\niila \dotfill}			& $<1$		& --		& --		 \\
    \vspace{-8pt}\\
    \hline
    \vspace{-8pt}\\
    \multicolumn{1}{l}{{\it line-to-\ltir ratio [in \%]}} & 
    \multicolumn{3}{c}{} \\
    \hline
    \vspace{-6pt}\\
	{\ciil/\ltir \dotfill}				& $0.25^{0.69}_{0.04}$~~(0.24~dex)		& $0.25$~~(0.22~dex)		& $0.13$~~(0.33~dex) \\
	{\oila/\ltir \dotfill}				& $0.17^{0.36}_{0.09}$~~(0.14~dex)		& $0.17$~~(0.14~dex)		& $0.10$~~(0.21~dex) \\
	{\oiiil/\ltir \dotfill}				& $0.50^{1.66}_{0.15}$~~(0.29~dex)		& $0.47$~~(0.26~dex)		& $0.080$~~(0.30~dex) \\
	{\oilb/\ltir \dotfill}				& $0.011^{0.023}_{0.006}$~~(0.19~dex)	& $0.010$~~(0.12~dex)		& $0.0065$~~(0.18~dex) \\
	{\niila/\ltir \dotfill}			& $0.0081^{0.0114}_{0.0028}$~~(0.21~dex)	& $0.0060$~~(0.25~dex)	& $0.016$~~(0.19~dex) \\
	{\niiil/\ltir \dotfill}				& $0.042^{0.111}_{0.026}$~~(0.31~dex)	& $0.038$~~(0.07~dex)		& $0.043$~~(0.17~dex) \\
	{(\ciil+\oila)/\ltir \dotfill}		& $0.47^{1.04}_{0.16}$~~(0.15~dex)		& $0.43$~~(0.17~dex)		& $0.26$~~(0.26~dex) \\
	{(\ciil+\oila+\oiiil)/\ltir \dotfill}	& $1.03^{2.14}_{0.40}$~~(0.18~dex)		& $0.97$~~(0.20~dex)		& $0.35$~~(0.26~dex) \\
    \vspace{-8pt}\\
    \hline \hline
    \vspace{-8pt}\\
  \end{tabular}
  \hfill{}
  \newline
  Median ratios (considering $>$3$\sigma$ detections) 
  and their dispersions in parenthesis, measured as 1.5 times 
  the median absolute deviation in logarithmic space. 
  The upper and lower script values correspond to the 
  maximum and minimum values, respectively. \\
  $(a)$~DGS sample excluding the LMC and SMC regions, 
  NGC\,4449, NGC\,6822, and IC\,10. 
  \label{table:dgsratios}
\end{table*}
\end{center}

\subsubsection{\oiiil \,/ \niila}
\oiiil and \niila are both tracers of the ionized gas, 
with similar critical densities ($\sim300-500$\,\cm), 
but different ionization potentials (35.5 and 14.5\,eV, respectively). 
It is remarkable how the dwarf galaxies occupy a completely different parameter 
space in \oiiil/\niila than the more metal-rich galaxies of the {B08} sample. 
The \oiiil/\niila ratio is observed to be high in the dwarfs, with a median 
value of $86$, which is $\sim$30 times higher than in {B08}. 
Our preferred interpretation for this high ratio is the presence of 
harder radiation fields in the dwarfs. This is supported by the fact 
that for those dwarf galaxies where we have observed \niii, the line 
is also enhanced compared to \nii. Secondly, the harder radiation fields 
naturally lead to warmer FIR color (high F60/F100), as is observed.
\oiiil/\niila is also sensitive to variations in the $N/O$ abundance ratio, 
which could be low at low metallicities. However, we find no trend of this ratio 
with metallicity (although \niila was not observed in the lowest metallicity galaxies). 

We do not find any statistically significant trend within the ensemble of 
dwarf galaxies between \oiiil/\niila and the parameters considered. With higher 
F60/F100 and lower \ltir on average, the dwarfs do not invalidate the weak 
trends observed in \cite{brauher-2008}, of \oiii/\nii increasing with FIR color 
(Fig.~\ref{fig:dgs_panel3}{\it a}) and decreasing with \ltir and \ltir/\lb. 
The fact that the \oiiil and \niila trends with FIR color are weak may indicate 
a combined effect of harder radiation fields and lower densities, with a contribution 
of low-density (high and low-excitation) ionized gas to these lines, in addition 
to the compact \hii regions.

\subsubsection{\niiil \,/ \niila}
The \niii57\,\mum line was observed in only four DGS sources 
(Haro\,11, Haro\,3, He\,2-10, and LMC-N11B), and three of them 
have both \niii and \niila detections. 
The \niiil/\niila ratios are high, with an average value of 8.1, 
which is 4 times higher than in {B08} (Table~\ref{table:dgsratios}). 
This ratio is not sensitive to variations in elemental abundances and 
an indicator of the effective stellar temperature \citep{rubin-1994}, 
as it requires 14.5\,eV to create $\rm{N^{+}}$, and 29.6\,eV to create $\rm{N^{++}}$. 
Thus higher \niiil/\niila ratios show the presence of harder radiation fields 
and are expected in dwarfs compared to normal galaxies. 

The \niiil/\oiiil ratio is on average 6 times lower in the DGS than in {B08}. 
This ratio is sensitive to ionization degree, $N/O$ abundance ratio, and 
density in the range $10^2-10^4$\,\cm (see Table~\ref{table:general}). 
Considering the high \niiil/\niila values, the data again suggest a high 
filling factor of low-density high-excitation material in dwarf galaxies.

\subsubsection{\oila \,/ \ciil}
\oila and \ciil both trace the neutral ISM, and the ratio of \oila/\ciil in the 
DGS galaxies falls in the range observed by {B08}, with a median value of 0.6. 
\oila/\ciil is not correlated with \ltir, but we see an increase of \oila/\ciil with FIR color 
($\rho=0.68$), and a weak correlation with \ltir/\lb ($\rho=0.36$) for the dwarfs 
(Fig.~\ref{fig:dgs_panel3}{\it b}, \ref{fig:dgs_panel3}{\it d}). 
Theoretical PDR studies have demonstrated that the \oila/\ciil ratio increases 
with density and radiation field strength \citep{kaufman-2006}, as the critical density 
and excitation energy of \oila are higher than for \cii, and therefore higher ratios 
are expected in compact star-forming regions (e.g., SBS\,0335-052), 
which have higher F60/F100 and \ltir/\lb \citep{malhotra-2001}.

\subsubsection{\oilb \,/ \oila}
The \oi line ratio is mainly sensitive to the density and temperature of the 
neutral gas, and to the optical depth of the cloud \citep{tielens-1985}. 
The median ratio of \oilb/\oila is 0.07 in the dwarfs. The range of values is narrow 
and a little higher than the median value found in the more metal-rich galaxies. 

We do not find statistically significant trends between the \oilb/\oila ratio and 
F60/F100, \ltir, or \ltir/\lb (Fig.~\ref{fig:dgs_panel3}{\it e}), contrary to 
\cite{hunter-2001} who noted a slight increase with \ltir/\lb (on a smaller dataset). 
The lack of correlation with FIR color is expected as the \oi line ratio stays constant 
for a constant $G_{\rm 0} /n_{\rm H}$ (provided $G_{\rm 0}\le10^4$). 
Ratios higher than 0.1 are only found in two galaxies (0.17 in VII\,Zw\,403, 
at a $3\sigma$ level, and 0.11 in LMC-N159), and are usually explained 
by noticeable optical depth effects on the \oi63\,\mum line 
\citep{malhotra-2001,hunter-2001,roellig-2006,kaufman-2006}. 
On integrated scales, optical depths are not high in the dwarfs (\av$\le$3\,mag; 
\citealt{abel-2007}).

\subsubsection{\oiiil \,/ \oila}
As observed in earlier studies of dwarf irregular galaxies by \cite{hunter-2001}, 
the \oiiil/\oila ratios are high in the dwarf galaxies, with median value 3. 
While \cite{brauher-2008} noted a slight decrease of \oiiil/\oila with \ltir/\lb, 
we find no correlation with FIR band ratio, \ltir, or \ltir/\lb within the 
ensemble of dwarf galaxies. 
The DGS galaxies as a group occupy a different parameter space than the {B08} 
galaxies (Fig.~\ref{fig:dgs_panel3}{\it f}). 

The \oiiil/\oila ratio, which is insensitive to variations of elemental abundances, 
may be used as an indicator of the filling factor of ionized gas, traced by \oiii, 
relative to the filling factor of PDRs, traced by \oi. 
In the extended sources, \oiii is detected over large spatial scales 
\citep[e.g., LMC-N11\,B, LMC-30\,Dor, NGC\,4449;][]{lebouteiller-2012,chevance-2015,karczewski-2015} 
indicating the presence of low density channels where the hard UV photons 
can travel far ($>$20\,pc) to excite the gas. The dwarf galaxies 
are probably mostly filled, by volume, with ionized gas, with densities likely 
below $\sim$510\,\cm ($n_{\rm crit}$ of \oiiil), and low filling factor of PDRs.

\subsubsection{\niila \,/ \ciil}
\label{sect:c2ion}
The \niila/\ciil ratio is on average $0.02$, $5$ times lower than in the {B08} galaxies. 
This is mostly a consequence of the high ionization fractions which 
favor the presence of $\rm{N^{++}}$ rather than $\rm{N^{+}}$. 
The \niila/\ciil ratio seems uncorrelated with FIR color, \ltir, or \ltir/\lb in the dwarfs 
(Fig.~\ref{fig:dgs_panel3}{\it c}). 
\cite{brauher-2008} also found an absence of correlation for all morphological types. 

This ratio can be used as an indicator of the \cii excitation mechanism, 
where high \nii/\cii ratios can mean that a non-negligible fraction of the \cii 
emission originates in the ionized gas and low ratios indicate that \cii 
originates in the PDR gas. 
\cite{oberst-2006} show that ionized gas alone yields at least \niila/\ciil$>$0.1. 
The conditions for \cii emission to originate in the ionized gas are low densities 
($n_{\rm crit, e^-}\sim50$\,\cm) and low ionization, as carbon would 
otherwise be doubly ionized (I.P. 24.5\,eV), i.e. the \niii emitting gas does not 
produce notable \cii emission. 
While \cite{brauher-2008} suggest that a significant fraction of the \cii emission 
in metal-rich galaxies originates in \hii regions where \nii comes from, our lower 
\niila/\ciil values favor a non-\hii region origin for the \cii line, 
from the neutral and/or very low-density ionized gas. 
From individual analyses, we have found that: 
(1)~for Haro\,11, where \niila/\ciil$=0.05$ is the highest ratio in the DGS galaxies, 
the low-density diffuse ionized gas contributes up to $\sim$40\% to the \cii emission; 
(2)~in the LMC regions, where \niila/\ciil$\simeq0.03$, \cii is associated with 
the ionized gas at $\le$15\% \citep{lebouteiller-2012,chevance-2015}; 
(3)~in NGC\,4449, \niila/\ciil$\simeq0.02$ and the fraction of \cii arising from 
the ionized gas is small \citep{karczewski-2015}. 
Therefore we estimate that, typically, less than $15\%$ of the \cii emission 
arises in the ionized gas. Although the ionized gas fills a large volume 
in the dwarf galaxies, its excitation state is too high for $\rm{C^+}$ to be present. 
Lower $N/O$ abundance at lower metallicity \citep[e.g.,][]{liang-2006} may 
change this estimate, but the fraction should remain low as we observe low \niila/\ciil ratios 
even at half-solar metallicity. \niila is not detected for metallicities $12+\log(O/H)<7.8$ 
and, where detected, \niila/\ciil does not show any trend with metallicity. 
More precise quantification of this fraction in the DGS galaxies requires 
multiphase modeling which is beyond the scope of this paper.

\subsection{FIR fine-structure line-to-\ltir ratios}
\label{sect:pacsltir}
 \begin{figure*}[!ht]
\centering
\includegraphics[clip,width=15cm]{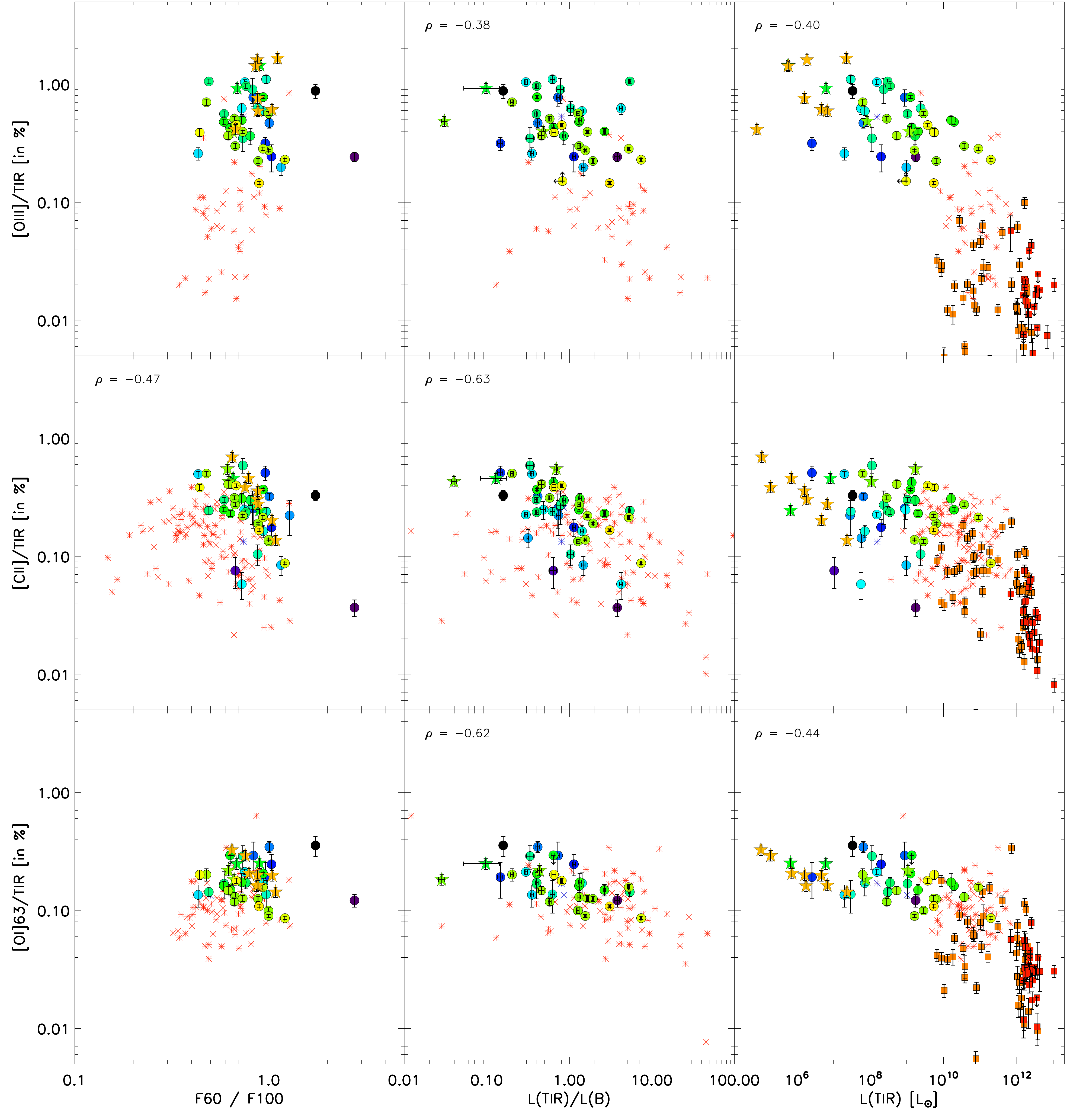}
\includegraphics[clip,width=6cm]{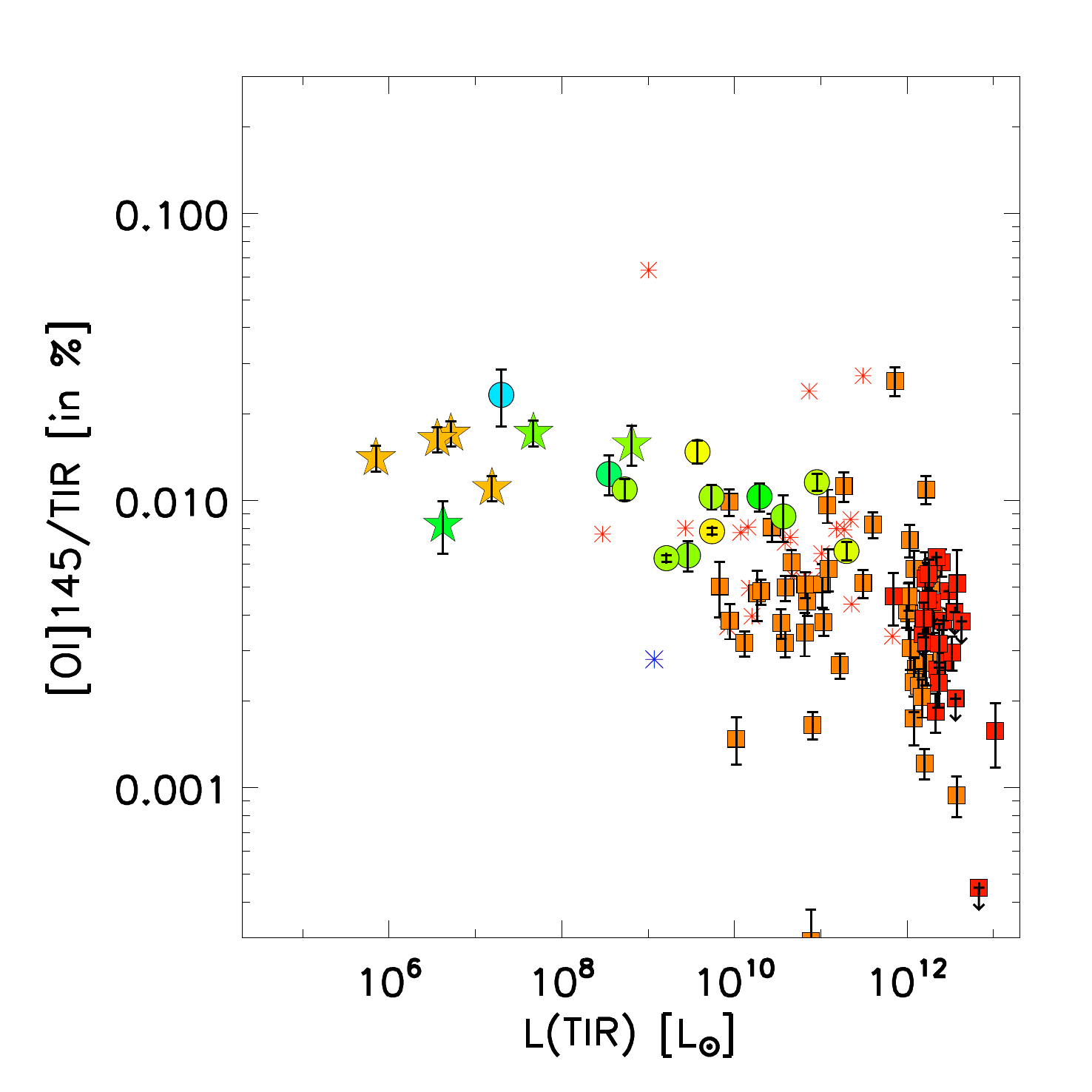}
\includegraphics[clip,width=6cm]{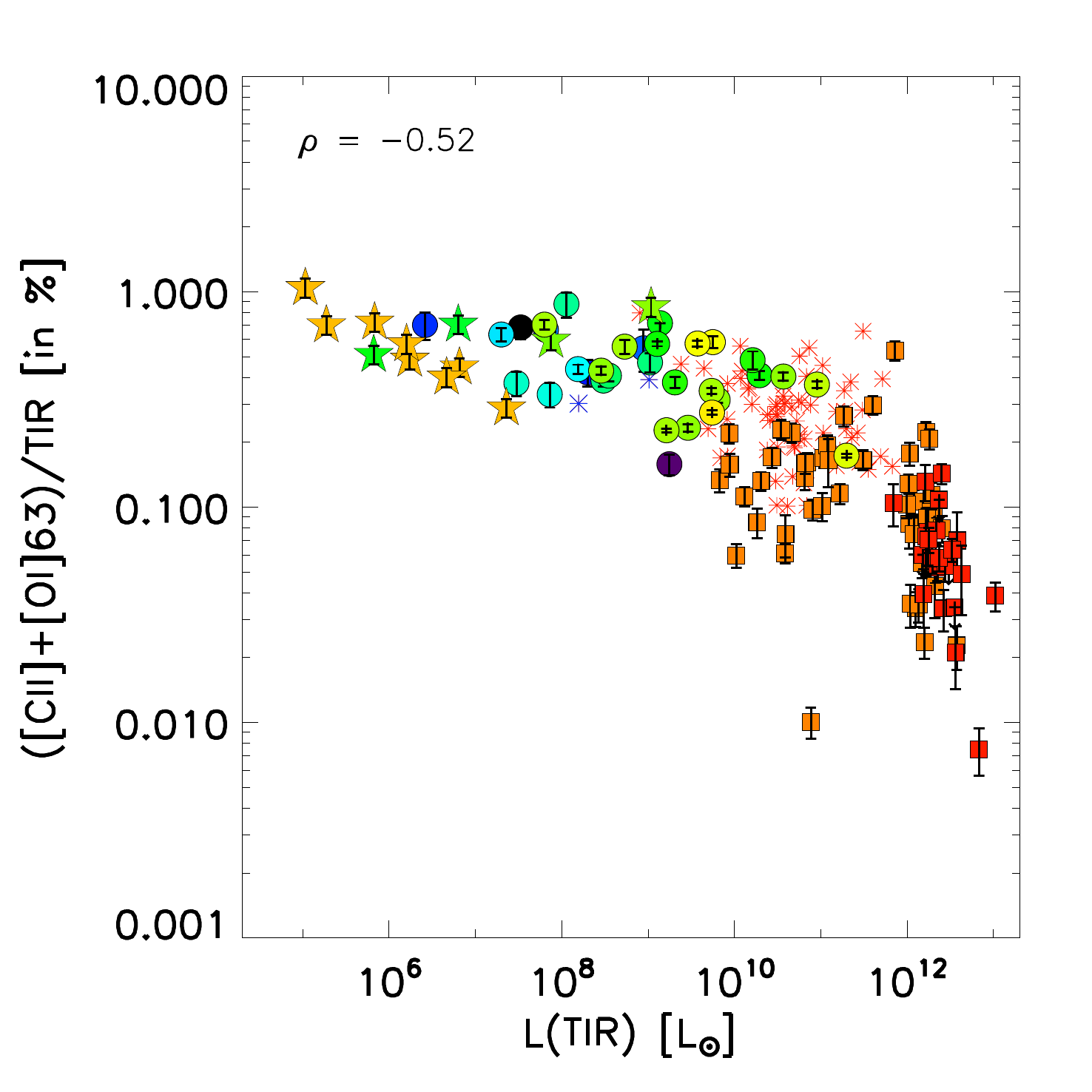}
\includegraphics[clip,width=6cm]{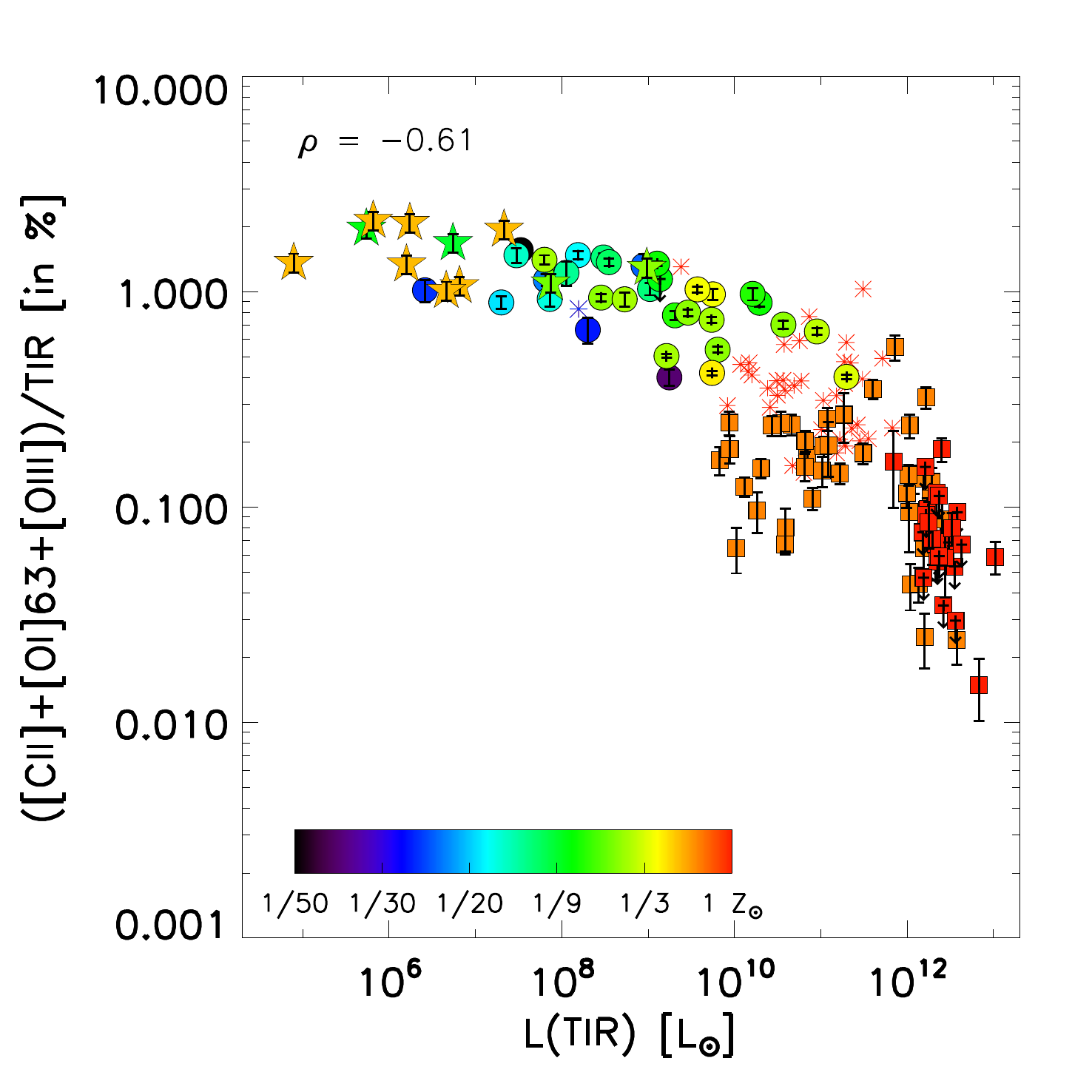}
\caption{
PACS line-to-\ltir ratios as a function of F60/F100, \ltir, and \ltir/\lb. 
Same legend as in the caption of Fig.~\ref{fig:dgs_panel1}.
}
\label{fig:dgs_panel4}
\end{figure*}

\subsubsection{General remarks}
In Fig.~\ref{fig:dgs_panel4}, we show ratios of PACS line-to-\ltir versus 
F60/F100, \ltir\footnote{Even though we plot line over \ltir versus \ltir, these 
diagrams are still useful as the x-axis is an indicator of galaxy size while 
the y-axis probes the heating and cooling physics of the ISM.}, 
and \ltir/\lb to investigate variations in the cooling by the gas and dust.

The line-to-\ltir ratios are found high in the dwarf galaxies, at least twice 
as high as in the {B08} galaxies for \cii/\ltir and \oiii/\ltir (see Table~\ref{table:dgsratios}). 
Among the highest points are the $extended$ galaxies, in which 
we have zoomed on the most active star-forming regions. 
Excitation by the young massive stars results in bright FIR lines, while 
the dust can still emit significantly further from the active regions. 
Line-to-\ltir ratios would probably be lower for the LMC in its entirety. 
In more quiescent regions of the $extended$ galaxies, line-to-\ltir ratios are 
lower in general. This is especially the case for \oiii and \oi, and less for \cii 
because of the contribution from diffuse emission to the total \cii flux. 
\cite{rubin-2009} estimate \cii/\ltir$\simeq$0.45\% for the entire LMC, which is 
close to our average $extended$ sample value. 
High ratios for the compact objects may therefore indicate that star-forming 
regions clearly dominate the overall emission of the galaxy.

\subsubsection{\oiiil \,/ \ltir}
\oiiil/\ltir values span an order of magnitude, from 0.1 to 1.7, with average 
value of 0.50\%. The dwarf galaxies are at the high end of the {B08} sample. 
The dwarfs also extend the trend observed in {B08} of \oiiil/\ltir decreasing 
with increasing \ltir and \ltir/\lb, which is weaker for the dwarfs ($\rho\simeq-0.4$) 
than for the {B08} galaxies (Fig.~\ref{fig:dgs_panel4}). 
This suggests enhanced \oiii emitting volume when the escape fractions are high. 
However, there are multiple possible interpretations of those correlations. 
\oiii/\ltir could also be reduced by a significant fraction of the \ltir emission 
unrelated to star formation. However similar behavior is observed in \oiii/\oi 
(see Fig.~\ref{fig:dgs_panel3}{\it f}). Because both of these lines 
are directly related to star formation, we rule out this interpretation. 
It is also possible that \ltir/\lb does not directly measure the covering factor 
of the star-forming regions if there is a significant contribution from 
field stars to the B-band luminosity. We have tested this possibility 
by making similar diagrams for galaxies with available K-band (field star dominated) 
and FUV (young stars) data. 
We find that \oiii/\ltir is actually more strongly correlated with \ltir/$L_{\rm FUV}$ than 
with \ltir/\lb and uncorrelated with \ltir/$L_{\rm K}$, which favors an interpretation 
in terms of UV escape fraction. Thus the correlations are best explained by 
the combination of high stellar effective temperature on galaxy-wide scales, 
resulting in high \oiii emission, and enhanced ISM porosity with lower \ltir and 
\ltir/\lb, which in turn allows UV photons to excite $\rm{O^{++}}$ on larger 
spatial scales. 
On the other hand, \oiiil/\ltir is uncorrelated with FIR band ratios in 
the dwarfs, while {B08} find an increase of \oiiil/\ltir with F60/F100. 
They attribute this increase in warmer galaxies to a higher density of 
\hii regions. As discussed in Sect.~\ref{sect:pacsratios}, \oiii may also originate 
from low-density ionized gas, hence flattening the relation in dwarfs.

\subsubsection{\ciil\,/ \ltir and \oila \,/ \ltir}
The ratio of \cii/\ltir spans more than an order of magnitude 
in the DGS galaxies, from 0.04 to 0.7\%, with a median value of 0.25\%. 
The ratio of \oila/\ltir does not vary significantly among the dwarfs, with 
a median value of 0.17\%. 
Both ratios have median values slightly higher in the dwarfs than in {B08}. 
Together, the median value of (\ciil+\oila)/\ltir is 0.47\%. 
Since \cii and \oila are the main coolants of the PDR, where the gas 
heating is dominated by photoelectric effect on dust grains, which cool 
through \ltir, the (\cii+\oila)/\ltir ratio is often used as a proxy for the 
photoelectric efficiency \citep[e.g.,][]{tielens-1985}. 
The larger spread in the \cii/\ltir values compared to \oila/\ltir (Fig.~\ref{fig:dgs_panel4}) 
may hint towards different origins of the \cii line (ionized/neutral gas) or 
different conditions of the \cii-emitting phases (as \cii saturates 
for densities $\ge10^4$\,\cm). Since the spread in (\oila+\cii)/\ltir is 
smaller compared to that seen in \cii/\ltir (0.15\,dex compared to 0.26\,dex), 
a mixture of dense and diffuse phases overall compensate for the differences 
in the \oila and \cii emission. 

The only, weak, trend identified in the dwarfs with FIR color is a decrease 
of \cii/\ltir ($\rho=-0.47$); this has been observed in the {B08} galaxies, 
although the trend is less pronounced. 
This can be interpreted as a compactness effect with reduced emission 
in denser and/or warmer environments \citep[e.g.,][]{malhotra-2001}. 
Moreover, \oila/\ltir is weakly anti-correlated with \ltir. 
The dwarfs seem to extend the correlation observed in {B08} of \ciil/\ltir 
and \oila/\ltir decreasing with increasing \ltir/\lb ($\rho\simeq-0.62$). 
The decrease in the \oila/\ltir values is small compared to the decrease 
in the \cii/\ltir values. As a result, (\ciil+\oila)/\ltir also decreases with increasing 
\ltir ($\rho\simeq-0.52$) and \ltir/\lb ($\rho\simeq-0.81$). 
We discuss those high line-to-\ltir and correlations further in Sect.~\ref{sect:pe}.

\subsubsection{(\cii+\oila+\oiii) \,/ \ltir}
The ratio of (\cii+\oila+\oiii)/\ltir spans values from 0.4 to 2.1, with a median of 1.03. 
This is about $3$ times higher than in the {B08} sample. 
The sum of \cii+\oila+\oiii may be regarded as the budget of the gas cooling 
(not a proxy for the photoelectric efficiency), which is clearly enhanced in the dwarfs, 
and \ltir as the total dust cooling. 

Following the previous remarks, this ratio is uncorrelated with FIR color, 
as \oiii dominates the FIR line cooling and no relation was found for \oiii/\ltir, 
but it clearly anti-correlates with \ltir and \ltir/\lb ($\rho\simeq-0.6, -0.7$). 
The DGS galaxies extend the trends observed in the {B08} sample.

\subsection{Summary of correlation analysis}
In summary, the dwarfs present high \oiii/\nii, \cii/\nii, \oiii/\oi, \cii/\ltir, \oi/\ltir, and \oiii/\ltir 
ratios, which set them apart as a group from metal-rich galaxies. Within the ensemble 
of dwarfs, we do not find evident correlations between their FIR emission and the considered 
parameters (metallicity, F60/F100, \ltir, \ltir/\lb), which indicates that their observed properties 
are not controlled by a single parameter.

The interpretation of the observed trends is not straightforward since several factors 
act and compete in the emission of the cooling lines and \ltir. These factors include 
cloud physical conditions, mixing of ISM phases, geometry, dust properties, and 
photoelectric efficiency. They all play a role to some extent since the observed ratios 
are values integrated over full-size galaxies.

\section{Trend analysis with radiative transfer models}
\label{sect:models}
In this section, we characterize the range of physical conditions 
describing the ISM, more specifically the \hii region and the PDR, 
of the DGS galaxies. We aim to interpret the observed group behavior 
in Sect.~\ref{sect:corr} with radiative transfer modeling, 
and in particular, to identify how the parameter space of {\it average} 
physical conditions change compared to metal-rich galaxies. 

We note that there is no study on large samples of galaxies 
that self-consistently models the \hii region and the PDR, besides the work 
on IR luminous galaxies by \cite{abel-2009} and \cite{gracia-carpio-2011} 
and on individual objects in \cite{kaufman-2006}. We aim for a general 
comparison of the metal-rich and metal-poor datasets, and leave the detailed 
ISM modeling of the dwarf galaxies for {\sc Paper~II}. 

\subsection{Line ratios and typical conditions in metal-rich galaxies}
For the PDR properties of ``normal'' metal-rich galaxies, we only use entries 
from \cite{brauher-2008} that have \ltir between $10^9$ and $10^{12}$\,\lsun 
($\sim$130 galaxies) to exclude extreme sources. More than half of that sample, 
comprising normal, starburst, and AGN galaxies was analyzed by \cite{malhotra-2001} 
and \cite{negishi-2001}, who constrained the physical conditions in the PDR 
using the \cite{kaufman-1999} models and found $n_{\rm H}\simeq10^{2-4.5}$\,\cm 
and $G_{\rm 0}\simeq10^{2-4.5}$. 
In the comparison to PDR models, \cite{malhotra-2001} and \cite{negishi-2001} 
correct for a non-PDR contribution to the \cii emission which is on the order 
of 50\%. In the dwarfs, we have estimated in Sect.~\ref{sect:c2ion} that this 
fraction is likely $<$15\%. We apply a correction of 10\% for non-PDR \cii emission 
to our sample. 

For the \hii region properties of metal-rich galaxies, we refer to the studies on 
MIR ionic lines of the \spit Infrared Nearby Galaxies Survey (SINGS; \citealt{dale-2009}) 
and of the luminous IR galaxies from GOALS \citep{inami-2013}. We consider neon 
and sulphur measurements for the galaxies in the luminosity range 
\ltir$\simeq10^9-10^{12}$\,\lsun ($\sim$190 sources). 
The \siiila/\siiilb ratio is traditionally used as a tracer of the electron density 
\citep[e.g.,][]{dudik-2007}, and the \neiiil/\neiil and \sivl/\siiila are diagnostics of the 
hardness of the radiation field \citep{giveon-2002,verma-2003,groves-2008}. 
A typical solution for the \hii region of those metal-rich galaxies has an electron 
density $n_{\rm e}\simeq350$\,\cm and ionization parameter $\log U\simeq-3$ 
\citep{dale-2009,inami-2013}. 
For comparable constraints on the \hii region properties in the dwarf galaxies, we use 
the IRS fluxes extracted for the $compact$ DGS galaxies in Sect.~\ref{sect:prepare}. 
Table~\ref{table:irsratios} lists the median line ratios for the metal-rich and metal-poor 
galaxies. 

We refer to the \hii region and PDR conditions for normal galaxies as the 
metal-rich fiducial model (Table~\ref{table:irsratios}).

\begin{center}
\begin{table}[!t]\small
  \caption{Observed and modeled properties of the dwarfs and normal galaxies.} 
  \hfill{}
\begin{tabular}{l c c c}
    \hline\hline
     \vspace{-8pt}\\
    \multicolumn{1}{l}{IRS line ratio} & 
    \multicolumn{1}{c}{Metal-rich sample} &
    \multicolumn{1}{c}{$Compact$ DGS sample} \\
    \hline
    \vspace{-8pt}\\
	{\siiilb/\siiila}	& 1.43~(0.14\,dex)		& 1.53~(0.12\,dex) \\
	{\sivl/\siiila}	& 0.18~(0.38\,dex)		& 1.30~(0.42\,dex) \\
	{\neiiil/\neiil}	& 0.17~(0.35\,dex)		& 4.90~(0.48\,dex) \\
    \hline
    \vspace{-8pt}\\
     {Metallicity}		& 2~$Z_{\rm ISM}$	& 0.25~$Z_{\rm ISM}$\\
    \multicolumn{3}{l}{Reference ISM abundances ($Z_{\rm ISM}$):} \\ 
    \multicolumn{3}{l}{~~~$O/H=3.2\times10^{-4}$, $C/H=1.4\times10^{-4}$, $N/H=8\times10^{-5}$}\\
    \multicolumn{3}{l}{~~~$Ne/H=1.2\times10^{-4}$, $S/H=3.2\times10^{-5}$}\\
    \hline
    \vspace{-8pt}\\
    \multicolumn{3}{l}{\hii region fiducial model} \\ 
    \hline
    \vspace{-8pt}\\
	{density~[\cm]}	 	& $10^{2.5}$	& $10^{2.0}$ \\
	{$\log U$}		 	& $-3.0$		& $-2.5$ \\
	{$R_{\rm S,\,eff}$~[pc]} 	& $16$		& $46$ \\
    \hline
    \vspace{-8pt}\\
    \multicolumn{3}{l}{PDR fiducial model} \\ 
    \hline
    \vspace{-8pt}\\
	{density~[\cm]}		& $10^{3.5}$	& $10^{4.0}$ \\
	{$\log G_{\rm 0}$}	& $3.3$		& $2.7$ \\
    \hline \hline
    \vspace{-5pt}\\
  \end{tabular}
  \hfill{}
  \newline
  Median IRS line ratios with their dispersions in parenthesis, 
  measured as 1.5 times the median absolute deviation in 
  logarithmic space (\citealt{dale-2009} and \citealt{inami-2013} for the 
  metal-rich galaxies; this work for the low-metallicity, DGS galaxies). 
  ISM abundances are taken from \cite{cowie-1986,savage-1996,meyer-1998}. 
  The model parameters of the metal-rich sample are from 
  \cite{dale-2009} and \cite{inami-2013} for the \hii region, and from 
  \cite{malhotra-2001} and \cite{negishi-2001} for the PDR. 
  $R_{\rm S,\,eff}$ is the effective radius of the \hii region. 
  \label{table:irsratios}
\end{table}
\end{center}

\subsection{Model parameters}
We model the \hii region and PDR simultaneously with the spectral synthesis 
code {Cloudy} v.13.03 \citep{ferland-2013}. The geometry is 1D spherical, 
consisting of the central source of energy surrounded by a cloud of material. 
The source of radiation is chosen as a young starburst that we simulate 
with Starburst99 \citep{leitherer-2010}. 
Following \cite{dale-2009}, we adopt a 10\,Myr continuous star formation scenario 
(with a Kroupa IMF and Geneva stellar tracks), and fix its luminosity to $10^9$\,\lsun. 

We vary the following physical conditions: the hydrogen density 
($n_{\rm H}$; from $10^{1.0}$ to $10^{5.0}$\,\cm in increments of 0.5\,dex) 
and the inner radius ($r_{\rm in}$; from $10^{19.7}$ to $10^{22.0}$\,cm, or 
15\,pc to 3,200\,pc, in increments of 0.3\,dex), which is the distance between 
the starburst and the surface of the modeled ISM shell. For each model, 
we compute $U$, $G_{\rm 0}$, and the IRAS F60/F100 band ratio.  
The FIR color is strongly correlated with the ionization parameter $U$ 
which is varied through the density and radius. It is defined as 
$\displaystyle U=\frac{Q({\rm H})}{4\pi r_{\rm in}^2 n_{\rm H} c}$, where 
$Q({\rm H})$ is the number of ionizing photons and $c$ is the speed of light. 
$G_{\rm 0}$ is the intensity of the FUV radiation field taken 
at the PDR front, in units of $1.6\times10^{-3}$\,erg\,cm$^{-2}$\,s$^{-1}$ 
\citep{habing-1968}. 

The gas-phase abundances are representative of the ISM (averages 
of observations in our galaxy), with $Z_{\rm ISM} = 12+\log(O/H) = 8.5$, 
and given in Table~\ref{table:irsratios}. 
The grain composition and size distribution are assumed typical 
of the galactic ISM, with $R_{\rm V}=3.1$ \citep{weingartner-2001}. 
The dust-to-gas mass ratio is $6.3\times10^{-3}$ and the PAH abundance 
(number of grains per hydrogen) is $PAH/H=1.5\times10^{-7}$ \citep{abel-2008}, 
corresponding in mass to $M(\rm{PAH})/M(\rm{dust})=0.02$. 
The mean metallicity measured in the DGS sample with the \cite{pt05} calibration 
is $12+\log(O/H)=7.9$. In the SINGS galaxies, it is $12+\log(O/H)=8.4$ with 
the \cite{pt05} calibration and $12+\log(O/H)=9.0$ with the \cite{kk04} calibration 
\citep{moustakas-2010}, with larger discrepancies in the methods for massive galaxies 
\citep{kewley-2008}. Therefore we take for representative metallicity of 
the metal-rich galaxies $12+\log(O/H)=8.7$. 
We run two grids of models, one at metallicity $0.25\,Z_{\rm ISM}$ for the 
dwarf galaxies, and one at metallicity $2\,Z_{\rm ISM}$ for the metal-rich galaxies. 
For the model grids, the metallicity of the stellar tracks as well as 
the metal, grain, and PAH abundances in Cloudy are scaled to the metallicity 
of the model ($0.25$ or $2$). 
We also include a turbulent velocity of 1.5\,\kms, as done in \cite{kaufman-1999}. 
All models are stopped at an \av of 10\,mag (corresponding to 
a column density of $N({\rm H}) = 10^{23}$ and $10^{22}$\,cm$^{-2}$ 
for the low and high-metallicity grids, respectively), to ensure 
that they go deep enough in the PDR. We note that the main PDR tracers 
are mostly formed at lower \av and changing the column density 
around those values does not change their emission significantly. 
The grids are generated assuming constant density.

\subsection{Model results}
\begin{figure*}[!t]
\centering
\includegraphics[clip,width=9.15cm]{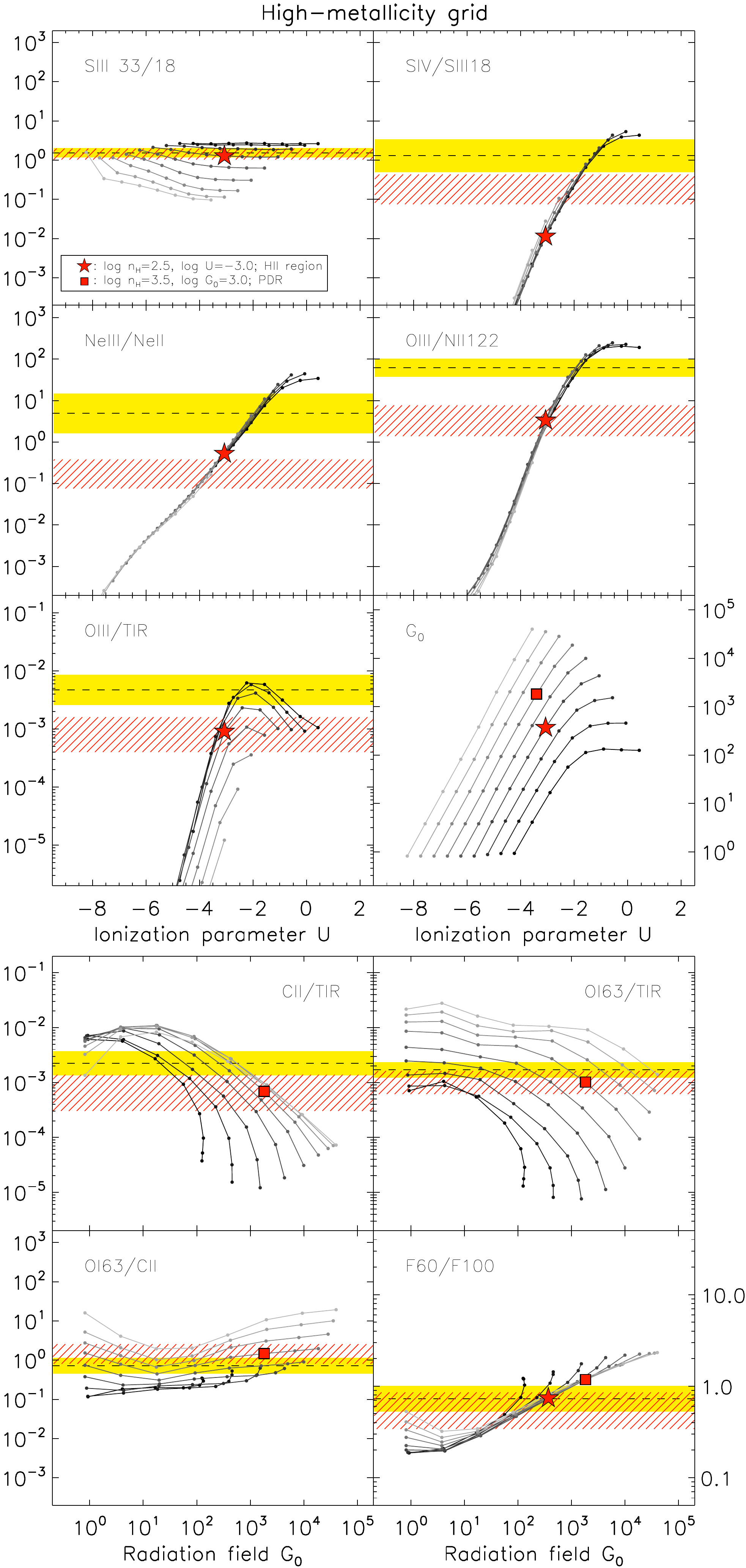}
\includegraphics[clip,width=9.15cm]{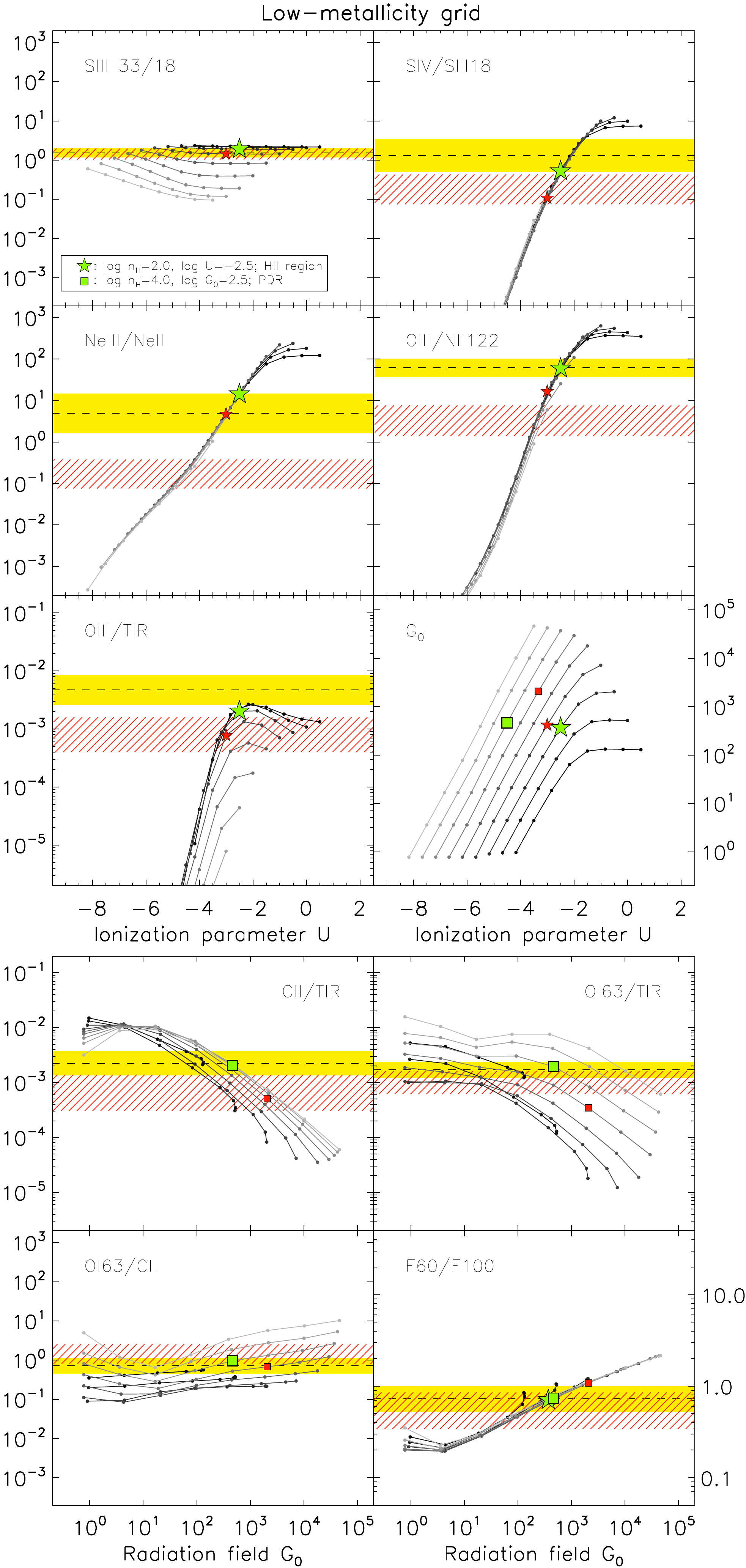}
\caption{
Cloudy predictions of the line intensities as a function of ionization parameter, 
$U$, and radiation field, $G_{\rm 0}$. Each connected track correspond to a given 
density (black = $10^{1.0}$\,\cm; light gray = $10^{5.0}$\,\cm, in steps of 0.5\,dex). 
The observed ratios are indicated by the yellow (dwarf galaxies) and 
red hashed (normal galaxies) bands. 
The \oi/\cii and \cii/\ltir values corrected for non-PDR \cii emission. 
The stars/squares show the fiducial \hii region/PDR models for 
the dwarfs (green symbols) and the normal galaxies (red symbols). 
\underline{\it Left:} results for the high-metallicity ($2\,Z_{\rm ISM}$) grid of models. 
\underline{\it Right:} results for the low-metallicity ($0.25\,Z_{\rm ISM}$) grid of models. 
The fiducial model for normal galaxies is added to the low-metallicity panel 
for comparison (smaller red symbols). 
}
\label{fig:grids}
\end{figure*}

Figure~\ref{fig:grids} shows the model results for the high 
and low-metallicity grids. To extract information on the ISM properties, 
we concentrate on the following ratios: 
\siiila/\siiilb, \sivl/\siiila, \neiiil/\neiil, \oiiil/\niila, \oiiil/\ltir, for the \hii region;
\oila/\ciil, \ciil/\ltir, \oila/\ltir for the PDR. The IRAS F60/F100 band ratio and 
the $G_{\rm 0}$ of the models are also plotted. To ease the comparison 
with the literature, we plot quantities in the \hii region as a function of $U$ 
and in the PDR as a function of $G_{\rm 0}$. 
First of all, we notice that all observed ratios (except \oiii/\ltir) are covered 
by the model grids, hence we can identify a parameter space that is favored 
for the dwarfs. 
Fiducial model parameters for the high and low-metallicity cases are 
given in Table~\ref{table:irsratios}.

\subsubsection{Conditions in the \hii region: high-metallicity case}
In the high-metallicity case ({\it left} panel of Fig.~\ref{fig:grids}), the fiducial 
model for the \hii region agrees partially with the observed IRS line ratios 
of metal-rich galaxies. 
The \siiilb/\siiila ratio is relatively independent of $U$ and the observed values 
agree best with densities between 100 and 1,000\,\cm. 
Both \neiiil/\neiil and \sivl/\siiila ratios are quite sensitive to $U$, the age of the burst, 
and less to density. 
The \neiiil/\neiil ratio is slightly over-predicted by the fiducial model. 
The median and dispersion values of this ratio reported in Table~\ref{table:irsratios} 
are driven by the GOALS galaxies, while a few SINGS galaxies 
clearly have higher \neiiil/\neiil ratios \citep{dale-2009}. 
\cite{inami-2013} have investigated the influence of metallicity and star-formation 
history (continuous versus bursty) on the predicted \neiiil/\neiil ratio. 
They demonstrate that their observed \neiiil/\neiil values are best match by 
models with young instantaneous bursts rather than continuous star formation 
models which systematically over-predict this ratio. 
The largest discrepancy between observations and model predictions is seen 
for the \sivl/\siiila ratio, which is under-predicted by one order of magnitude. 
Part of this discrepancy is due to an observational bias because the detection rate 
of the \sivl line is lower (about 4 times less than the neon lines). 
For the galaxies where \sivl/\siiila is detected, the \neiiil/\neiil 
ratios are higher than the reported median value. The \neiiil/\neiil and \sivl/\siiila 
ratios are well correlated, and extrapolating this correlation to all galaxies 
where \neiiil/\neiil is detected would lower the observed median ratio of 
\sivl/\siiila by a factor of $\sim$3. 
Despite this correction, it seems difficult to reproduce the \neiiil/\neiil and 
\sivl/\siiila ratios simultaneously with our high-metallicity grid. 

Concerning the PACS lines, both \oiii/\ltir and \oiii/\niila are well matched by the 
fiducial model. This suggest that \oiiil/\niila may be used as an indicator 
of the hardness of the radiation field as an alternative to the IRS ratios. 
This model also predicts that about 10\% of the \cii emission 
is originating in the ionized gas, while \cite{malhotra-2001} and \cite{negishi-2001} 
estimated a $\sim$50\% contribution using the \niila line alone. 
Since we have taken into account multiple ionic lines, our derivation 
of the \hii region properties is more robust and the previous ionized gas \cii fractions 
may have been over-estimated. Alternatively, 
an unmodeled low-density, low-ionization component may increase this fraction 
by emitting more \cii (and \nii), but is not necessary here to match the observations.

\subsubsection{Conditions in the \hii region: low-metallicity case}
The observed \siiilb/\siiila ratios are similar in dwarf and normal galaxies, 
and again agree well with densities between 100 and 1,000\,\cm. 
From observations, the \neiiil/\neiil, \sivl/\siiila, and \oiiil/\niila ratios are 
higher in the dwarf galaxies than in metal-rich galaxies, which is a clear 
indication of harder radiation fields \citep[e.g.,][]{madden-2006}. 
At low metallicity, the model predictions for those line ratios are higher by a factor 
of $\sim$5 than in the high-metallicity grid (for a given set of $n_{\rm H}$ and $U$) 
because the radiation field of the starburst is harder. This can be seen by the shift 
of position of the red star from the {\it left} to the {\it right} panels in Fig.~\ref{fig:grids}. 
Even with the change of the starburst metallicity, the observations still require 
higher $U$ to better match the observed \oiii/\niila ratio. 
While $\log U=-3$ agrees well with the observed ratios in normal galaxies, 
$\log U=-2.5$ agrees better with observations for the dwarf galaxies. 
The \neiiil/\neiil and \sivl/\siiila ratios are slightly over and under-predicted, 
respectively, by the model, as in the high-metallicity case. 
Going to higher $U$, the \oiii/\ltir, which is also higher in the dwarfs than in 
the normal galaxies, is then better reproduced by marginally lower densities 
($n_{\rm H}\simeq10^2$\,\cm), but at the expense of a less well predicted 
\siii line ratio. As a best compromise, we take $\log U=-2.5$ and $n_{\rm H}=10^2$\,\cm 
as parameters for our fiducial low-metallicity \hii region model. 
In those conditions, our fiducial model predicts that only 1\% of the \cii 
emission is originating in the ionized gas for the dwarf galaxies. 
However, the \oiii/\ltir ratio remains under-predicted by all the grid models. 
We discuss the origin of this major discrepancy in Sect.~\ref{sect:discuss}.

\subsubsection{Conditions in the PDR: high-metallicity case}
In the PDR, the ratios of \oi/\cii, \cii/\ltir, and \oi/\ltir are sensitive 
to both density and $G_{\rm 0}$. With $n_{\rm H}\simeq10^{3.5}$\,\cm and 
$\log G_{\rm 0}\simeq3.3$, the fiducial PDR model for metal-rich galaxies 
matches well all three ratios ({\it bottom, left} panels of Fig.~\ref{fig:grids}). 
This diagram also shows that, if more \cii emission were originating 
from the PDR than the 50\% assumed here, reducing the density and $G_{\rm 0}$ 
values by 0.5~dex would suffice. 

However, the F60/F100 band ratio is over-predicted by a factor of $\sim$2 
for the metal-rich galaxies. Contrary to the dwarfs, a substantial fraction 
of the dust emission at wavelengths $\ge$100\,\mum may not be related 
to the star-forming regions in normal galaxies 
\citep[e.g.,][]{dale-2007,bendo-2012b,bendo-2014,delooze-2014b}. 
If old stars contribute to dust heating, and thus to the FIR emission, 
the line-to-\ltir ratios associated with the star formation would be 
higher than observed, which would imply a lower $G_{\rm 0}$ than 
deduced currently. Alternatively, since F60/F100 is sensitive to \av and 
all models are stopped at \av of 10\,mag, the IRAS band ratios 
of the normal galaxies may be under-estimated by the models.

\subsubsection{Conditions in the PDR: low-metallicity case}
Considering the low-metallicity case, the predictions for \oi/\cii decrease by 
a factor of $\sim$2 ({\it bottom, right} panel of Fig.~\ref{fig:grids}) owing to 
a decrease in the PDR temperature \citep{roellig-2006}. 
Moreover, in the models with $U\ge10^{-2}$, we observe two effects. 
First, those models have low densities and as a consequence the thickness 
of the cloud becomes larger than the inner radius. As a result, there is 
significant geometric dilution of the UV radiation field. The $\rm{C^+}$ and 
$\rm{O^0}$ emitting layers are larger, and thus \cii/\ltir and \oi/\ltir ratios are high. 
Second, regardless of metallicity, there is more dust absorption in 
the \hii region at those high-$U$ values \citep{abel-2009}, and thus 
little increase of $G_{\rm 0}$ in the PDR for increasing $U$. 
As a consequence, the predicted ratios of \oi/\cii, \cii/\ltir and \oi/\ltir flatten 
out with $U$. 

The observed \oi/\cii ratios in the dwarfs and normal galaxies are similar, 
while \cii/\ltir and \oi/\ltir are higher in the dwarfs. 
The ratios of the dwarfs indicate lower $G_{\rm 0}$ or higher $n_{\rm H}$ values. 
We find that a solution with $n_{\rm H}\simeq10^4$\,\cm and $\log G_{\rm 0}\simeq2.5$ 
is adequate to reproduce all three ratios. 
This PDR solution agrees with the observed F60/F100 band ratio for the dwarfs.

\section{Discussion}
\label{sect:discuss}
\subsection{Interpretation in terms of ISM structure}
The fiducial models imply different sizes for the low-metallicity 
and high-metallicity components (Table~\ref{table:irsratios}). 
The ionic lines are emitted by a modeled shell, for which the size can be converted 
to an effective Str\"omgren radius $R_{\rm S,~eff}$. 
which can be thought of as a cumulative volume of \hii regions in a galaxy. 
As an example, for the high-metallicity case, $R_{\rm S,~eff}$ of 16\,pc 
corresponds to about one hundred \hii regions if we consider each to have 
a typical radius of 3-4\,pc. 
Because the fiducial low-metallicity \hii region model has harder radiation field 
and lower density, we find that its effective radius is three 
times larger than that of the high-metallicity \hii region. 
The depth of the PDR, defined as the location where the gas is half atomic 
half molecular, is similar in the low-metallicity and high-metallicity ISM. 
However, the low-metallicity PDR is located further away. 

Discrepancies in the predicted \oiii/\ltir ratios between the high and low-metallicity 
grids are mainly due to a reduced gas-phase metallicity. This reduces 
the absolute \oiii emission produced in the \hii region, while both grids of models 
have the same, fixed input luminosity which is fully reprocessed in the PDR, exiting 
as \ltir (no escaping UV). 
To increase the predicted ratios of ionic lines to \ltir and better match the observations, 
we would need to reduce significantly the mean \av of the models. This can be 
achieved by accounting for an escape fraction, reducing the covering factor of 
the PDR (hence \ltir and the average column density). In other words, the emission 
from the PDR must be reduced relative to that of the \hii region, by a factor 
of approximately $3$ in our case (for the green star to fall on the observed median 
\oiii/\ltir value in the {\it right} panel of Fig.~\ref{fig:grids}).
Lower PDR covering factors were also required 
in the detailed, multi-phase modeling of Haro\,11 \citep{cormier-2012}. 
The fact that the median value of \ltir/\lb, which represents the fraction of 
obscured/unobscured emission, is $\sim$2 in the {B08} galaxies and $0.7$ 
in the DGS galaxies corroborates our finding of lower PDR covering factors 
for the low-metallicity models. 
This also confirms the observed trend of decreasing \oiii/\ltir with increasing \ltir/\lb. 
The high escape fraction in dwarf galaxies is also observed in their SEDs 
\citep[e.g.,][]{galliano-2008a}.

The higher ratios of \oi and \cii to \ltir indicate lower $U$ in the low-metallicity 
PDR, which can be interpreted as a change of ISM configuration. 
Since the \hii region and PDR low-metallicity fiducial models have the same 
$G_{\rm 0}$ (see Fig.~\ref{fig:grids}), placing the PDR adjacent to the \hii region 
and increasing the density would result in lowering $U$. 
In the high-metallicity fiducial model, $G_{\rm 0}$ is higher in the PDR than in the 
\hii region, which would require to place the PDR inside the \hii region. 
Combined with the fact that the low-metallicity \hii region fills a larger volume 
than the high-metallicity \hii region, this implies that the minimum 
distance of the low-metallicity PDR to the starburst should be $6$ times larger 
than in the high-metallicity setting. 
The increase in the typical PDR distance is a natural consequence of a 
lower covering factor. 

These considerations highlight a change in the ISM structure and PDR distribution 
of low-metallicity galaxies, where the filling factor of ionized gas relative to the PDRs 
is higher than in metal-rich galaxies. The extent of the \hii region is bigger 
and the UV photons travel larger distance before they reach a PDR clump.

\subsection{Interpretation of the \cii/\ltir and \oi/\ltir trends}
\label{sect:pe}
In Sect.~\ref{sect:pacsltir}, we observed that the line-to-\ltir ratios decrease 
with increasing \ltir and \ltir/\lb, and that \cii/\ltir is anti-correlated with 
F60/F100. Using the PDR models, we discuss which physical effects 
dominate in producing these trends. 

Line-to-\ltir ratios have been studied extensively in massive galaxies 
(normal, star-forming galaxies and ULIRGs). 
They give the budget of the gas cooling over the dust cooling. 
The (\cii+\oila)/\ltir ratio, a common measure of the photoelectric efficiency 
in the PDR, is observed to vary (see Fig.~\ref{fig:dgs_panel4}), 
with a clear deficit of \cii/\ltir in ULIRGs \citep{luhman-2003,diaz-santos-2013}, 
and FIR lines in general \citep{gracia-carpio-2011}, as well as a decrease 
with FIR color in star-forming galaxies \citep{malhotra-2001}. 
Several hypotheses to explain the apparent line deficit in galaxies include 
(1)~more dust screening at high ionization parameter \citep{luhman-2003,abel-2009}; 
(2)~lower photoelectric efficiency due to charged small dust grains 
under intense radiation fields \citep{malhotra-2001,croxall-2012}; 
(3)~lower photoelectric efficiency due to a decrease in the PAH abundance 
under hard radiation fields \citep{madden-2006,rubin-2009}; 
and (4)~large dust optical depths leading to extinction of the emission lines 
\citep{papadopoulos-2010,rangwala-2011}.
The dwarf galaxies extend the relative importance of the gas cooling 
to high values. Here, we aim to investigate which of those interpretations 
can explain the difference from the normal to the dwarf galaxies. 

Because of their low dust content, variations in grain abundances 
and charging could affect the photoelectric efficiency dramatically 
in the dwarf galaxies. 
The quantity \ltir sums emission from several grain populations: 
from the big grains (predominantly) to the very small grains and PAHs, 
which are the most important for the photoelectric heating \citep{bakes-1994}. 
Those grains emit from all ISM phases, and the fraction originating 
in the PDRs themselves is unknown. Star-forming dwarf galaxies 
generally exhibit enhanced small grain abundances \citep{galliano-2003}, 
which correlate with \hii regions, and reduced PAH abundances, 
which mostly trace the PDR, as a result of the hard radiation fields 
\citep{madden-2006}. Thus those two populations may balance 
their relative contribution in the photoelectric effect. 
Moreover, the mean free path of UV photons in low-metallicity 
environments is larger, resulting in a UV field dilution over larger 
spatial scales, which reduces grain charging \citep{israel-2011}. 
\cite{sandstrom-2012} find that, even though their abundance 
is reduced, PAHs are generally more neutral in the star-forming regions 
of the SMC than in typical spiral galaxies. 
\cite{lebouteiller-2012} examine both spatial distributions of (\ciil+\oila) 
with \ltir and PAH emission in the star-forming region N11\,B of the LMC, 
and find that PAH emission traces best the gas heating as compared to \ltir. 
The net result of those considerations is that, overall, the observed 
(\ciil+\oila)/\ltir ratios are high in the dwarfs (even if we subtract a small 
contribution of the ionized gas to the \cii emission), indicating either 
relatively {\it efficient} photoelectric heating or non-photoelectric 
excitation of these lines (cosmic rays, X-rays, etc.). The ratios tend 
to flatten out to 1\% at low luminosity, which could indicate a ceiling 
on the photoelectric efficiency. 
The influence of more involved changes to the modeling (decrease 
of PAH abundance, grain size distribution, or X-ray illumination) will be 
examined in {\sc Paper~II}. Here we conclude that a simple change 
in geometry, by putting the cloud further away, with viable density and 
illumination conditions can suffice to explain the high observed 
line-to-\ltir ratios. 

\cii/\ltir and, to a lesser extent, \oila/\ltir decrease as $G_{\rm 0}$ increases 
(Fig.~\ref{fig:grids}). 
Two main effects are responsible for that: the increased grain charging and 
increased dust-to-gas opacity which cause a larger fraction of the radiation to go 
into heating the dust. The dwarfs show high \cii/\ltir and \oi/\ltir because 
$G_{\rm 0}$ and $G_{\rm 0}$/$n_{\rm H}$ (which traces grain charging) are lower 
in the low-metallicity PDR than found in normal galaxies \citep{malhotra-2001}. 
Since \oila increases faster with $G_{\rm 0}$ than \cii, \oila/\ltir is less affected, 
except at the highest $G_{\rm 0}$ values. We also note that, \oila being more sensitive 
to density than \cii, \oila/\ltir increases with $n_{\rm H}$ for a given $G_{\rm 0}$. 
In the models, F60/F100 follows immediately from $G_{\rm 0}$. 
The anti-correlation between \cii/\ltir and the FIR band ratio, which 
we interpreted as an effect of compactness, is evident in Fig.~\ref{fig:grids}, 
and the lack of correlation between \oila/\ltir and FIR color agrees 
with the models. The stronger dependency of \cii/\ltir on $G_{\rm 0}$ 
compared to \oila/\ltir may account for the larger scatter in the 
observed \cii/\ltir values. The positive correlation between \oila/\ciil 
and FIR color is also reproduced by the models, which may be 
understood as an aging effect of the star-forming regions that expand 
as they age. 
The models do not show significant optical depth effects. 
The fact that the \oilb line would become optically thick at higher column 
densities than \oila, and that \oilb and \oila behave similarly with \ltir and 
FIR color in the observations also confirms the absence of optical depth effects 
at those physical scales. Nevertheless, given the low statistics that we have for 
the \oilb line and that \oilb is uncorrelated with \ltir/\lb, we cannot confidently 
rule out optical depth effects. 

Overall, the low-metallicity PDRs are described by a low-$G_{\rm 0}$ regime 
accompanied by a density increase. This low $G_{\rm 0}$/$n_{\rm H}$ 
causes UV field dilution and reduces grain charging \citep{israel-2011}. 
The proposed changes in ISM structure can therefore explain the higher line-to-\ltir 
ratios found at lower \ltir and \ltir/\lb (i.e. increased porosity with lower filling factor 
of dense phases), as noted in Sect.~\ref{sect:pacsltir}.

\section{Conclusions}
\label{sect:concl}
We present observations of the FIR fine-structure cooling lines 
\cii157\,\mum, \oi63 and 145\,\mum, \oiii88\,\mum, \niii57\,\mum, and 
\nii122 and 205\,\mum obtained with the PACS spectrometer onboard \hers 
in 48 low-metallicity galaxies from the Dwarf Galaxy Survey. 

\begin{itemize}
\item
The brightest FIR line in the dwarfs is \oiii, on average twice brighter than \cii, 
the \oiiil/\ciil ratio varying from 0.5 to 13. 
The FIR lines are bright relative to the TIR luminosity, together accounting 
for several percents of \ltir. \cii/\ltir range from 0.04 to 0.7\% and (\ciil+\oila)/\ltir 
range from 0.2 to 1.0\%, which is about twice as high as in metal-rich galaxies. 
The gas cooling is enhanced on galaxy-wide scales. 
When zooming in onto the active regions of the LMC, line-to-\ltir ratios are generally 
higher than galaxy-integrated values. This matches our interpretation that the dwarf 
galaxies are dominated by emission from their star-forming regions even on galaxy 
integrated scales. 

\item
When compared to the \cite{brauher-2008} ISO sample, the dwarf galaxies 
occupy a completely different parameter space than the more metal-rich galaxies 
in several central tracers. In particular, the \oiiil/\niila, \niiil/\niila, 
\ciil/\niila, and \oiiil/\oila line ratios are very high in the dwarfs, which we interpret 
in terms of radiation field hardness and phase filling factor, with prominent 
high-excitation ionized regions. 
In spite of the possible influence of abundance variations, the \oiii/\nii 
ratio appears to be a good tracer of the radiation field hardness.
The low \niila/\ciil ratios indicate that there is no substantial 
contribution of the ionized gas to the \cii emission (typically $<$15\%). 

\item
With the help of radiative transfer models, the FIR line ratios are interpreted 
as a change in the ISM properties. The high \cii/\ltir and \oiii/\ltir ratios indicate 
a combination of moderate FUV fields and low covering factor of the PDR. 
The structure of the low-metallicity ISM is qualitatively and quantitatively different 
than found in metal-rich environments, harboring only a low filling factor of dense 
gas surrounding the star formation sites, embedded in   
a large volume filling factor of diffuse gas. In this leaky structure, 
UV photons escape from \hii regions out to large distances. 
Overall, the filling factor of the ISM affected by the star-forming regions is high.
\end{itemize}

Given the prospects for observing FIR lines at high redshift (ALMA, NOEMA, CCAT), 
the \ciil, \oiiil (twice brighter than \ciil), and \oila (as bright as \ciil) lines 
appear as promising diagnostics of the low-metallicity ISM conditions. 
Each line account for about 0.3\% of the TIR luminosity.
\ciil is a reliable tracer of the PDR, with little contribution from the ionized gas. 
\oiiil traces the high-excitation ionized gas while \oiii/\oi indicates, independently 
of elemental abundance variations, the filling factor of ionized/neutral phases, 
which we find high in the low-metallicity star-forming galaxies. \\

\begin{acknowledgements}
We are grateful to A. Poglitsch, A. Contursi, J. Graci{\'a} Carpio, 
B. Vandenbussche, and P. Royer for their help with the PACS data. 
We thank D. Dale for sharing the B-band data of the normal galaxies 
with us and the referee for careful reading of the manuscript. 
We acknowledge support from the Agence Nationale de la Recherche (ANR) 
through the programme SYMPATICO (Program Blanc Projet ANR-11-BS56-0023). 
IDL is a postdoctoral researcher of the FWO-Vlaanderen (Belgium). 
PACS has been developed by a consortium of institutes led by MPE (Germany) 
and including UVIE (Austria); KU Leuven, CSL, IMEC (Belgium); CEA, LAM (France); 
MPIA (Germany); INAF-IFSI/OAA/OAP/OAT, LENS, SISSA (Italy); IAC (Spain). 
This development has been supported by the funding agencies BMVIT (Austria), 
ESA-PRODEX (Belgium), CEA/CNES (France), DLR (Germany), ASI/INAF (Italy), 
and CICYT/MCYT (Spain). 
This research has made use of the NASA/ IPAC Infrared Science Archive, which 
is operated by the Jet Propulsion Laboratory, California Institute of Technology, 
under contract with the National Aeronautics and Space Administration.
\end{acknowledgements}

\bibliographystyle{aa}
\bibliography{../../../BIB/references}

\begin{thebibliography}{98}
\expandafter\ifx\csname natexlab\endcsname\relax\def\natexlab#1{#1}\fi

\bibitem[{{Abel} {et~al.}(2009){Abel}, {Dudley}, {Fischer}, {Satyapal}, \& {van
  Hoof}}]{abel-2009}
{Abel}, N.~P., {Dudley}, C., {Fischer}, J., {Satyapal}, S., \& {van Hoof},
  P.~A.~M. 2009, \apj, 701, 1147

\bibitem[{{Abel} {et~al.}(2007){Abel}, {Sarma}, {Troland}, \&
  {Ferland}}]{abel-2007}
{Abel}, N.~P., {Sarma}, A.~P., {Troland}, T.~H., \& {Ferland}, G.~J. 2007,
  \apj, 662, 1024

\bibitem[{{Abel} {et~al.}(2008){Abel}, {van Hoof}, {Shaw}, {Ferland}, \&
  {Elwert}}]{abel-2008}
{Abel}, N.~P., {van Hoof}, P.~A.~M., {Shaw}, G., {Ferland}, G.~J., \& {Elwert},
  T. 2008, \apj, 686, 1125

\bibitem[{{Bakes} \& {Tielens}(1994)}]{bakes-1994}
{Bakes}, E.~L.~O. \& {Tielens}, A.~G.~G.~M. 1994, \apj, 427, 822

\bibitem[{{Bendo} {et~al.}(2014){Bendo}, {Baes}, {Bianchi}, {Boquien},
  {Boselli}, {Cooray}, {Cortese}, {De Looze}, {di Serego Alighieri}, {Fritz},
  {Gentile}, {Hughes}, {Lu}, {Pappalardo}, {Smith}, {Spinoglio}, {Viaene}, \&
  {Vlahakis}}]{bendo-2014}
{Bendo}, G.~J., {Baes}, M., {Bianchi}, S., {et~al.} 2014, ArXiv e-prints

\bibitem[{{Bendo} {et~al.}(2012{\natexlab{a}}){Bendo}, {Boselli}, {Dariush},
  {Pohlen}, {Roussel}, {Sauvage}, {Smith}, {Wilson}, {Baes}, {Cooray},
  {Clements}, {Cortese}, {Foyle}, {Galametz}, {Gomez}, {Lebouteiller}, {Lu},
  {Madden}, {Mentuch}, {O'Halloran}, {Page}, {Remy}, {Schulz}, \&
  {Spinoglio}}]{bendo-2012b}
{Bendo}, G.~J., {Boselli}, A., {Dariush}, A., {et~al.} 2012{\natexlab{a}},
  \mnras, 419, 1833

\bibitem[{{Bendo} {et~al.}(2012{\natexlab{b}}){Bendo}, {Galliano}, \&
  {Madden}}]{bendo-2012}
{Bendo}, G.~J., {Galliano}, F., \& {Madden}, S.~C. 2012{\natexlab{b}}, \mnras,
  423, 197

\bibitem[{{Bennett} {et~al.}(1994){Bennett}, {Fixsen}, {Hinshaw}, {Mather},
  {Moseley}, {Wright}, {Eplee}, {Gales}, {Hewagama}, {Isaacman}, {Shafer}, \&
  {Turpie}}]{bennett-1994}
{Bennett}, C.~L., {Fixsen}, D.~J., {Hinshaw}, G., {et~al.} 1994, \apj, 434, 587

\bibitem[{{Bernard-Salas} {et~al.}(2012){Bernard-Salas}, {Habart}, {Arab},
  {Abergel}, {Dartois}, {Martin}, {Bontemps}, {Joblin}, {White}, {Bernard}, \&
  {Naylor}}]{bernard-salas-2012}
{Bernard-Salas}, J., {Habart}, E., {Arab}, H., {et~al.} 2012, \aap, 538, A37

\bibitem[{{Boselli} {et~al.}(2002){Boselli}, {Gavazzi}, {Lequeux}, \&
  {Pierini}}]{boselli-2002}
{Boselli}, A., {Gavazzi}, G., {Lequeux}, J., \& {Pierini}, D. 2002, \aap, 385,
  454

\bibitem[{{Brauher} {et~al.}(2008){Brauher}, {Dale}, \& {Helou}}]{brauher-2008}
{Brauher}, J.~R., {Dale}, D.~A., \& {Helou}, G. 2008, \apjs, 178, 280

\bibitem[{{Chevance} {et~al.}(in prep.)}]{chevance-2015}
{Chevance}, M. {et~al.} in prep.

\bibitem[{{Cigan} {et~al.}(submitted to ApJ)}]{cigan-2015}
{Cigan}, P. {et~al.} submitted to ApJ

\bibitem[{{Cormier} {et~al.}(2012){Cormier}, {Lebouteiller}, {Madden}, {Abel},
  {Hony}, {Galliano}, {Baes}, {Barlow}, {Cooray}, {De Looze}, {Galametz},
  {Karczewski}, {Parkin}, {R{\'e}my}, {Sauvage}, {Spinoglio}, {Wilson}, \&
  {Wu}}]{cormier-2012}
{Cormier}, D., {Lebouteiller}, V., {Madden}, S.~C., {et~al.} 2012, \aap, 548,
  A20

\bibitem[{{Cormier} {et~al.}(2014){Cormier}, {Madden}, {Lebouteiller}, {Hony},
  {Aalto}, {Costagliola}, {Hughes}, {R{\'e}my-Ruyer}, {Abel}, {Bayet},
  {Bigiel}, {Cannon}, {Cumming}, {Galametz}, {Galliano}, {Viti}, \&
  {Wu}}]{cormier-2014}
{Cormier}, D., {Madden}, S.~C., {Lebouteiller}, V., {et~al.} 2014, \aap, 564,
  A121

\bibitem[{{Cowie} \& {Songaila}(1986)}]{cowie-1986}
{Cowie}, L.~L. \& {Songaila}, A. 1986, \araa, 24, 499

\bibitem[{{Croxall} {et~al.}(2012){Croxall}, {Smith}, {Wolfire}, {Roussel},
  {Sandstrom}, {Draine}, {Aniano}, {Dale}, {Armus}, {Beir{\~a}o}, {Helou},
  {Bolatto}, {Appleton}, {Brandl}, {Calzetti}, {Crocker}, {Galametz}, {Groves},
  {Hao}, {Hunt}, {Johnson}, {Kennicutt}, {Koda}, {Krause}, {Li}, {Meidt},
  {Murphy}, {Rahman}, {Rix}, {Sauvage}, {Schinnerer}, {Walter}, \&
  {Wilson}}]{croxall-2012}
{Croxall}, K.~V., {Smith}, J.~D., {Wolfire}, M.~G., {et~al.} 2012, \apj, 747,
  81

\bibitem[{{Dale} {et~al.}(2007){Dale}, {Gil de Paz}, {Gordon}, {Hanson},
  {Armus}, {Bendo}, {Bianchi}, {Block}, {Boissier}, {Boselli}, {Buckalew},
  {Buat}, {Burgarella}, {Calzetti}, {Cannon}, {Engelbracht}, {Helou},
  {Hollenbach}, {Jarrett}, {Kennicutt}, {Leitherer}, {Li}, {Madore}, {Martin},
  {Meyer}, {Murphy}, {Regan}, {Roussel}, {Smith}, {Sosey}, {Thilker}, \&
  {Walter}}]{dale-2007}
{Dale}, D.~A., {Gil de Paz}, A., {Gordon}, K.~D., {et~al.} 2007, \apj, 655, 863

\bibitem[{{Dale} \& {Helou}(2002)}]{dale-2002}
{Dale}, D.~A. \& {Helou}, G. 2002, \apj, 576, 159

\bibitem[{{Dale} {et~al.}(2009){Dale}, {Smith}, {Schlawin}, {Armus},
  {Buckalew}, {Cohen}, {Helou}, {Jarrett}, {Johnson}, {Moustakas}, {Murphy},
  {Roussel}, {Sheth}, {Staudaher}, {Bot}, {Calzetti}, {Engelbracht}, {Gordon},
  {Hollenbach}, {Kennicutt}, \& {Malhotra}}]{dale-2009}
{Dale}, D.~A., {Smith}, J.~D.~T., {Schlawin}, E.~A., {et~al.} 2009, \apj, 693,
  1821

\bibitem[{{De Breuck} {et~al.}(2014){De Breuck}, {Williams}, {Swinbank},
  {Caselli}, {Coppin}, {Davis}, {Maiolino}, {Nagao}, {Smail}, {Walter},
  {Wei{\ss}}, \& {Zwaan}}]{de-breuck-2014}
{De Breuck}, C., {Williams}, R.~J., {Swinbank}, M., {et~al.} 2014, \aap, 565,
  A59

\bibitem[{{De Looze} {et~al.}(2011){De Looze}, {Baes}, {Bendo}, {Cortese}, \&
  {Fritz}}]{delooze-2011}
{De Looze}, I., {Baes}, M., {Bendo}, G.~J., {Cortese}, L., \& {Fritz}, J. 2011,
  \mnras, 416, 2712

\bibitem[{{De Looze} {et~al.}(2014{\natexlab{a}}){De Looze}, {Cormier},
  {Lebouteiller}, {Madden}, {Baes}, {Bendo}, {Boquien}, {Boselli}, {Clements},
  {Cortese}, {Cooray}, {Galametz}, {Galliano}, {Graci{\'a}-Carpio}, {Isaak},
  {Karczewski}, {Parkin}, {Pellegrini}, {R{\'e}my-Ruyer}, {Spinoglio}, {Smith},
  \& {Sturm}}]{delooze-2014}
{De Looze}, I., {Cormier}, D., {Lebouteiller}, V., {et~al.} 2014{\natexlab{a}},
  \aap, 568, A62

\bibitem[{{De Looze} {et~al.}(2014{\natexlab{b}}){De Looze}, {Fritz}, {Baes},
  {Bendo}, {Cortese}, {Boquien}, {Boselli}, {Camps}, {Cooray}, {Cormier},
  {Davies}, {De Geyter}, {Hughes}, {Jones}, {Karczewski}, {Lebouteiller}, {Lu},
  {Madden}, {R{\'e}my-Ruyer}, {Spinoglio}, {Smith}, {Viaene}, \&
  {Wilson}}]{delooze-2014b}
{De Looze}, I., {Fritz}, J., {Baes}, M., {et~al.} 2014{\natexlab{b}}, \aap,
  571, A69

\bibitem[{{D{\'{\i}}az-Santos} {et~al.}(2013){D{\'{\i}}az-Santos}, {Armus},
  {Charmandaris}, {Stierwalt}, {Murphy}, {Haan}, {Inami}, {Malhotra},
  {Meijerink}, {Stacey}, {Petric}, {Evans}, {Veilleux}, {van der Werf}, {Lord},
  {Lu}, {Howell}, {Appleton}, {Mazzarella}, {Surace}, {Xu}, {Schulz},
  {Sanders}, {Bridge}, {Chan}, {Frayer}, {Iwasawa}, {Melbourne}, \&
  {Sturm}}]{diaz-santos-2013}
{D{\'{\i}}az-Santos}, T., {Armus}, L., {Charmandaris}, V., {et~al.} 2013, \apj,
  774, 68

\bibitem[{{Dudik} {et~al.}(2007){Dudik}, {Weingartner}, {Satyapal}, {Fischer},
  {Dudley}, \& {O'Halloran}}]{dudik-2007}
{Dudik}, R.~P., {Weingartner}, J.~C., {Satyapal}, S., {et~al.} 2007, \apj, 664,
  71

\bibitem[{{Engelbracht} {et~al.}(2008){Engelbracht}, {Rieke}, {Gordon},
  {Smith}, {Werner}, {Moustakas}, {Willmer}, \& {Vanzi}}]{engelbracht-2008}
{Engelbracht}, C.~W., {Rieke}, G.~H., {Gordon}, K.~D., {et~al.} 2008, \apj,
  678, 804

\bibitem[{{Farrah} {et~al.}(2013){Farrah}, {Lebouteiller}, {Spoon},
  {Bernard-Salas}, {Pearson}, {Rigopoulou}, {Smith}, {Gonz{\'a}lez-Alfonso},
  {Clements}, {Efstathiou}, {Cormier}, {Afonso}, {Petty}, {Harris}, {Hurley},
  {Borys}, {Verma}, {Cooray}, \& {Salvatelli}}]{farrah-2013}
{Farrah}, D., {Lebouteiller}, V., {Spoon}, H.~W.~W., {et~al.} 2013, \apj, 776,
  38

\bibitem[{{Ferland} {et~al.}(2013){Ferland}, {Porter}, {van Hoof}, {Williams},
  {Abel}, {Lykins}, {Shaw}, {Henney}, \& {Stancil}}]{ferland-2013}
{Ferland}, G.~J., {Porter}, R.~L., {van Hoof}, P.~A.~M., {et~al.} 2013, \rmxaa,
  49, 137

\bibitem[{{Galliano} {et~al.}(2008){Galliano}, {Dwek}, \&
  {Chanial}}]{galliano-2008a}
{Galliano}, F., {Dwek}, E., \& {Chanial}, P. 2008, \apj, 672, 214

\bibitem[{{Galliano} {et~al.}(2011){Galliano}, {Hony}, {Bernard}, {Bot},
  {Madden}, {Roman-Duval}, {Galametz}, {Li}, {Meixner}, {Engelbracht},
  {Lebouteiller}, {Misselt}, {Montiel}, {Panuzzo}, {Reach}, \&
  {Skibba}}]{galliano-2011}
{Galliano}, F., {Hony}, S., {Bernard}, J.-P., {et~al.} 2011, \aap, 536, A88

\bibitem[{{Galliano} {et~al.}(2003){Galliano}, {Madden}, {Jones}, {Wilson},
  {Bernard}, \& {Le Peintre}}]{galliano-2003}
{Galliano}, F., {Madden}, S.~C., {Jones}, A.~P., {et~al.} 2003, \aap, 407, 159

\bibitem[{{Genzel} \& {Cesarsky}(2000)}]{genzel-2000}
{Genzel}, R. \& {Cesarsky}, C.~J. 2000, \araa, 38, 761

\bibitem[{{Giveon} {et~al.}(2002){Giveon}, {Sternberg}, {Lutz}, {Feuchtgruber},
  \& {Pauldrach}}]{giveon-2002}
{Giveon}, U., {Sternberg}, A., {Lutz}, D., {Feuchtgruber}, H., \& {Pauldrach},
  A.~W.~A. 2002, \apj, 566, 880

\bibitem[{{Graci{\'a}-Carpio} {et~al.}(2011){Graci{\'a}-Carpio}, {Sturm},
  {Hailey-Dunsheath}, {Fischer}, {Contursi}, {Poglitsch}, {Genzel},
  {Gonz{\'a}lez-Alfonso}, {Sternberg}, {Verma}, {Christopher}, {Davies},
  {Feuchtgruber}, {de Jong}, {Lutz}, \& {Tacconi}}]{gracia-carpio-2011}
{Graci{\'a}-Carpio}, J., {Sturm}, E., {Hailey-Dunsheath}, S., {et~al.} 2011,
  \apjl, 728, L7

\bibitem[{{Griffin} {et~al.}(2010){Griffin}, {Abergel}, {Abreu}, {Ade},
  {Andr{\'e}}, {Augueres}, {Babbedge}, {Bae}, {Baillie}, {Baluteau}, {Barlow},
  {Bendo}, {Benielli}, {Bock}, {Bonhomme}, {Brisbin}, {Brockley-Blatt},
  {Caldwell}, {Cara}, {Castro-Rodriguez}, {Cerulli}, {Chanial}, {Chen},
  {Clark}, {Clements}, {Clerc}, {Coker}, {Communal}, {Conversi}, {Cox},
  {Crumb}, {Cunningham}, {Daly}, {Davis}, {de Antoni}, {Delderfield}, {Devin},
  {di Giorgio}, {Didschuns}, {Dohlen}, {Donati}, {Dowell}, {Dowell}, {Duband},
  {Dumaye}, {Emery}, {Ferlet}, {Ferrand}, {Fontignie}, {Fox}, {Franceschini},
  {Frerking}, {Fulton}, {Garcia}, {Gastaud}, {Gear}, {Glenn}, {Goizel},
  {Griffin}, {Grundy}, {Guest}, {Guillemet}, {Hargrave}, {Harwit}, {Hastings},
  {Hatziminaoglou}, {Herman}, {Hinde}, {Hristov}, {Huang}, {Imhof}, {Isaak},
  {Israelsson}, {Ivison}, {Jennings}, {Kiernan}, {King}, {Lange}, {Latter},
  {Laurent}, {Laurent}, {Leeks}, {Lellouch}, {Levenson}, {Li}, {Li},
  {Lilienthal}, {Lim}, {Liu}, {Lu}, {Madden}, {Mainetti}, {Marliani}, {McKay},
  {Mercier}, {Molinari}, {Morris}, {Moseley}, {Mulder}, {Mur}, {Naylor},
  {Nguyen}, {O'Halloran}, {Oliver}, {Olofsson}, {Olofsson}, {Orfei}, {Page},
  {Pain}, {Panuzzo}, {Papageorgiou}, {Parks}, {Parr-Burman}, {Pearce},
  {Pearson}, {P{\'e}rez-Fournon}, {Pinsard}, {Pisano}, {Podosek}, {Pohlen},
  {Polehampton}, {Pouliquen}, {Rigopoulou}, {Rizzo}, {Roseboom}, {Roussel},
  {Rowan-Robinson}, {Rownd}, {Saraceno}, {Sauvage}, {Savage}, {Savini},
  {Sawyer}, {Scharmberg}, {Schmitt}, {Schneider}, {Schulz}, {Schwartz},
  {Shafer}, {Shupe}, {Sibthorpe}, {Sidher}, {Smith}, {Smith}, {Smith},
  {Spencer}, {Stobie}, {Sudiwala}, {Sukhatme}, {Surace}, {Stevens}, {Swinyard},
  {Trichas}, {Tourette}, {Triou}, {Tseng}, {Tucker}, {Turner}, {Vaccari},
  {Valtchanov}, {Vigroux}, {Virique}, {Voellmer}, {Walker}, {Ward}, {Waskett},
  {Weilert}, {Wesson}, {White}, {Whitehouse}, {Wilson}, {Winter}, {Woodcraft},
  {Wright}, {Xu}, {Zavagno}, {Zemcov}, {Zhang}, \& {Zonca}}]{griffin-2010}
{Griffin}, M.~J., {Abergel}, A., {Abreu}, A., {et~al.} 2010, \aap, 518, L3

\bibitem[{{Groves} {et~al.}(2008){Groves}, {Dopita}, {Sutherland}, {Kewley},
  {Fischera}, {Leitherer}, {Brandl}, \& {van Breugel}}]{groves-2008}
{Groves}, B., {Dopita}, M.~A., {Sutherland}, R.~S., {et~al.} 2008, \apjs, 176,
  438

\bibitem[{{Habing}(1968)}]{habing-1968}
{Habing}, H.~J. 1968, \bain, 19, 421

\bibitem[{{Hailey-Dunsheath} {et~al.}(2010){Hailey-Dunsheath}, {Nikola},
  {Stacey}, {Oberst}, {Parshley}, {Benford}, {Staguhn}, \&
  {Tucker}}]{hailey-dunsheath-2010}
{Hailey-Dunsheath}, S., {Nikola}, T., {Stacey}, G.~J., {et~al.} 2010, \apjl,
  714, L162

\bibitem[{{Houck} {et~al.}(2004){Houck}, {Roellig}, {Van Cleve}, {Forrest},
  {Herter}, {Lawrence}, {Matthews}, {Reitsema}, {Soifer}, {Watson}, {Weedman},
  {Huisjen}, {Troeltzsch}, {Barry}, {Bernard-Salas}, {Blacken}, {Brandl},
  {Charmandaris}, {Devost}, {Gull}, {Hall}, {Henderson}, {Higdon}, {Pirger},
  {Schoenwald}, {Sloan}, {Uchida}, {Appleton}, {Armus}, {Burgdorf},
  {Fajardo-Acosta}, {Grillmair}, {Ingalls}, {Morris}, \&
  {Teplitz}}]{houck-2004}
{Houck}, J.~R., {Roellig}, T.~L., {Van Cleve}, J., {et~al.} 2004, in Society of
  Photo-Optical Instrumentation Engineers (SPIE) Conference Series, Vol. 5487,
  Society of Photo-Optical Instrumentation Engineers (SPIE) Conference Series,
  ed. {J.~C.~Mather}, 62--76

\bibitem[{{Hunter} {et~al.}(2001){Hunter}, {Kaufman}, {Hollenbach}, {Rubin},
  {Malhotra}, {Dale}, {Brauher}, {Silbermann}, {Helou}, {Contursi}, \&
  {Lord}}]{hunter-2001}
{Hunter}, D.~A., {Kaufman}, M., {Hollenbach}, D.~J., {et~al.} 2001, \apj, 553,
  121

\bibitem[{{Inami} {et~al.}(2013){Inami}, {Armus}, {Charmandaris}, {Groves},
  {Kewley}, {Petric}, {Stierwalt}, {D{\'{\i}}az-Santos}, {Surace}, {Rich},
  {Haan}, {Howell}, {Evans}, {Mazzarella}, {Marshall}, {Appleton}, {Lord},
  {Spoon}, {Frayer}, {Matsuhara}, \& {Veilleux}}]{inami-2013}
{Inami}, H., {Armus}, L., {Charmandaris}, V., {et~al.} 2013, \apj, 777, 156

\bibitem[{{Israel} \& {Maloney}(2011)}]{israel-2011}
{Israel}, F.~P. \& {Maloney}, P.~R. 2011, \aap, 531, A19

\bibitem[{{Israel} {et~al.}(1996){Israel}, {Maloney}, {Geis}, {Herrmann},
  {Madden}, {Poglitsch}, \& {Stacey}}]{israel-1996}
{Israel}, F.~P., {Maloney}, P.~R., {Geis}, N., {et~al.} 1996, \apj, 465, 738

\bibitem[{{Karczewski et al.}(in prep.)}]{karczewski-2015}
{Karczewski et al.} in prep.

\bibitem[{{Kaufman} {et~al.}(2006){Kaufman}, {Wolfire}, \&
  {Hollenbach}}]{kaufman-2006}
{Kaufman}, M.~J., {Wolfire}, M.~G., \& {Hollenbach}, D.~J. 2006, \apj, 644, 283

\bibitem[{{Kaufman} {et~al.}(1999){Kaufman}, {Wolfire}, {Hollenbach}, \&
  {Luhman}}]{kaufman-1999}
{Kaufman}, M.~J., {Wolfire}, M.~G., {Hollenbach}, D.~J., \& {Luhman}, M.~L.
  1999, \apj, 527, 795

\bibitem[{{Kewley} \& {Ellison}(2008)}]{kewley-2008}
{Kewley}, L.~J. \& {Ellison}, S.~L. 2008, \apj, 681, 1183

\bibitem[{{Kobulnicky} \& {Kewley}(2004)}]{kk04}
{Kobulnicky}, H.~A. \& {Kewley}, L.~J. 2004, \apj, 617, 240

\bibitem[{{Lebouteiller} {et~al.}(2011){Lebouteiller}, {Barry}, {Spoon},
  {Bernard-Salas}, {Sloan}, {Houck}, \& {Weedman}}]{lebouteiller-2011}
{Lebouteiller}, V., {Barry}, D.~J., {Spoon}, H.~W.~W., {et~al.} 2011, \apjs,
  196, 8

\bibitem[{{Lebouteiller} {et~al.}(2012){Lebouteiller}, {Cormier}, {Madden},
  {Galliano}, {Indebetouw}, {Abel}, {Sauvage}, {Hony}, {Contursi}, {Poglitsch},
  {R{\'e}my}, {Sturm}, \& {Wu}}]{lebouteiller-2012}
{Lebouteiller}, V., {Cormier}, D., {Madden}, S.~C., {et~al.} 2012, \aap, 548,
  A91

\bibitem[{{Lebouteiller} {et~al.}(in prep.)}]{lebouteiller-2015}
{Lebouteiller}, V. {et~al.} in prep.

\bibitem[{{Leitherer} {et~al.}(2010){Leitherer}, {Ortiz Ot{\'a}lvaro},
  {Bresolin}, {Kudritzki}, {Lo Faro}, {Pauldrach}, {Pettini}, \&
  {Rix}}]{leitherer-2010}
{Leitherer}, C., {Ortiz Ot{\'a}lvaro}, P.~A., {Bresolin}, F., {et~al.} 2010,
  \apjs, 189, 309

\bibitem[{{Liang} {et~al.}(2006){Liang}, {Yin}, {Hammer}, {Deng}, {Flores}, \&
  {Zhang}}]{liang-2006}
{Liang}, Y.~C., {Yin}, S.~Y., {Hammer}, F., {et~al.} 2006, \apj, 652, 257

\bibitem[{{Liseau} {et~al.}(2006){Liseau}, {Justtanont}, \&
  {Tielens}}]{liseau-2006}
{Liseau}, R., {Justtanont}, K., \& {Tielens}, A.~G.~G.~M. 2006, \aap, 446, 561

\bibitem[{{Luhman} {et~al.}(1998){Luhman}, {Satyapal}, {Fischer}, {Wolfire},
  {Cox}, {Lord}, {Smith}, {Stacey}, \& {Unger}}]{luhman-1998}
{Luhman}, M.~L., {Satyapal}, S., {Fischer}, J., {et~al.} 1998, \apjl, 504, L11

\bibitem[{{Luhman} {et~al.}(2003){Luhman}, {Satyapal}, {Fischer}, {Wolfire},
  {Sturm}, {Dudley}, {Lutz}, \& {Genzel}}]{luhman-2003}
{Luhman}, M.~L., {Satyapal}, S., {Fischer}, J., {et~al.} 2003, \apj, 594, 758

\bibitem[{{Madden}(2000)}]{madden-2000}
{Madden}, S.~C. 2000, New Astronomy Review, 44, 249

\bibitem[{{Madden} {et~al.}(2006){Madden}, {Galliano}, {Jones}, \&
  {Sauvage}}]{madden-2006}
{Madden}, S.~C., {Galliano}, F., {Jones}, A.~P., \& {Sauvage}, M. 2006, \aap,
  446, 877

\bibitem[{{Madden} {et~al.}(1997){Madden}, {Poglitsch}, {Geis}, {Stacey}, \&
  {Townes}}]{madden-1997}
{Madden}, S.~C., {Poglitsch}, A., {Geis}, N., {Stacey}, G.~J., \& {Townes},
  C.~H. 1997, \apj, 483, 200

\bibitem[{{Madden} {et~al.}(2013){Madden}, {R{\'e}my-Ruyer}, {Galametz},
  {Cormier}, {Lebouteiller}, {Galliano}, {Hony}, {Bendo}, {Smith}, {Pohlen},
  {Roussel}, {Sauvage}, {Wu}, {Sturm}, {Poglitsch}, {Contursi}, {Doublier},
  {Baes}, {Barlow}, {Boselli}, {Boquien}, {Carlson}, {Ciesla}, {Cooray},
  {Cortese}, {de Looze}, {Irwin}, {Isaak}, {Kamenetzky}, {Karczewski}, {Lu},
  {MacHattie}, {O''Halloran}, {Parkin}, {Rangwala}, {Schirm}, {Schulz},
  {Spinoglio}, {Vaccari}, {Wilson}, \& {Wozniak}}]{madden-2013}
{Madden}, S.~C., {R{\'e}my-Ruyer}, A., {Galametz}, M., {et~al.} 2013, \pasp,
  125, 600

\bibitem[{{Madden} {et~al.}(in prep.)}]{madden-2015}
{Madden}, S.~C. {et~al.} in prep.

\bibitem[{{Maiolino} {et~al.}(2009){Maiolino}, {Caselli}, {Nagao}, {Walmsley},
  {De Breuck}, \& {Meneghetti}}]{maiolino-2009}
{Maiolino}, R., {Caselli}, P., {Nagao}, T., {et~al.} 2009, \aap, 500, L1

\bibitem[{{Malhotra} {et~al.}(1997){Malhotra}, {Helou}, {Stacey}, {Hollenbach},
  {Lord}, {Beichman}, {Dinerstein}, {Hunter}, {Lo}, {Lu}, {Rubin},
  {Silbermann}, {Thronson}, \& {Werner}}]{malhotra-1997}
{Malhotra}, S., {Helou}, G., {Stacey}, G., {et~al.} 1997, \apjl, 491, L27

\bibitem[{{Malhotra} {et~al.}(2001){Malhotra}, {Kaufman}, {Hollenbach},
  {Helou}, {Rubin}, {Brauher}, {Dale}, {Lu}, {Lord}, {Stacey}, {Contursi},
  {Hunter}, \& {Dinerstein}}]{malhotra-2001}
{Malhotra}, S., {Kaufman}, M.~J., {Hollenbach}, D., {et~al.} 2001, \apj, 561,
  766

\bibitem[{{Markwardt}(2009)}]{markwardt-2009}
{Markwardt}, C.~B. 2009, in Astronomical Society of the Pacific Conference
  Series, Vol. 411, Astronomical Data Analysis Software and Systems XVIII, ed.
  D.~A. {Bohlender}, D.~{Durand}, \& P.~{Dowler}, 251

\bibitem[{{Meyer} {et~al.}(1998){Meyer}, {Jura}, \& {Cardelli}}]{meyer-1998}
{Meyer}, D.~M., {Jura}, M., \& {Cardelli}, J.~A. 1998, \apj, 493, 222

\bibitem[{{Moustakas} {et~al.}(2010){Moustakas}, {Kennicutt}, {Tremonti},
  {Dale}, {Smith}, \& {Calzetti}}]{moustakas-2010}
{Moustakas}, J., {Kennicutt}, Jr., R.~C., {Tremonti}, C.~A., {et~al.} 2010,
  \apjs, 190, 233

\bibitem[{{Negishi} {et~al.}(2001){Negishi}, {Onaka}, {Chan}, \&
  {Roellig}}]{negishi-2001}
{Negishi}, T., {Onaka}, T., {Chan}, K., \& {Roellig}, T.~L. 2001, \aap, 375,
  566

\bibitem[{{Oberst} {et~al.}(2006){Oberst}, {Parshley}, {Stacey}, {Nikola},
  {L{\"o}hr}, {Harnett}, {Tothill}, {Lane}, {Stark}, \& {Tucker}}]{oberst-2006}
{Oberst}, T.~E., {Parshley}, S.~C., {Stacey}, G.~J., {et~al.} 2006, \apjl, 652,
  L125

\bibitem[{{Ott}(2010)}]{ott-2010}
{Ott}, S. 2010, in Astronomical Society of the Pacific Conference Series, Vol.
  434, Astronomical Data Analysis Software and Systems XIX, ed. {Y.~Mizumoto,
  K.-I.~Morita, \& M.~Ohishi}, 139

\bibitem[{{Papadopoulos} {et~al.}(2010){Papadopoulos}, {Isaak}, \& {van der
  Werf}}]{papadopoulos-2010}
{Papadopoulos}, P.~P., {Isaak}, K., \& {van der Werf}, P. 2010, \apj, 711, 757

\bibitem[{{P{\'e}quignot}(2008)}]{pequignot-2008}
{P{\'e}quignot}, D. 2008, \aap, 478, 371

\bibitem[{{Pierini} {et~al.}(2003){Pierini}, {Leech}, \&
  {V{\"o}lk}}]{pierini-2003}
{Pierini}, D., {Leech}, K.~J., \& {V{\"o}lk}, H.~J. 2003, \aap, 397, 871

\bibitem[{{Pilbratt} {et~al.}(2010){Pilbratt}, {Riedinger}, {Passvogel},
  {Crone}, {Doyle}, {Gageur}, {Heras}, {Jewell}, {Metcalfe}, {Ott}, \&
  {Schmidt}}]{pilbratt-2010}
{Pilbratt}, G.~L., {Riedinger}, J.~R., {Passvogel}, T., {et~al.} 2010, \aap,
  518, L1+

\bibitem[{{Pilyugin} \& {Thuan}(2005)}]{pt05}
{Pilyugin}, L.~S. \& {Thuan}, T.~X. 2005, \apj, 631, 231

\bibitem[{{Pineda} {et~al.}(2013){Pineda}, {Langer}, {Velusamy}, \&
  {Goldsmith}}]{pineda-2013}
{Pineda}, J.~L., {Langer}, W.~D., {Velusamy}, T., \& {Goldsmith}, P.~F. 2013,
  \aap, 554, A103

\bibitem[{{Poglitsch} {et~al.}(1995){Poglitsch}, {Krabbe}, {Madden}, {Nikola},
  {Geis}, {Johansson}, {Stacey}, \& {Sternberg}}]{poglitsch-1995}
{Poglitsch}, A., {Krabbe}, A., {Madden}, S.~C., {et~al.} 1995, \apj, 454, 293

\bibitem[{{Poglitsch} {et~al.}(2010){Poglitsch}, {Waelkens}, {Geis},
  {Feuchtgruber}, {Vandenbussche}, {Rodriguez}, {Krause}, {Renotte}, {van
  Hoof}, {Saraceno}, {Cepa}, {Kerschbaum}, {Agn{\`e}se}, {Ali}, {Altieri},
  {Andreani}, {Augueres}, {Balog}, {Barl}, {Bauer}, {Belbachir}, {Benedettini},
  {Billot}, {Boulade}, {Bischof}, {Blommaert}, {Callut}, {Cara}, {Cerulli},
  {Cesarsky}, {Contursi}, {Creten}, {De Meester}, {Doublier}, {Doumayrou},
  {Duband}, {Exter}, {Genzel}, {Gillis}, {Gr{\"o}zinger}, {Henning},
  {Herreros}, {Huygen}, {Inguscio}, {Jakob}, {Jamar}, {Jean}, {de Jong},
  {Katterloher}, {Kiss}, {Klaas}, {Lemke}, {Lutz}, {Madden}, {Marquet},
  {Martignac}, {Mazy}, {Merken}, {Montfort}, {Morbidelli}, {M{\"u}ller},
  {Nielbock}, {Okumura}, {Orfei}, {Ottensamer}, {Pezzuto}, {Popesso},
  {Putzeys}, {Regibo}, {Reveret}, {Royer}, {Sauvage}, {Schreiber}, {Stegmaier},
  {Schmitt}, {Schubert}, {Sturm}, {Thiel}, {Tofani}, {Vavrek}, {Wetzstein},
  {Wieprecht}, \& {Wiezorrek}}]{poglitsch-2010}
{Poglitsch}, A., {Waelkens}, C., {Geis}, N., {et~al.} 2010, \aap, 518, L2+

\bibitem[{{Rangwala} {et~al.}(2011){Rangwala}, {Maloney}, {Glenn}, {Wilson},
  {Rykala}, {Isaak}, {Baes}, {Bendo}, {Boselli}, {Bradford}, {Clements},
  {Cooray}, {Fulton}, {Imhof}, {Kamenetzky}, {Madden}, {Mentuch}, {Sacchi},
  {Sauvage}, {Schirm}, {Smith}, {Spinoglio}, \& {Wolfire}}]{rangwala-2011}
{Rangwala}, N., {Maloney}, P.~R., {Glenn}, J., {et~al.} 2011, \apj, 743, 94

\bibitem[{{R{\'e}my-Ruyer} {et~al.}(2014){R{\'e}my-Ruyer}, {Madden},
  {Galliano}, {Galametz}, {Takeuchi}, {Asano}, {Zhukovska}, {Lebouteiller},
  {Cormier}, {Jones}, {Bocchio}, {Baes}, {Bendo}, {Boquien}, {Boselli},
  {DeLooze}, {Doublier-Pritchard}, {Hughes}, {Karczewski}, \&
  {Spinoglio}}]{remy-2014}
{R{\'e}my-Ruyer}, A., {Madden}, S.~C., {Galliano}, F., {et~al.} 2014, \aap,
  563, A31

\bibitem[{{R{\'e}my-Ruyer} {et~al.}(2013){R{\'e}my-Ruyer}, {Madden},
  {Galliano}, {Hony}, {Sauvage}, {Bendo}, {Roussel}, {Pohlen}, {Smith},
  {Galametz}, {Cormier}, {Lebouteiller}, {Wu}, {Baes}, {Barlow}, {Boquien},
  {Boselli}, {Ciesla}, {De Looze}, {Karczewski}, {Panuzzo}, {Spinoglio},
  {Vaccari}, \& {Wilson}}]{remy-2013}
{R{\'e}my-Ruyer}, A., {Madden}, S.~C., {Galliano}, F., {et~al.} 2013, \aap,
  557, A95

\bibitem[{{R{\'e}my-Ruyer} {et~al.}(in prep.)}]{remy-2015}
{R{\'e}my-Ruyer}, A. {et~al.} in prep.

\bibitem[{{R{\"o}llig} {et~al.}(2006){R{\"o}llig}, {Ossenkopf}, {Jeyakumar},
  {Stutzki}, \& {Sternberg}}]{roellig-2006}
{R{\"o}llig}, M., {Ossenkopf}, V., {Jeyakumar}, S., {Stutzki}, J., \&
  {Sternberg}, A. 2006, \aap, 451, 917

\bibitem[{{Rubin} {et~al.}(2009){Rubin}, {Hony}, {Madden}, {Tielens},
  {Meixner}, {Indebetouw}, {Reach}, {Ginsburg}, {Kim}, {Mochizuki}, {Babler},
  {Block}, {Bracker}, {Engelbracht}, {For}, {Gordon}, {Hora}, {Leitherer},
  {Meade}, {Misselt}, {Sewilo}, {Vijh}, \& {Whitney}}]{rubin-2009}
{Rubin}, D., {Hony}, S., {Madden}, S.~C., {et~al.} 2009, \aap, 494, 647

\bibitem[{{Rubin} {et~al.}(1994){Rubin}, {Simpson}, {Lord}, {Colgan},
  {Erickson}, \& {Haas}}]{rubin-1994}
{Rubin}, R.~H., {Simpson}, J.~P., {Lord}, S.~D., {et~al.} 1994, \apj, 420, 772

\bibitem[{{Sandstrom} {et~al.}(2012){Sandstrom}, {Bolatto}, {Bot}, {Draine},
  {Ingalls}, {Israel}, {Jackson}, {Leroy}, {Li}, {Rubio}, {Simon}, {Smith},
  {Stanimirovi{\'c}}, {Tielens}, \& {van Loon}}]{sandstrom-2012}
{Sandstrom}, K.~M., {Bolatto}, A.~D., {Bot}, C., {et~al.} 2012, \apj, 744, 20

\bibitem[{{Sargsyan} {et~al.}(2014){Sargsyan}, {Samsonyan}, {Lebouteiller},
  {Weedman}, {Barry}, {Bernard-Salas}, {Houck}, \& {Spoon}}]{sargsyan-2014}
{Sargsyan}, L., {Samsonyan}, A., {Lebouteiller}, V., {et~al.} 2014, \apj, 790,
  15

\bibitem[{{Savage} \& {Sembach}(1996)}]{savage-1996}
{Savage}, B.~D. \& {Sembach}, K.~R. 1996, \araa, 34, 279

\bibitem[{{Schlegel} {et~al.}(1998){Schlegel}, {Finkbeiner}, \&
  {Davis}}]{schlegel-1998}
{Schlegel}, D.~J., {Finkbeiner}, D.~P., \& {Davis}, M. 1998, \apj, 500, 525

\bibitem[{{Stacey} {et~al.}(1991){Stacey}, {Geis}, {Genzel}, {Lugten},
  {Poglitsch}, {Sternberg}, \& {Townes}}]{stacey-1991}
{Stacey}, G.~J., {Geis}, N., {Genzel}, R., {et~al.} 1991, \apj, 373, 423

\bibitem[{{Stacey} {et~al.}(2010){Stacey}, {Hailey-Dunsheath}, {Ferkinhoff},
  {Nikola}, {Parshley}, {Benford}, {Staguhn}, \& {Fiolet}}]{stacey-2010}
{Stacey}, G.~J., {Hailey-Dunsheath}, S., {Ferkinhoff}, C., {et~al.} 2010, \apj,
  724, 957

\bibitem[{{Swinbank} {et~al.}(2012){Swinbank}, {Karim}, {Smail}, {Hodge},
  {Walter}, {Bertoldi}, {Biggs}, {de Breuck}, {Chapman}, {Coppin}, {Cox},
  {Danielson}, {Dannerbauer}, {Ivison}, {Greve}, {Knudsen}, {Menten},
  {Simpson}, {Schinnerer}, {Wardlow}, {Wei{\ss}}, \& {van der
  Werf}}]{swinbank-2012}
{Swinbank}, A.~M., {Karim}, A., {Smail}, I., {et~al.} 2012, \mnras, 427, 1066

\bibitem[{{Tielens} \& {Hollenbach}(1985)}]{tielens-1985}
{Tielens}, A.~G.~G.~M. \& {Hollenbach}, D. 1985, \apj, 291, 722

\bibitem[{{Verma} {et~al.}(2003){Verma}, {Lutz}, {Sturm}, {Sternberg},
  {Genzel}, \& {Vacca}}]{verma-2003}
{Verma}, A., {Lutz}, D., {Sturm}, E., {et~al.} 2003, \aap, 403, 829

\bibitem[{{Weingartner} \& {Draine}(2001)}]{weingartner-2001}
{Weingartner}, J.~C. \& {Draine}, B.~T. 2001, \apj, 548, 296

\bibitem[{{Wolfire} {et~al.}(2010){Wolfire}, {Hollenbach}, \&
  {McKee}}]{wolfire-2010}
{Wolfire}, M.~G., {Hollenbach}, D., \& {McKee}, C.~F. 2010, \apj, 716, 1191

\bibitem[{{Wolfire} {et~al.}(1990){Wolfire}, {Tielens}, \&
  {Hollenbach}}]{wolfire-1990}
{Wolfire}, M.~G., {Tielens}, A.~G.~G.~M., \& {Hollenbach}, D. 1990, \apj, 358,
  116

\end{thebibliography}

\onecolumn
\begin{landscape}\scriptsize 
\begin{longtab}
\begin{longtable}{l c c c c c c c c c}
  \caption{Line fluxes of the DGS galaxies.\newline
Notes. Fluxes are in 10$^{-18}$\,W\,m$^{-2}$ for compact and fully mapped extended 
galaxies ($compact$ sample), and surface brightness in W\,m$^{-2}$\,sr$^{-1}$ 
for partially mapped extended galaxies ($extended$ sample). 
Upper limits are 3$\sigma$ levels. 
Beam sizes are 9.5\arcs at 57, 63, and 88\,\mum, 10\arcs at 122\,\mum, 
11\arcs at 145\,\mum, 11.5\arcs at 157\,\mum, and 14.5\arcs at 205\,\mum. \\
$(a)$~Extraction methods for the $compact$ sample: $central$, $3\times3$, 
$5\times5$, aperture photometry (radius in \arcs), or total of the line map 
(see Sect.~\ref{sect:fluxextract} for details). 
For the $extended$ sample, we indicate the peak surface brightness 
of the region mapped. The peak position is shown by a dark cross on the line maps 
in Appendix~\ref{app:append-b}). \\
$(b)$~Correction factor derived from the PACS 100\,\mum maps to apply 
in order to account for the fraction of line emission outside of the coverage 
of our observations. 
All galaxies with emission at the edges of the PACS observations are 
indicated with the symbol (*), and we report correction factors if missing more 
than 15\% of the total flux. For NGC\,1140, this factor has to be applied 
for all fluxes but \cii because it corrects for a different line coverage 
(see Sect.~\ref{sect:incomplete} for details). \\
We note that for UM\,311 and NGC\,4861, the PACS fluxes and correction 
factors are measured at the position of the main \hii complexes. The \cii maps 
cover more of the system/galaxy, with a total flux of $929$ 
and $516\times10^{-18}$\,W\,m$^{-2}$, respectively.
} \\
    \hline\hline
     \vspace{-8pt}\\
    \multicolumn{1}{l}{Source Name} & 
    \multicolumn{1}{c}{Extract.$^{(a)}$} & 
    \multicolumn{7}{c}{Observed Spectral Line} & 
    \multicolumn{1}{c}{Corr. factor$^{(b)}$} \\ \cline{3-9}
    \multicolumn{1}{l}{} &
    \multicolumn{1}{c}{} &
    \multicolumn{1}{c}{\cii157\,\mum} &
    \multicolumn{1}{c}{\oi63\,\mum} &
    \multicolumn{1}{c}{\oiii88\,\mum} &
    \multicolumn{1}{c}{\oi145\,\mum} &
    \multicolumn{1}{c}{\nii122\,\mum} &
    \multicolumn{1}{c}{\nii205\,\mum} &
    \multicolumn{1}{c}{\niii57\,\mum} &
    \multicolumn{1}{c}{} \\
    \hline
    \vspace{-8pt}\\
\endfirsthead

  \caption{Line fluxes of the DGS galaxies.} \\
    \hline\hline
     \vspace{-8pt}\\
    \multicolumn{1}{l}{Source Name} & 
    \multicolumn{1}{c}{Extraction$^{(a)}$} & 
    \multicolumn{7}{c}{Observed Spectral Line} & 
    \multicolumn{1}{c}{Corr. factor$^{(b)}$} \\ \cline{3-9}
    \multicolumn{1}{l}{} &
    \multicolumn{1}{c}{} &
    \multicolumn{1}{c}{\cii157\,\mum} &
    \multicolumn{1}{c}{\oi63\,\mum} &
    \multicolumn{1}{c}{\oiii88\,\mum} &
    \multicolumn{1}{c}{\oi145\,\mum} &
    \multicolumn{1}{c}{\nii122\,\mum} &
    \multicolumn{1}{c}{\nii205\,\mum} &
    \multicolumn{1}{c}{\niii57\,\mum} &
    \multicolumn{1}{c}{} \\
    \hline
    \vspace{-8pt}\\
\endhead

\hline \multicolumn{10}{l}{\textit{continued.}}
\endfoot

\endlastfoot

    \hline
    \multicolumn{10}{l}{$Compact$ sample} \\
    \hline
Haro\,11   & $3\times3$ & $6.55\pm0.07~(\times 10^{2})$ &    $6.44\pm0.12~(\times 10^{2})$ &    $1.72\pm0.03~(\times 10^{3})$ &    $50.00\pm3.53$ &    $35.10\pm2.75$ &    $\le 25.44$ & $2.83\pm0.08~(\times 10^{2})$ &    $   -  $ \\
Haro\,2   & $5\times5$ & $8.24\pm0.11~(\times 10^{2})$ &    $5.44\pm0.20~(\times 10^{2})$ &    $9.72\pm0.35~(\times 10^{2})$ &    $ -  $ &  $ -  $ &  $ -  $ &  $ -  $ &  $   -  $ \\
Haro\,3   & $5\times5$ & $1.03\pm0.01~(\times 10^{3})$ &    $5.90\pm0.27~(\times 10^{2})$ &    $1.85\pm0.04~(\times 10^{3})$ &    $48.20\pm4.53$ &    $21.40\pm3.87$ &    $ -  $ &  $1.23\pm0.17~(\times 10^{2})$ &    $   -  $ \\
He\,2-10   & $5\times5$ & $3.88\pm0.05~(\times 10^{3})$ &    $2.52\pm0.04~(\times 10^{3})$ &    $3.38\pm0.05~(\times 10^{3})$ &    $1.82\pm0.03~(\times 10^{2})$ &    $ -  $ &  $ -  $ &  $9.85\pm1.09~(\times 10^{2})$ &    $   -  $ \\
HS\,0017+1055   & $3\times3$ & $4.19\pm0.75$ &    $ -  $ &  $9.83\pm1.38$ &    $ -  $ &  $ -  $ &  $ -  $ &  $ -  $ &  $   -  $ \\
HS\,0052+2536   & $central$ & $43.90\pm3.89$ &    $24.70\pm4.60$ &    $70.70\pm4.04$ &    $ -  $ &  $ -  $ &  $ -  $ &  $ -  $ &  $   -  $ \\
HS\,0822+3542   & $central$ & $1.96\pm0.55$ &    $ -  $ &  $ -  $ &  $ -  $ &  $ -  $ &  $ -  $ &  $ -  $ &  $   -  $ \\
HS\,1222+3741   & $central$ & $2.43\pm0.45$ &    $ -  $ &  $14.60\pm1.54$ &    $ -  $ &  $ -  $ &  $ -  $ &  $ -  $ &  $   -  $ \\
HS\,1236+3937   & $central$ & $\le 3.45$ & $ -  $ &  $ -  $ &  $ -  $ &  $ -  $ &  $ -  $ &  $ -  $ &  $   -  $ \\
HS\,1304+3529   & $3\times3$ & $26.10\pm2.19$ &    $ -  $ &  $31.90\pm4.86$ &    $ -  $ &  $ -  $ &  $ -  $ &  $ -  $ &  $   -  $ \\
HS\,1319+3224   & $central$ & $2.78\pm0.83$ &    $ -  $ &  $9.24\pm2.06$ &    $ -  $ &  $ -  $ &  $ -  $ &  $ -  $ &  $   -  $ \\
HS\,1330+3651   & $3\times3$ & $29.20\pm2.96$ &    $\le 20.28$ & $29.80\pm2.08$ &    $ -  $ &  $ -  $ &  $ -  $ &  $ -  $ &  $   -  $ \\
HS\,1442+4250   & $3\times3$ & $12.90\pm2.25$ &    $ -  $ &  $ -  $ &  $ -  $ &  $ -  $ &  $ -  $ &  $ -  $ &  $   -  $ \\
HS\,2352+2733   & $central$ & $\le 3.84$ & $ -  $ &  $6.94\pm1.97$ &    $ -  $ &  $ -  $ &  $ -  $ &  $ -  $ &  $   -  $ \\
II\,Zw\,40   & $5\times5$ & $8.49\pm0.10~(\times 10^{2})$ &    $6.31\pm0.26~(\times 10^{2})$ &    $3.59\pm0.04~(\times 10^{3})$ &    $40.80\pm4.78$ &    $\le 26.52$ & $ -  $ &  $ -  $ &  $   -  $ \\
I\,Zw\,18   & $3\times3$ & $10.60\pm0.82$ &    $11.50\pm2.14$ &    $28.40\pm3.43$ &    $ -  $ &  $ -  $ &  $ -  $ &  $ -  $ &  $   -  $ \\
Mrk\,1089~(*)   & $5\times5$ & $8.59\pm0.08~(\times 10^{2})$ &    $4.07\pm0.32~(\times 10^{2})$ &    $9.38\pm0.33~(\times 10^{2})$ &    $27.60\pm5.00$ &    $35.10\pm6.04$ &    $ -  $ &  $ -  $ &  $ 0.15 $ \\
Mrk\,1450   & $3\times3$ & $60.50\pm4.28$ &    $35.50\pm2.77$ &    $2.62\pm0.09~(\times 10^{2})$ &    $ -  $ &  $2.56\pm0.65$ &    $ -  $ &  $ -  $ &  $   -  $ \\
Mrk\,153~(*)   & $3\times3$ & $53.10\pm1.43$ &    $29.70\pm7.57$ &    $98.90\pm3.80$ &    $ -  $ &  $ -  $ &  $ -  $ &  $ -  $ &  $ 0.15 $ \\
Mrk\,209~(*)   & $5\times5$ & $68.20\pm3.55$ &    $38.90\pm11.60$ &    $3.13\pm0.12~(\times 10^{2})$ &    $ -  $ &  $ -  $ &  $ -  $ &  $ -  $ &  $   -  $ \\
Mrk\,930   & $3\times3$ & $2.18\pm0.06~(\times 10^{2})$ &    $1.42\pm0.12~(\times 10^{2})$ &    $4.21\pm0.20~(\times 10^{2})$ &    $9.07\pm0.99$ &    $\le 8.31$ & $ -  $ &  $ -  $ &  $ 0.15 $ \\
NGC\,1140   & $total$ & $1.19\pm0.02~(\times 10^{3})$ &    $4.27\pm0.20~(\times 10^{2})$ &    $1.08\pm0.03~(\times 10^{3})$ &    $35.40\pm3.08$ &    $20.90\pm3.20$ &    $ -  $ &  $ -  $ &  $ [0.20] $ \\
NGC\,1569~(*)   & $3\times3$ & $1.33\pm0.01~(\times 10^{4})$ &    $7.39\pm0.05~(\times 10^{3})$ &    $2.80\pm0.01~(\times 10^{4})$ &    $ -  $ &  $ -  $ &  $ -  $ &  $ -  $ &  $ 0.15 $ \\
NGC\,1705~(*)   & $5\times5/total$ & $3.90\pm0.10~(\times 10^{2})$ &    $1.56\pm0.17~(\times 10^{2})$ &    $5.46\pm0.26~(\times 10^{2})$ &    $ -  $ &  $ -  $ &  $ -  $ &  $ -  $ &  $   -  $ \\
NGC\,2366~(*)   & $5\times5/total$ & $4.90\pm0.09~(\times 10^{2})$ &    $4.60\pm0.16~(\times 10^{2})$ &    $2.26\pm0.01~(\times 10^{3})$ &    $ -  $ &  $ -  $ &  $ -  $ &  $ -  $ &  $0.55$ \\
NGC\,4214~(*)   & $total$ & $5.42\pm0.03~(\times 10^{3})$ &    $1.95\pm0.05~(\times 10^{3})$ &    $4.85\pm0.07~(\times 10^{3})$ &    $1.45\pm0.06~(\times 10^{2})$ &    $78.90\pm12.00$ &    $\le 77.98$ &    $ -  $ &  $0.35$ \\
NGC\,4861~(*)   & $5\times5/total$ & $3.08\pm0.05~(\times 10^{2})$ &    $2.23\pm0.24~(\times 10^{2})$ &    $1.26\pm0.02~(\times 10^{3})$ &    $16.10\pm2.48$ &    $ -  $ &  $ -  $ &  $ -  $ &  $0.35$ \\
NGC\,5253   & $5\times5$ & $4.51\pm0.01~(\times 10^{3})$ &    $2.93\pm0.05~(\times 10^{3})$ &    $9.01\pm0.04~(\times 10^{3})$ &    $2.06\pm0.04~(\times 10^{2})$ &    $92.50\pm12.10$ &    $ -  $ &  $ -  $ &  $   -  $ \\
NGC\,625~(*)   & $5\times5/total$ & $1.50\pm0.01~(\times 10^{3})$ &    $5.69\pm0.30~(\times 10^{2})$ &    $2.45\pm0.03~(\times 10^{3})$ &    $ -  $ &  $ -  $ &  $ -  $ &  $ -  $ &  $ 0.20 $ \\
Pox\,186   & $central$ & $3.14\pm0.81$ &    $ -  $ &  $33.70\pm3.33$ &    $ -  $ &  $ -  $ &  $ -  $ &  $ -  $ &  $   -  $ \\
SBS\,0335-052   & $central/3\times3$ & $6.61\pm0.97$ &    $21.90\pm2.25$ &    $43.60\pm2.59$ &    $ -  $ &  $ -  $ &  $ -  $ &  $ -  $ &  $   -  $ \\
SBS\,1159+545   & $central$ & $3.48\pm0.57$ &    $4.86\pm0.97$ &    $4.81\pm1.22$ &    $ -  $ &  $ -  $ &  $ -  $ &  $ -  $ &  $   -  $ \\
SBS\,1211+540   & $3\times3$ & $5.48\pm1.77$ &    $ -  $ &  $ -  $ &  $ -  $ &  $ -  $ &  $ -  $ &  $ -  $ &  $   -  $ \\
SBS\,1249+493   & $3\times3$ & $6.06\pm1.01$ &    $ -  $ &  $ -  $ &  $ -  $ &  $ -  $ &  $ -  $ &  $ -  $ &  $   -  $ \\
SBS\,1415+437   & $3\times3$ & $36.40\pm1.78$ &    $39.10\pm3.72$ &    $53.20\pm4.39$ &    $ -  $ &  $ -  $ &  $ -  $ &  $ -  $ &  $   -  $ \\
SBS\,1533+574   & $3\times3$ & $51.40\pm2.40$ &    $33.20\pm3.57$ &    $88.60\pm5.06$ &    $ -  $ &  $ -  $ &  $ -  $ &  $ -  $ &  $   -  $ \\
Tol\,1214-277   & $central$ & $4.98\pm1.58$ &    $5.70\pm1.70$ &    $15.10\pm2.23$ &    $ -  $ &  $ -  $ &  $ -  $ &  $ -  $ &  $   -  $ \\
UGC\,4483   & $3\times3/5\times5$ & $41.50\pm5.14$ &    $15.60\pm5.11$ &    $25.70\pm2.67$ &    $ -  $ &  $ -  $ &  $ -  $ &  $ -  $ &  $   -  $ \\
UM\,133   & $3\times3$ & $41.40\pm4.13$ &    $20.30\pm4.10$ &    $24.50\pm5.03$ &    $ -  $ &  $ -  $ &  $ -  $ &  $ -  $ &  $   -  $ \\
UM\,311~(*)   & $5\times5$ & $4.99\pm0.08~(\times 10^{2})$ &    $2.62\pm0.20~(\times 10^{2})$ &    $5.08\pm0.27~(\times 10^{2})$ &    $ -  $ &  $\le 11.31$ & $ -  $ &  $ -  $ &  $0.60$ \\
UM\,448   & $central$ & $8.00\pm0.19~(\times 10^{2})$ &    $5.89\pm0.27~(\times 10^{2})$ &    $1.06\pm0.02~(\times 10^{3})$ &    $43.40\pm2.60$ &    $12.20\pm2.14$ &    $ -  $ &  $ -  $ &  $   -  $ \\
UM\,461   & $5\times5$ & $21.80\pm2.57$ &    $22.70\pm4.97$ &    $79.00\pm5.91$ &    $ -  $ &  $\le 2.04$ & $ -  $ &  $ -  $ &  $   -  $ \\
VII\,Zw\,403   & $3\times3/30$\arcs & $1.57\pm0.05~(\times 10^{2})$ &    $42.70\pm8.82$ &    $81.60\pm8.92$ &    $7.36\pm1.62$ &    $\le 5.88$ & $ -  $ &  $ -  $ &  $   -  $ \\
    \hline\hline
    \multicolumn{10}{l}{$Extended$ sample} \\
    \hline
IC\,10   & $n.a.$ & $3.63\pm0.02~(\times 10^{-7})$ &    $3.40\pm0.06~(\times 10^{-7})$ &    $9.28\pm0.11~(\times 10^{-7})$ &    $1.73\pm0.08~(\times 10^{-8})$ &    $8.05\pm1.05~(\times 10^{-9})$ &    $ -  $ &  $ -  $ &  $   -  $ \\
LMC-30Dor   & $n.a.$ & $2.29\pm0.01~(\times 10^{-6})$ &    $3.58\pm0.02~(\times 10^{-6})$ &    $2.12\pm0.02~(\times 10^{-5})$ &    $3.50\pm0.04~(\times 10^{-7})$ &    $5.23\pm0.22~(\times 10^{-8})$ &    $ -  $ &  $ -  $ &  $   -  $ \\
LMC-N11A   & $n.a.$ & $3.53\pm0.04~(\times 10^{-7})$ &    $7.94\pm0.06~(\times 10^{-7})$ &    $ -  $ &    $ -  $ &  $ -  $ &  $ -  $ &  $ -  $ &  $   -  $ \\
LMC-N11B   & $n.a.$ & $7.92\pm0.10~(\times 10^{-7})$ &    $1.66\pm0.02~(\times 10^{-6})$ &    $2.12\pm0.04~(\times 10^{-6})$ &    $8.33\pm0.30~(\times 10^{-8})$ &    $1.48\pm0.26~(\times 10^{-8})$ &    $\le1.10~(\times 10^{-8})$ &    $1.53\pm0.13~(\times 10^{-7})$ &    $   -  $ \\
LMC-N11C   & $n.a.$ & $4.78\pm0.04~(\times 10^{-7})$ &    $4.28\pm0.04~(\times 10^{-7})$ &    $1.89\pm0.01~(\times 10^{-6})$ &    $ -  $ &  $ -  $ &  $ -  $ &  $ -  $ &  $   -  $ \\
LMC-N11I   & $n.a.$ & $2.12\pm0.02~(\times 10^{-7})$ &    $2.81\pm0.03~(\times 10^{-7})$ &    $1.25\pm0.02~(\times 10^{-7})$ &    $ -  $ &  $ -  $ &  $ -  $ &  $ -  $ &  $   -  $ \\
LMC-N158   & $n.a.$ & $9.90\pm0.07~(\times 10^{-7})$ &    $1.03\pm0.01~(\times 10^{-6})$ &    $2.75\pm0.02~(\times 10^{-6})$ &    $ -  $ &  $ -  $ &  $ -  $ &  $ -  $ &  $   -  $ \\
LMC-N159   & $n.a.$ & $1.30\pm0.01~(\times 10^{-6})$ &    $1.56\pm0.01~(\times 10^{-6})$ &    $5.06\pm0.03~(\times 10^{-6})$ &    $2.18\pm0.03~(\times 10^{-7})$ &    $2.92\pm0.15~(\times 10^{-8})$ &    $\le1.73~(\times 10^{-8})$ &    $ -  $ &  $   -  $ \\
LMC-N160   & $n.a.$ & $1.35\pm0.01~(\times 10^{-6})$ &    $3.40\pm0.02~(\times 10^{-6})$ &    $9.79\pm0.05~(\times 10^{-6})$ &    $3.49\pm0.03~(\times 10^{-7})$ &    $ -  $ &  $ -  $ &  $ -  $ &  $   -  $ \\
NGC\,4449   & $n.a.$ & $1.60\pm0.01~(\times 10^{-7})$ &    $8.74\pm0.19~(\times 10^{-8})$ &    $2.61\pm0.03~(\times 10^{-7})$ &    $6.49\pm0.85~(\times 10^{-9})$ &    $5.34\pm1.20~(\times 10^{-9})$ &    $ -  $ &  $ -  $ &  $   -  $ \\
NGC\,6822   & $n.a.$ & $1.43\pm0.01~(\times 10^{-7})$ &    $1.36\pm0.04~(\times 10^{-7})$ &    $7.33\pm0.09~(\times 10^{-7})$ &    $7.76\pm1.03~(\times 10^{-9})$ &    $\le6.00~(\times 10^{-9})$ &    $ -  $ &  $ -  $ &  $   -  $ \\
SMC-N66   & $n.a.$ & $1.86\pm0.01~(\times 10^{-7})$ &    $5.14\pm0.04~(\times 10^{-7})$ &    $6.61\pm0.08~(\times 10^{-7})$ &    $ -  $ &  $ -  $ &  $ -  $ &  $ -  $ &  $   -  $ \\
    \hline \hline
    \vspace{-5pt}
\label{table:totfluxes}
\end{longtable}
\end{longtab}
\end{landscape}

\onecolumn
\appendix{
\section{Details of the PACS observing strategy}
\label{app:append-a}
\small
\begin{longtab}
\begin{longtable}[!ht]{l c c l l c l}
  \caption{Technical details about the DGS observations.\newline
Notes. $(a)$ Coordinates of the map center. 
$(b)$ Observation Identification number. 
$(c)$ Observing mode: UN=Unchopped Line Scan, CN=Chopnod Line Scan, WS=Wavelength Switching. 
When left blank, the coordinates, observing mode, and map size are identical to the previous line. }\\
    \hline\hline
     \vspace{-8pt}\\
    \multicolumn{1}{l}{Source Name} & 
    \multicolumn{2}{c}{Coordinates$^{(a)}$} & 
    \multicolumn{1}{c}{OBSID$^{(b)}$} &
    \multicolumn{1}{c}{Mode$^{(c)}$} & 
    \multicolumn{1}{c}{Map size} &
    \multicolumn{1}{l}{Line} \\ \cline{2-3}
   \multicolumn{1}{l}{} & 
    \multicolumn{2}{c}{(J2000)} &
    \multicolumn{1}{l}{} &
    \multicolumn{1}{l}{} & 
    \multicolumn{1}{c}{($^{\prime\prime}$)$^2$} &
    \multicolumn{1}{l}{} \\
    \hline 
    \vspace{-8pt}\\
\endfirsthead
  \caption[]{Technical details about the DGS observations.} \\
    \hline\hline
     \vspace{-8pt}\\
    \multicolumn{1}{l}{Source Name} & 
    \multicolumn{2}{c}{Coordinates$^{(a)}$} & 
    \multicolumn{1}{c}{OBSID$^{(b)}$} &
    \multicolumn{1}{c}{Mode$^{(c)}$} & 
    \multicolumn{1}{c}{Map size} &
    \multicolumn{1}{l}{Line} \\ \cline{2-3}
   \multicolumn{1}{l}{} & 
    \multicolumn{2}{c}{(J2000)} &
    \multicolumn{1}{l}{} &
    \multicolumn{1}{l}{} & 
    \multicolumn{1}{c}{($^{\prime\prime}$)$^2$} &
    \multicolumn{1}{l}{} \\
    \hline 
    \vspace{-8pt}\\
\endhead

\hline \multicolumn{7}{l}{\textit{continued.}}
\endfoot

\endlastfoot

Haro\,11		&	0h36m52.50s	&	-33d33m19.0s		&	1342199236	&	CN	&	51$\times$51	&	\ciil, \oilb, \oiiil	\\
			&	0h36m52.50s	&	-33d33m19.0s		&	1342199237	&		&			&	\oila, \niiil, \niila	\\
			&	0h36m52.36s	&	-33d33m19.7s		&	1342234063	&		&	47$\times$47	&	\niilb	\\
Haro\,2		&	10h32m31.88s	&	+54d24m03.7s		&	1342221392	&	CN	&	47$\times$47	&	\oila	\\
			&				&					&	1342230081	&		&	71$\times$71	&	\ciil	\\
			&				&					&	1342230082	&		&	47$\times$47	&	\oiiil	\\
Haro\,3		&	10h45m22.41s	&	+55d57m37.4s		&	1342221890	&	CN	&	79$\times$79	&	\oiiil	\\
			&				&					&	1342221891	&		&	79$\times$79	&	\oila	\\
			&				&					&	1342221892	&		&	95$\times$95	&	\ciil	\\
			&				&					&	1342221893	&		&	47$\times$47	&	\niila	\\
			&				&					&	1342221894	&		&	71$\times$71	&	\oilb	\\
			&				&					&	1342221895	&		&	63$\times$63	&	\niiil	\\
He\,2-10		&	8h36m15.18s	&	-26d24m33.9s		&	1342221971	&	CN	&	47$\times$47	&	\niiil	\\
			&				&					&	1342221972	&		&	47$\times$47	&	\oilb	\\
			&				&					&	1342221973	&		&	53$\times$53	&	\oiiil	\\
			&				&					&	1342221974	&		&	53$\times$53	&	\oila	\\
			&				&					&	1342221975	&		&	51$\times$51	&	\ciil	\\
HS\,0017+1055	&	0h20m21.40s	&	+11d12m21.0s		&	1342234065	&	CN	&	47$\times$47	&	\ciil	\\
			&				&					&	1342212588	&		&			&	\oiiil	\\
HS\,0052+2536	&	0h54m56.00s	&	+25d53m23.0s		&	1342213134	&	CN	&	47$\times$47	&	\ciil	\\
			&				&					&	1342213135	&		&			&	\oila	\\
			&				&					&	1342213136	&		&			&	\oiiil	\\
HS\,0822+3542	&	8h25m55.52s	&	+35d32m32.0s		&	1342220753	&	CN	&	47$\times$47	&	\ciil	\\
HS\,1222+3741	&	12h24m36.72s	&	+37d24m36.6s		&	1342232306	&	CN	&	47$\times$47	&	\ciil	\\
			&				&					&	1342199399	&		&			&	\oiiil	\\
HS\,1236+3937	&	12h39m20.14s	&	+39d21m04.8s		&	1342199400	&	CN	&	47$\times$47	&	\ciil	\\
HS\,1304+3529	&	13h21m19.92s	&	+32d08m23.0s		&	1342199736	&	CN	&	47$\times$47	&	\ciil	\\
			&	13h06m24.24s	&	+35d13m41.6s		&	1342232552	&		&			&	\oiiil	\\
HS\,1319+3224	&	13h21m19.71s	&	+32d08m25.2s		&	1342199227	&	CN	&	47$\times$47	&	\ciil	\\
			&				&					&	1342232551	&		&			&	\oiiil	\\
HS\,1330+3651	&	13h33m08.28s	&	+36d36m33.1s		&	1342199734	&	CN	&	47$\times$47	&	\ciil	\\
			&				&					&	1342199735	&		&			&	\oiiil	\\
			&				&					&	1342232549	&		&			&	\oila	\\
HS\,1442+4250	&	14h44m12.81s	&	+42d37m44.0s		&	1342208927	&	CN	&	47$\times$47	&	\ciil	\\
HS\,2352+2733	&	23h54m56.70s	&	+27d49m59.0s		&	1342213133	&	CN	&	47$\times$47	&	\ciil	\\
			&				&					&	1342234066	&		&			&	\oiiil	\\
II\,Zw\,40		&	5h55m43.10s	&	+3d23m20.0s		&	1342228249	&	CN	&	79$\times$79	&	\oila	\\
			&	5h55m43.10s	&	+3d23m20.0s		&	1342228250	&		&	71$\times$71	&	\oiiil	\\
			&	5h55m43.10s	&	+3d23m20.0s		&	1342228251	&		&	85$\times$85	&	\oilb	\\
			&	5h55m43.00s	&	+3d23m31.4s		&	1342228252	&		&	47$\times$47	&	\niila	\\
			&	5h55m43.10s	&	+3d23m20.0s		&	1342228253	&		&	95$\times$95	&	\ciil	\\
I\,Zw\,18		&	9h34m01.44s	&	+55d14m34.8s		&	1342220973	&	CN	&	47$\times$47	&	\ciil	\\
			&	9h34m02.03s	&	+55d14m28.1s		&	1342253757	&		&		&	\oila	\\
			&	9h34m02.03s	&	+55d14m28.1s		&	1342253758	&		&		&	\oiiil	\\
IC\,10		&	0h20m26.10s	&	+59d17m17.6s		&	1342214364	&	UN	&	81$\times$81	&	\oila	\\
			&	0h20m52.12s	&	+59d17m42.4s		&	1342214365	&		&	47$\times$47	&	\niila	\\
			&	0h20m56.70s	&	+59d17m46.8s		&	1342214366	&		&	47$\times$47	&	\oilb	\\
			&	0h20m26.10s	&	+59d17m17.6s		&	1342214367	&		&	81$\times$81	&	\oiiil	\\
			&	0h20m48.32s	&	+59d18m09.0s		&	1342214368	&		&	115$\times$183	&	\ciil	\\
			&	0h20m18.70s	&	+59d18m45.2s		&	1342214369	&		&	81$\times$47	&	\oiiil	\\
			&	0h20m14.00s	&	+59d19m53.0s		&	1342214370	&		&	149$\times$81	&	\ciil	\\
			&	0h20m18.70s	&	+59d18m45.2s		&	1342214371	&		&	81$\times$47	&	\oila	\\
			&	0h20m13.30s	&	+59d19m57.5s		&	1342214372	&		&	115$\times$47	&	\oila	\\
			&	0h20m13.30s	&	+59d19m57.5s		&	1342214273	&		&	115$\times$47	&	\oiiil	\\
			&	0h20m18.00s	&	+59d18m44.0s		&	1342223371	&		&	47$\times$47	&	\niila	\\
			&	0h20m28.25s	&	+59d16m58.0s		&	1342223372	&		&	47$\times$47	&	\niila	\\
			&	0h20m28.25s	&	+59d16m58.0s		&	1342223373	&		&	47$\times$47	&	\oilb	\\
			&	0h20m18.00s	&	+59d18m44.0s		&	1342223374	&		&	47$\times$47	&	\oilb	\\
LMC-30Dor	&	5h38m35.00s	&	-69d05m39.0s		&	1342222085	&	UN	&	143$\times$47	&	\oiiil, \ciil	\\
			&	5h38m48.00s	&	-69d06m37.0s		&	1342222086	&		&	71$\times$47	&	\oila	\\
			&	5h38m58.00s	&	-69d04m43.0s		&	1342222087	&		&	143$\times$47	&	\oila	\\
			&	5h38m58.00s	&	-69d04m43.0s		&	1342222088	&		&	143$\times$47	&	\oiiil, \ciil	\\
			&	5h38m48.00s	&	-69d06m37.0s		&	1342222089	&		&	71$\times$47	&	\oiiil, \ciil	\\
			&	5h38m56.66s	&	-69d04m56.9s		&	1342222090	&		&	47$\times$47	&	\oilb	\\
			&	5h38m34.92s	&	-69d06m07.0s		&	1342222091	&		&	47$\times$47	&	\niila	\\
			&	5h38m45.00s	&	-69d05m23.0s		&	1342222092	&		&	191$\times$47	&	\oila	\\
			&	5h38m35.00s	&	-69d05m39.0s		&	1342222093	&		&	143$\times$47	&	\oilb	\\
			&	5h38m45.00s	&	-69d05m23.0s		&	1342222094	&		&	191$\times$47	&	\oiiil, \ciil	\\
			&	5h38m46.10s	&	-69d04m58.8s		&	1342222095	&		&	47$\times$47	&	\niila	\\
			&	5h38m45.00s	&	-69d05m23.0s		&	1342222096	&		&	191$\times$47	&	\oilb	\\
			&	5h38m35.00s	&	-69d05m39.0s		&	1342222097	&		&	143$\times$47	&	\oila	\\
			&	5h38m38.00s	&	-69d06m00.0s		&	1342231279	&		&	123$\times$47	&	\ciil	\\
			&	5h38m30.00s	&	-69d06m07.0s		&	1342231280	&		&	123$\times$47	&	\oiiil	\\
			&	5h38m30.00s	&	-69d06m07.0s		&	1342231281	&		&	123$\times$47	&	\ciil	\\
			&	5h38m55.00s	&	-69d03m49.0s		&	1342231282	&		&	85$\times$47	&	\oila, \ciil	\\
			&	5h38m30.00s	&	-69d06m07.0s		&	1342231283	&		&	123$\times$47	&	\oila, \ciil	\\
			&	5h38m56.00s	&	-69d04m50.0s		&	1342231284	&		&	85$\times$47	&	\niila	\\
			&	5h38m40.00s	&	-69d04m38.0s		&	1342231285	&		&	85$\times$47	&	\oiiil	\\
LMC-N11A	&	4h57m16.00s	&	-66d23m24.0s		&	1342214638	&	UN	&	47$\times$47	&	\ciil, \oila	\\
LMC-N11B	&	4h56m95.14s	&	-66d24m23.8s		&	1342188940	&	WS	&	127$\times$127	&	\oiiil	\\
			&	4h56m95.14s	&	-66d24m23.8s		&	1342188941	&		&	127$\times$127	&	\ciil, \oila, \oilb, \niila, \niilb, \niiil	\\
			&	4h56m47.00s	&	-66d24m32.0s		&	1342219439	&	UN	&	47$\times$47	&	\niila	\\
			&	4h56m57.20s	&	-66d25m13.0s		&	1342225175	&		&	47$\times$47	&	\niila, \oilb	\\
LMC-N11C	&	4h57m48.00s	&	-66d28m30.0s		&	1342221976	&	UN	&	47$\times$47	&	\oiiil, \ciil	\\
			&	4h57m48.00s	&	-66d28m30.0s		&	1342221977	&		&	47$\times$47	&	\oila	\\
			&	4h57m40.30s	&	-66d26m59.0s		&	1342221978	&		&	81$\times$47	&	\oila	\\
			&	4h57m50.59s	&	-66d29m49.7s		&	1342221979	&		&	81$\times$81	&	\ciil	\\
			&	4h57m50.59s	&	-66d29m49.7s		&	1342221980	&		&	81$\times$81	&	\oila	\\
			&	4h57m40.30s	&	-66d26m59.0s		&	1342221981	&		&	81$\times$47	&	\oiiil, \ciil	\\
LMC-N11I		&	4h55m44.00s	&	-66d34m24.0s		&	1342222769	&	UN	&	115$\times$47	&	\oila	\\
			&				&					&	1342222770	&		&	115$\times$47	&	\oiiil, \ciil	\\
LMC-N158	&	5h39m13.00s	&	-69d30m18.0s		&	1342214032	&	UN	&	81$\times$81	&	\oiiil, \ciil	\\
			&				&					&	1342214033	&		&	81$\times$81	&	\oila	\\
LMC-N159	&	5h39m37.53s	&	-69d45m27.0s		&	1342222075	&	UN	&	47$\times$47	&	\oilb	\\
			&	5h40m08.01s	&	-69d44m53.0s		&	1342222076	&		&	47$\times$47	&	\oilb, \niilb	\\
			&	5h39m39.50s	&	-69d46m10.7s		&	1342222077	&		&	47$\times$47	&	\oilb	\\
			&	5h39m38.00s	&	-69d45m48.0s		&	1342222078	&		&	81$\times$81	&	\oiiil, \ciil	\\
			&	5h39m52.00s	&	-69d45m12.0s		&	1342222079	&		&	115$\times$47	&	\oila, \niila	\\
			&	5h39m38.00s	&	-69d45m48.0s		&	1342222080	&		&	81$\times$81	&	\oila	\\
			&	5h39m52.00s	&	-69d45m12.0s		&	1342222081	&		&	115$\times$47	&	\oiiil, \ciil	\\
			&	5h40m08.00s	&	-69d44m46.0s		&	1342222082	&		&	81$\times$81	&	\oila, \niila	\\
			&	5h40m08.00s	&	-69d44m46.0s		&	1342222083	&		&	81$\times$81	&	\oiiil, \ciil	\\
			&	5h39m53.70s	&	-69d45m13.0s		&	1342222084	&		&	81$\times$81	&	\oilb, \niilb	\\
LMC-N159S	&	5h40m00.00s	&	-69d50m33.4s		&	1342222767	&	UN	&	81$\times$81	&	\oila	\\
			&	5h40m00.00s	&	-69d50m33.4s		&	1342222768	&		&	81$\times$81	&	\ciil	\\
LMC-N160	&	5h39m38.00s	&	-69d39m04.7s		&	1342222071	&	UN	&	47$\times$47	&	\oilb	\\
			&	5h39m41.30s	&	-69d38m50.0s		&	1342222072	&		&	75$\times$103	&	\oila	\\
			&	5h39m44.21s	&	-69d38m40.2s		&	1342222073	&		&	47$\times$47	&	\oilb	\\
			&	5h39m41.30s	&	-69d38m50.0s		&	1342222074	&		&	76$\times$105	&	\oiiil, \ciil	\\
Mrk\,153		&	10h49m05.04s	&	+52d20m07.8s		&	1342209014	&	CN	&	47$\times$47	&	\oila	\\
			&				&					&	1342209015	&		&			&	\ciil	\\
			&				&					&	1342232267	&	CN	&	47$\times$47	&	\oiiil	\\
Mrk\,209		&	12h26m16.02s	&	+48d29m36.6s		&	1342199423	&	CN	&	47$\times$47	&	\oiiil, \ciil	\\
			&				&					&	1342199424	&		&			&	\oila	\\
Mrk\,930		&	23h31m58.29s	&	+28d56m49.9s		&	1342212518	&	CN	&	47$\times$47	&	\oila	\\
			&				&					&	1342212519	&		&	47$\times$47	&	\oilb	\\
			&				&					&	1342212520	&		&	51$\times$51	&	\ciil	\\
			&				&					&	1342212521	&		&	47$\times$47	&	\niila	\\
			&				&					&	1342212522	&		&	53$\times$53	&	\oiiil	\\
Mrk\,1089		&	5h01m37.76s	&	-4d15m28.4s		&	1342217857	&	CN	&	47$\times$47	&	\oiiil	\\
			&				&					&	1342217858	&		&	47$\times$47	&	\oilb	\\
			&				&					&	1342217859	&		&	51$\times$51	&	\ciil	\\
			&				&					&	1342217860	&		&	51$\times$51	&	\niila	\\
			&				&					&	1342217861	&		&	53$\times$53	&	\oila	\\
Mrk\,1450		&	11h38m35.62s	&	+57d52m27.2s		&	1342222067	&	CN	&	47$\times$47	&	\oila	\\
			&				&					&	1342222068	&		&	47$\times$47	&	\oiiil	\\
			&				&					&	1342222069	&		&	47$\times$47	&	\niila	\\
			&				&					&	1342222070	&		&	51$\times$51	&	\ciil	\\
NGC\,625		&	1h35m06.00s	&	-41d26m10.3s		&	1342222216	&	CN	&	63$\times$63	&	\oiiil	\\
			&				&					&	1342222217	&		&	47$\times$47	&	\oila	\\
			&				&					&	1342222218	&		&	85$\times$85	&	\ciil	\\
NGC\,1140	&	2h54m33.20s	&	-10d01m50.0s		&	1342214034	&	CN	&	95$\times$95	&	\ciil	\\
			&	2h54m33.54s	&	-10d01m42.2s		&	1342214035	&		&	63$\times$63	&	\oiiil	\\
			&	2h54m33.54s	&	-10d01m42.2s		&	1342214036	&		&	47$\times$47	&	\oila	\\
			&	2h54m33.54s	&	-10d01m42.2s		&	1342225171	&		&	47$\times$47	&	\niila	\\
			&	2h54m33.54s	&	-10d01m42.2s		&	1342224588	&		&	47$\times$47	&	\oilb	\\
NGC\,1569	&	4h30m49.06s	&	+64d50m52.6s		&	1342225753	&	UN	&	95$\times$143	&	\ciil	\\
			&				&					&	1342225754	&		&	95$\times$79	&	\oila	\\
			&				&					&	1342225755	&		&	119$\times$71	&	\oiiil	\\
NGC\,1705	&	4h54m13.50s	&	-53d21m39.8s		&	1342222213	&	CN	&	63$\times$63	&	\oiiil	\\
			&				&					&	1342222214	&		&	47$\times$47	&	\oila	\\
			&				&					&	1342222215	&		&	71$\times$71	&	\ciil	\\
NGC\,2366	&	7h28m42.60s	&	+69d11m19.0s		&	1342220604	&		&	47$\times$47	&	\oiiil	\\
			&				&					&	1342220605	&		&	71$\times$71	&	\ciil	\\
			&				&					&	1342220606	&		&	63$\times$63	&	\oila	\\
NGC\,4214	&	12h15m39.17s	&	+36d19m36.8s		&	1342187843	&	CN	&	111$\times$111	&	\oila	\\
			&				&					&	1342187844	&		&			&	\oiiil	\\
			&				&					&	1342187845	&		&	95$\times$95	&	\ciil	\\
			&				&					&	1342188034	&		&			&	\niilb	\\
			&				&					&	1342188035	&		&			&	\oilb	\\
			&				&					&	1342188036	&		&			&	\niila	\\
NGC\,4449	&	12h28m11.90s	&	+44d05m39.6s		&	1342197813	&	UN	&	119$\times$191	&	\ciil	\\
			&				&					&	1342197814	&		&			&	\oila	\\
			&				&					&	1342197815	&	CN	&	47$\times$47	&	\ciil	\\
			&				&					&	1342197816	&		&			&	\oila	\\
			&				&					&	1342197817	&		&			&	\ciil	\\
			&				&					&	1342197818	&		&			&	\oila	\\
			&				&					&	1342197819	&		&			&	\ciil	\\
			&				&					&	1342197820	&		&			&	\oila	\\
			&	12h28m11.30s	&	44d05m30.6s		&	1342223138	&	UN	&	81$\times$47	&	\oilb	\\
			&	12h28m11.30s	&	44d05m30.6s		&	1342223139	&	CN	&	81$\times$81	&	\oiiil	\\
			&	12h28m11.30s	&	44d05m30.6s		&	1342223140	&	UN	&	81$\times$47	&	\niila	\\
			&	12h28m15.40s	&	44d06m57.6s		&	1342223141	&		&	81$\times$115	&	\oila	\\
			&	12h28m15.40s	&	44d06m57.6s		&	1342223142	&		&	81$\times$115	&	\oiiil	\\
			&	12h28m15.40s	&	44d06m57.6s		&	1342223143	&		&	81$\times$115	&	\ciil	\\
NGC\,4861	&	12h59m02.34s	&	+34d51m34.0s		&	1342208901	&	CN	&	47$\times$47	&	\oila, \oilb, \niila, \niilb, \niiil	\\
			&	12h59m02.34s	&	+34d51m34.0s		&	1342208902	&		&	85$\times$215	&	\ciil	\\
			&	12h59m02.34s	&	+34d51m34.0s		&	1342208903	&		&	47$\times$47	&	\oiiil	\\
			&	12h59m00.30s	&	+34d50m43.6s		&	1342221887	&		&	47$\times$47	&	\oila, \oilb	\\
			&	12h59m00.30s	&	+34d50m43.6s		&	1342221888	&		&	47$\times$47	&	\oiiil	\\
NGC\,5253	&	13h39m55.96s	&	-31d38m24.4s		&	1342202125	&	CN	&	78$\times$78	&	\oila	\\
			&				&					&	1342202126	&		&	94$\times$141	&	\ciil	\\
			&				&					&	1342214026	&	CN	&	63$\times$63	&	\oiiil	\\
			&				&					&	1342214027	&		&	47$\times$47	&	\niila	\\
			&				&					&	1342214028	&		&	62$\times$47	&	\oilb	\\
NGC\,6822	&	19h45m05.00s	&	-14d43m22.1s		&	1342216633	&	UN	&	47$\times$47	&	\ciil	\\
			&	19h44m50.00s	&	14d52m46.9s		&	1342216634	&		&	71$\times$71	&	\oila	\\
			&	19h45m05.00s	&	-14d43m22.1s		&	1342216635	&		&	47$\times$47	&	\oiiil	\\
			&	19h44m50.00s	&	14d52m46.9s		&	1342216636	&		&	77$\times$47	&	\ciil	\\
			&	19h44m50.00s	&	14d52m46.9s		&	1342216637	&		&	77$\times$47	&	\oiiil	\\
			&	19h44m52.80s	&	-14d43m04.9s		&	1342216638	&		&	71$\times$71	&	\oiiil, \ciil	\\
			&	19h44m52.80s	&	-14d43m04.9s		&	1342216639	&		&	71$\times$71	&	\niila	\\
			&	19h44m52.80s	&	-14d43m04.9s		&	1342230147	&	UN	&	81$\times$81	&	\oila	\\
			&	19h44m52.80s	&	-14d43m04.9s		&	1342230148	&		&	47$\times$47	&	\oilb	\\
			&	19h45m05.00s	&	-14d43m22.1s		&	1342230149	&		&	47$\times$47	&	\oila	\\
Pox\,186		&	13h25m48.66s	&	-11d36m37.8s		&	1342213284	&	CN	&	47$\times$47	&	\oiiil	\\
			&				&					&	1342234995	&		&			&	\ciil	\\
SBS\,0335-052	&	3h37m44.06s	&	-5d02m40.2s		&	1342214221	&	CN	&	47$\times$47	&	\ciil	\\
			&				&					&	1342248295	&		&			&	\oiiil	\\
			&				&					&	1342249197	&		&			&	\oila	\\
SBS\,1159+545	&	12h02m02.37s	&	+54d15m49.5s		&	1342199228	&	CN	&	47$\times$47	&	\ciil	\\
			&				&					&	1342232309	&		&			&	\oiiil	\\
			&				&					&	1342232310	&		&			&	\oila	\\
SBS\,1211+540	&	12h14m02.48s	&	+53d45m17.4s		&	1342199422	&	CN	&	47$\times$47	&	\ciil	\\
SBS\,1249+493&	12h51m52.53s	&	+49d03m26.9s		&	1342232266	&	CN	&	47$\times$47	&	\ciil	\\
SBS\,1415+437&	14h17m01.40s	&	+43d30m04.5s		&	1342199731	&	CN	&	47$\times$47	&	\oila	\\
			&				&					&	1342199732	&		&			&	\oiiil	\\
			&				&					&	1342199733	&		&			&	\ciil	\\
SBS\,1533+574&	15h34m13.80s	&	+57d17m06.0s		&	1342199230	&	CN	&	47$\times$47	&	\ciil	\\
			&				&					&	1342199231	&		&			&	\oiiil	\\
			&				&					&	1342199232	&		&			&	\oila	\\
			&				&					&	1342199399	&		&			&	\oiiil	\\
SMC-N66		&	0h59m08.00s	&	-72d10m46.4s		&	1342214029	&	UN	&	149$\times$47	&	\oiiil, \ciil	\\
			&	0h59m08.00s	&	-72d10m46.4s		&	1342214030	&		&	149$\times$47	&	\oila	\\
			&	0h59m09.73s	&	-72d10m03.4s		&	1342214031	&		&	47$\times$47	&	\oila, \ciil	\\
Tol\,1214-277	&	12h17m17.09s	&	-28d02m32.7s		&	1342199408	&	CN	&	47$\times$47	&	\ciil 	\\	
			&				&					&	1342234059	&		&			&	\oiiil	\\
			&				&					&	1342234060	&		&			&	\oila	\\
UGC\,4483	&	8h37m03.00s	&	+69d46m54.0s		&	1342203684	&	CN	&	47$\times$47	&	\ciil	\\
			&	8h37m03.00s	&	+69d46m54.0s		&	1342203685	&		&			&	\oiiil	\\
			&	8h37m03.00s	&	+69d46m31.0s		&	1342203686	&		&			&	\ciil, \oila	\\
UM\,133		&	1h44m41.28s	&	+4d53m25.9s		&	1342212533	&	CN	&	47$\times$47	&	\ciil	\\
			&				&					&	1342235699	&		&			&	\oiiil	\\
			&				&					&	1342235700	&		&			&	\oila	\\
UM\,311		&	1h15m32.00s	&	-0d51m38.0s		&	1342213288	&	CN	&	71$\times$71	&	\ciil	\\
			&				&					&	1342213289	&		&	63$\times$63	&	\oiiil	\\
			&				&					&	1342213290	&		&	47$\times$47	&	\niila	\\
			&				&					&	1342213291	&		&	63$\times$63	&	\oila	\\
UM\,448		&	11h42m12.40s	&	+0d20m02.7s		&	1342222199	&	CN	&	53$\times$53	&	\oiiil	\\
			&				&					&	1342222200	&		&	47$\times$47	&	\niila	\\
			&				&					&	1342222201	&		&	51$\times$51	&	\ciil	\\
			&				&					&	1342222202	&		&	51$\times$51	&	\oilb	\\
			&				&					&	1342222203	&		&	53$\times$53	&	\oila	\\
UM\,461		&	11h51m33.35s	&	-2d22m21.9s		&	1342222204	&	CN	&	47$\times$47	&	\oila	\\
			&				&					&	1342222205	&		&	51$\times$51	&	\ciil	\\
			&				&					&	1342222206	&		&	47$\times$47	&	\niila	\\
			&				&					&	1342222207	&		&	53$\times$53	&	\oiiil	\\
VII\,Zw\,403	&	11h27m59.90s	&	+78d59m39.0s		&	1342199286	&	CN	&	47$\times$47	&	\niila, \oilb	\\
			&				&					&	1342199287	&		&	63$\times$63	&	\oiiil	\\
			&				&					&	1342199288	&		&	63$\times$63	&	\oila	\\
			&				&					&	1342199289	&		&	71$\times$71	&	\ciil	\\
    \hline \hline
  \label{table:obsdetails}
\end{longtable}
\end{longtab}

\clearpage
\section{Spectral maps and line profiles of the \hers Dwarf Galaxy Survey}
\label{app:append-b}
Display example of the PACS data for NGC\,1140. 
The full atlas is available in the online material. 
For each spectral line, the flux map and spectrum 
of the brightest spatial pixel, which position is indicated 
by a black cross on the map, are shown. 
White circles represent the PACS beam sizes. 
We note that the maps are not PSF-deconvolved. 
%
\begin{figure}[h]
\centering
\begin{minipage}{8.8cm}
 \includegraphics[clip,trim=0 1.35cm 0 0,width=8.8cm]{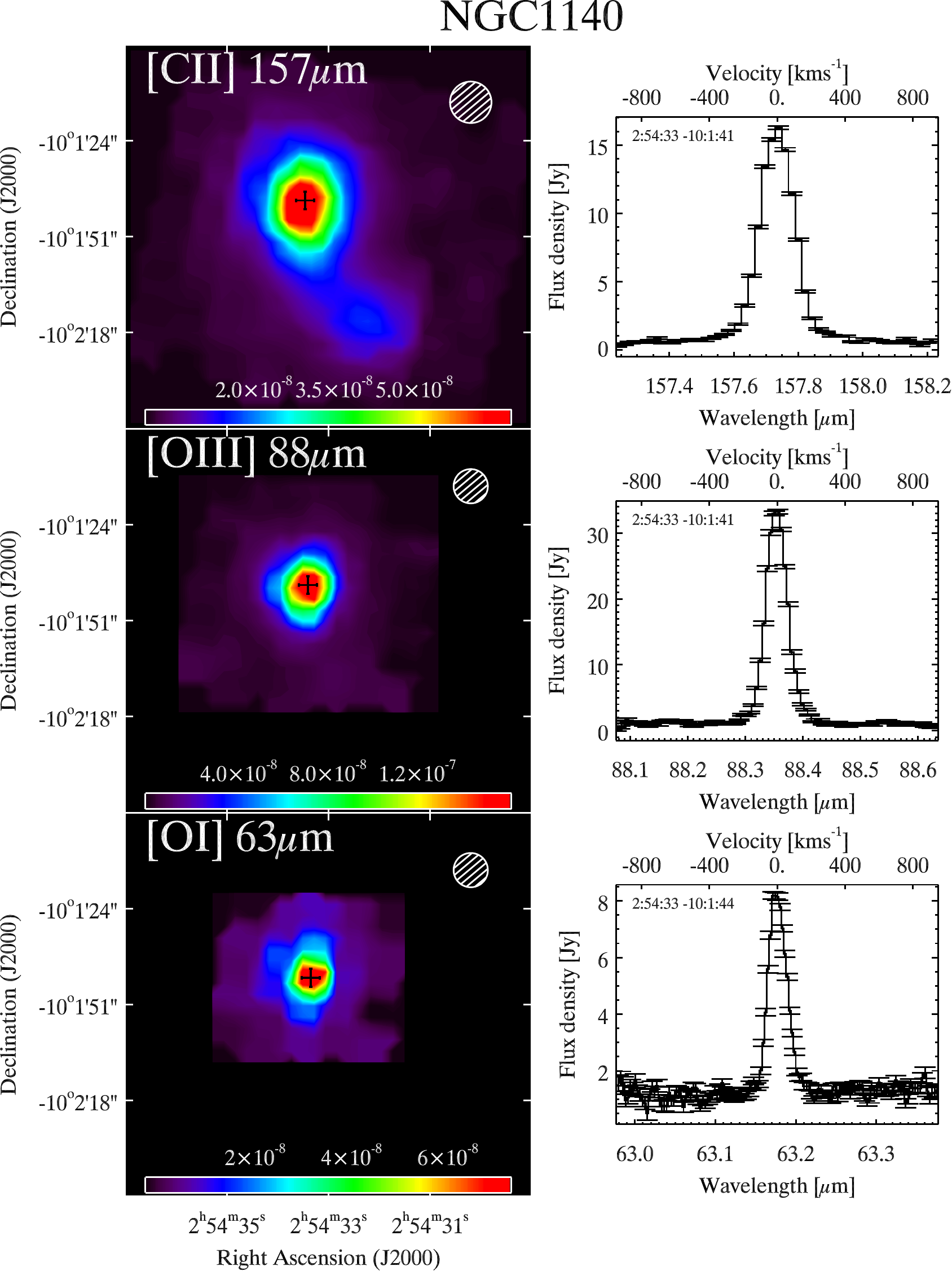} 
 \includegraphics[clip,trim=0 0 1.5mm 9mm,width=8.8cm]{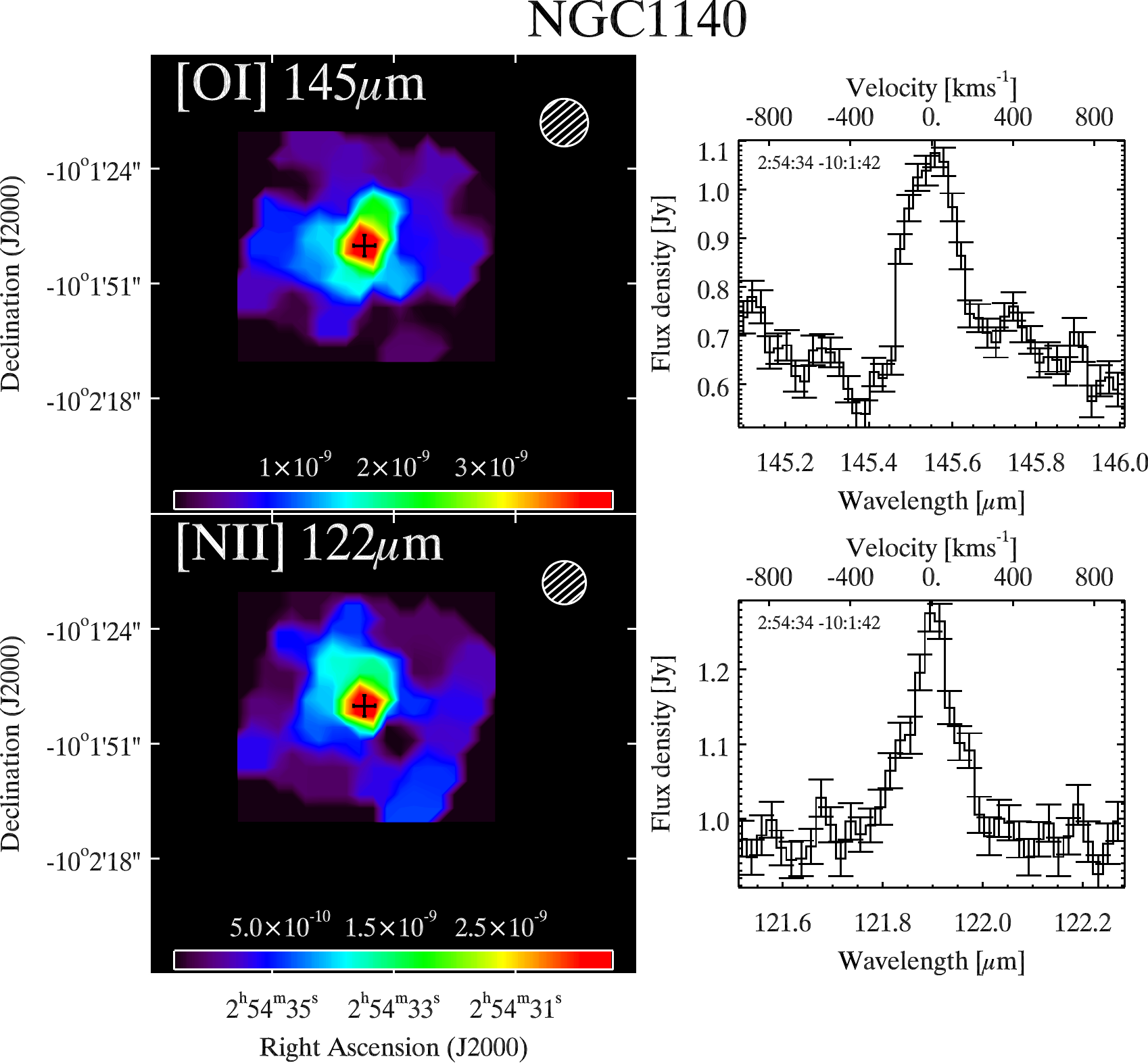}
 \label{fig:ngc1140-1}
\end{minipage}
\end{figure}

\clearpage
\section{\spit IRS fluxes of the compact DGS galaxies}
\label{app:append-c}
We have extracted IRS data of the compact DGS galaxies, 
as described in Sect.~\ref{sect:irsdata}. Fluxes of the main MIR 
fine-structure cooling lines are listed here (Table~\ref{table:irsfluxes} 
for the low-resolution data and Table~\ref{table:irsfluxes_hires} 
for the high-resolution data). 

The spectra of the short and long wavelength modules of the IRS 
were calibrated using a correction based on the spatial extent of 
the source. The correction, which is wavelength-dependent, assumes 
a spherical symmetry for the source shape to account for the light 
that falls outside the slit. Using this correction, there is no significant 
flux offsets between the spectra of the short and long wavelength modules 
for the low-resolution data. For the high-resolution data, 
extended sources are extracted in different apertures which lead to 
offsets between the spectra of the short and long wavelength modules. 
In those cases, the short wavelength spectra are stitched to the long 
wavelength spectra. 

We have compared the overall continuum level of the IRS data to the 
IRAC and MIPS photometry to verify if all emission is recovered 
by our reduction (i.e. total flux for the galaxy). 
This is generally the case, except for NGC\,1569 which is clearly 
more extended than the IRS slit. 
For reference, we indicate in Table~\ref{table:irsfluxes} the ratio of 
the observed-to-synthetic MIPS 24\,\mum flux, denoted $R_{MIPS24}$. 
The observed photometry is from \cite{bendo-2012} and the synthetic 
photometry is measured from the IRS data using the MIPS filter profile. 
This scaling factor should be applied to obtain {\it total} IRS fluxes. 
We opt for the 24\,\mum emission because it arises from active star-forming 
regions and it is less affected by IRS module stitching than the shorter 
wavelengths. We note that observed-to-synthetic IRAC 8\,\mum fluxes 
are further offset by a factor $\sim$1.8 in 3 galaxies with strong PAH emission 
(He\,2-10, Mrk\,1089, NGC\,1140). This could be due to PAH (and PDR line) 
emission outside of the IRS slit as those galaxies show extended structures 
or to calibration uncertainties. 

In Table~\ref{table:irsfluxes_hires}, we also indicate the average ratio between 
the high-resolution and low-resolution fluxes of the main spectral lines, 
denoted $HR/LR$. Line fluxes generally agree within 20\% around that ratio. 
Differences are mainly due to the extraction method and source extent. 
\neii and the PAH feature at 12.7\,\mum, as well as [O\,{\sc iv}] and 
[Fe\,{\sc ii}] at 25.9\,\mum, are resolved and thus more accurately measured 
with the high-resolution mode.

\begin{center}
\begin{table*}[!ht]\small
  \caption{Low-resolution MIR line fluxes in the compact galaxies of the DGS.\newline
Notes. Fluxes are in 10$^{-18}$\,W\,m$^{-2}$. 
Upper limits are given at 2$\sigma$. 
} 
\begin{tabular}{l c c c c c c}
    \hline\hline
     \vspace{-8pt}\\
    \multicolumn{1}{l}{Source Name} & 
    \multicolumn{5}{c}{Main Observed Spectral Lines} &
    \multicolumn{1}{c}{$R_{MIPS24}$} \\ \cline{2-6}
    \vspace{-8pt}\\
    \multicolumn{1}{l}{} &
    \multicolumn{1}{c}{\siv10.5\,\mum} &
    \multicolumn{1}{c}{\neii12.8\,\mum} &
    \multicolumn{1}{c}{\neiii15.6\,\mum} &
    \multicolumn{1}{c}{\siii18.7\,\mum} &
    \multicolumn{1}{c}{\siii33.5\,\mum} &
    \multicolumn{1}{c}{} \\
    \hline
    \vspace{-8pt}\\
Haro\,3   & $3.44\pm0.16~(\times 10^{2})$ &    $3.65\pm0.12~(\times 10^{2})$ &    $ -  $ &  $ -  $ &  $ -  $ & - \\
Haro\,11   & $5.28\pm0.12~(\times 10^{2})$ &    $2.86\pm0.09~(\times 10^{2})$ &    $1.26\pm0.14~(\times 10^{3})$ &    $5.54\pm0.68~(\times 10^{2})$ &    $6.56\pm0.81~(\times 10^{2})$ & $1.13\pm0.05$  \\
He\,2-10   & $4.25\pm0.77~(\times 10^{2})$ &    $4.17\pm0.24~(\times 10^{3})$ &    $1.50\pm0.16~(\times 10^{3})$ &    $3.24\pm0.29~(\times 10^{3})$ &    $3.79\pm0.52~(\times 10^{3})$ & $1.13\pm0.05$  \\
HS\,0052+2536   & $13.50\pm3.07$ &    $8.55\pm1.32$ &    $14.97\pm4.51$ &    $21.06\pm6.28$ &    $17.27\pm4.13$ & $1.15\pm0.09$  \\
HS\,0822+3542   & $5.23\pm0.47$ &    $\le 0.50$ & $4.35\pm1.54$ &    $\le 1.84$ & $\le 2.33$ & $0.88\pm0.10$  \\
HS\,1222+3741   & $7.80\pm1.27$ &    $\le 1.81$ & $\le 7.89$ & $\le 6.79$ & $\le 26.26$ & -  \\
HS\,1304+3529   & $11.10\pm1.58$ &    $\le 1.63$ & $16.24\pm4.76$ &    $16.34\pm4.99$ &    $\le 25.44$ & $1.07\pm0.10$  \\
HS\,1319+3224   & $\le 6.94$ & $\le 1.09$ & $\le 3.76$ & $\le 7.97$ & $\le 11.17$ & -  \\
HS\,1330+3651   & $8.49\pm1.10$ &    $\le 0.85$ & $16.45\pm6.49$ &    $\le 17.24$ & $\le 16.75$ & -  \\
HS\,1442+4250   & $6.92\pm1.58$ &    $\le 1.12$ & $ -  $ &  $ -  $ &  $ -  $ & -  \\
II\,Zw\,40   & $1.80\pm0.03~(\times 10^{3})$ &    $83.49\pm5.83$ &    $1.48\pm0.04~(\times 10^{3})$ &    $5.79\pm0.44~(\times 10^{2})$ &    $7.81\pm0.63~(\times 10^{2})$ & $1.18\pm0.05$  \\
I\,Zw\,18   & $4.34\pm0.38$ &    $\le 0.50$ & $7.42\pm1.40$ &    $4.44\pm1.19$ &    $3.36\pm1.66$ & $1.22\pm0.08$  \\
Mrk\,153   & $18.91\pm2.00$ &    $5.44\pm1.33$ &    $29.92\pm2.54$ &    $19.15\pm1.77$ &    $24.33\pm8.66$ & $1.33\pm0.06$  \\ 
Mrk\,209   & $79.92\pm4.46$ &    $\le 8.95$ & $ -  $ &  $ -  $ &  $ -  $ & -  \\
Mrk\,930   & $1.29\pm0.20~(\times 10^{2})$ &    $30.19\pm4.27$ &    $2.11\pm0.18~(\times 10^{2})$ &    $76.56\pm8.33$ &    $1.41\pm0.17~(\times 10^{2})$ & $1.24\pm0.05$  \\
Mrk\,1089   & $1.48\pm0.10~(\times 10^{2})$ &    $1.80\pm0.08~(\times 10^{2})$ &    $4.71\pm0.29~(\times 10^{2})$ &    $3.05\pm0.18~(\times 10^{2})$ &    $5.27\pm0.83~(\times 10^{2})$ & $1.27\pm0.06$  \\
Mrk\,1450   & $69.77\pm3.19$ &    $10.69\pm1.86$ &    $94.68\pm3.29$ &    $49.60\pm4.48$ &    $62.92\pm6.03$ & $1.07\pm0.05$  \\
NGC\,1140   & $1.81\pm0.10~(\times 10^{2})$ &    $1.59\pm0.07~(\times 10^{2})$ &    $4.78\pm0.20~(\times 10^{2})$ &    $2.47\pm0.19~(\times 10^{2})$ &    $4.36\pm0.96~(\times 10^{2})$ & $1.26\pm0.05$  \\
NGC\,1569   & $ -  $ &    $ -  $ &    $2.73\pm0.16~(\times 10^{3})$ &    $1.38\pm0.07~(\times 10^{3})$ &    $2.50\pm0.37~(\times 10^{3})$ & $3.16\pm0.14$  \\
Pox\,186   & $36.29\pm3.66$ &    $0.95\pm0.25$ &    $13.81\pm3.33$ &    $7.37\pm1.25$ &    $\le 4.64$ & $0.86\pm0.09$  \\
SBS\,0335-052   & $20.04\pm2.10$ &    $1.35\pm0.56$ &    $17.35\pm2.71$ &    $4.24\pm1.60$ &    $\le 5.81$ & $1.02\pm0.05$  \\
SBS\,1159+545  & $3.91\pm1.78$ &    $\le 2.51$ & $\le 3.32$ & $7.06\pm1.96$ &    $\le 6.92$ & $1.10\pm0.13$  \\
SBS\,1211+540   & $3.10\pm0.66$ &    $\le 1.15$ & $4.94\pm0.73$ &    $3.44\pm1.32$ &    $\le 16.72$ & $1.25\pm0.41$  \\
SBS\,1249+493   & $4.74\pm0.91$ &    $\le 0.57$ & $9.85\pm3.89$ &    $\le 11.72$ & $\le 16.27$ & $0.85\pm0.17$  \\
SBS\,1415+437   & $13.95\pm0.60$ &    $3.10\pm1.23$ &    $ -  $ &  $ -  $ &  $ -  $ & -  \\
SBS\,1533+574   & $28.93\pm2.87$ &    $4.28\pm0.51$ &    $49.49\pm9.09$ &    $14.10\pm1.53$ &    $25.28\pm5.94$ & -  \\
Tol\,1214-277   & $10.53\pm2.41$ &    $\le 1.52$ & $5.89\pm2.48$ &    $\le 5.10$ & $\le 7.11$ & $1.19\pm0.11$  \\
UM\,448   & $2.21\pm0.11~(\times 10^{2})$ &    $3.76\pm0.22~(\times 10^{2})$ &    $6.93\pm0.83~(\times 10^{2})$ &    $3.24\pm0.19~(\times 10^{2})$ &    $6.63\pm0.48~(\times 10^{2})$ & $1.03\pm0.04$  \\
UM\,461   & $45.89\pm4.23$ &    $2.53\pm0.62$ &    $25.61\pm6.50$ &    $11.91\pm1.88$ &    $11.96\pm3.76$ & $1.03\pm0.10$  \\ 
    \hline
  \end{tabular}
\begin{tabular}{l c c c c c}
    \hline
     \vspace{-8pt}\\
    \multicolumn{1}{l}{Source Name} & 
    \multicolumn{5}{c}{Other Observed Spectral Lines} \\ \cline{2-6}
    \vspace{-8pt}\\
    \multicolumn{1}{l}{} &
    \multicolumn{1}{c}{[Ar\,{\sc ii}]~6.99\,\mum} &
    \multicolumn{1}{c}{[Ne\,{\sc v}]~14.3\,\mum} &
    \multicolumn{1}{c}{[O\,{\sc iv}]~25.9\,\mum} &
    \multicolumn{1}{c}{[Si\,{\sc ii}]~34.8\,\mum} &
    \multicolumn{1}{c}{[Ne\,{\sc iii}]~36.0\,\mum} \\
    \hline
    \vspace{-8pt}\\
Haro\,3   & $\le 56.13$ & $ -  $ &  $ -  $ &  $ -  $ &  $ -  $ \\
Haro\,11   & $37.83\pm11.61$ &    $\le 91.56$ & $\le 35.97$ & $5.16\pm1.31~(\times 10^{2})$ &    $\le 3.75~(\times 10^{2})$ \\ 
He\,2-10   & $4.94\pm0.43~(\times 10^{2})$ &    $1.72\pm0.73~(\times 10^{2})$ &    $\le 9.42~(\times 10^{2})$ & $2.41\pm0.42~(\times 10^{3})$ &    $\le 1.24~(\times 10^{3})$ \\
HS\,0052+2536   & $\le 4.04$ & $\le 7.35$ & $\le 3.11$ & $25.71\pm7.21$ &    $ -  $ \\
HS\,0822+3542   & $\le 2.44$ & $\le 2.14$ & $\le 0.85$ & $2.45\pm0.92$ &    $\le 6.18$ \\
HS\,1222+3741   & $\le 3.04$ & $\le 3.99$ & $\le 7.75$ & $\le 53.57$ & $ -  $ \\
HS\,1304+3529   & $\le 8.49$ & $\le 6.02$ & $\le 6.11$ & $\le 24.22$ & $\le 16.16$ \\
HS\,1319+3224   & $ -  $ & $\le 5.44$ & $\le 1.34$ & $9.70\pm3.12$ &    $\le 24.74$ \\
HS\,1330+3651   & $\le 3.38$ & $\le 16.31$ & $\le 4.68$ & $\le 25.65$ & $\le 35.09$ \\
HS\,1442+4250   & $\le 4.40$ & $ -  $ &  $ -  $ &  $ -  $ &  $ -  $ \\
II\,Zw\,40   & $\le 34.98$ & $\le 62.44$ & $96.47\pm39.45$ &    $4.51\pm0.49~(\times 10^{2})$ &    $\le 1.99~(\times 10^{2})$ \\
I\,Zw\,18   & $\le 1.36$ & $\le 5.05$ & $\le 2.96$ & $8.10\pm2.76$ &    $\le 10.18$ \\
Mrk\,153   & $\le 14.23$ & $\le 5.44$ & $\le 4.89$ & $18.98\pm7.45$ &    $\le 21.97$ \\
Mrk\,209   & $\le 26.98$ & $ -  $ &  $ -  $ &  $ -  $ &  $ -  $ \\
Mrk\,930   & $\le 28.00$ & $\le 9.00$ & $\le 10.31$ & $1.08\pm0.08~(\times 10^{2})$ &    $\le 68.98$ \\
Mrk\,1089   & $24.62\pm7.03$ &    $21.44\pm7.14$ &    $\le 13.54$ & $2.64\pm0.51~(\times 10^{2})$ &    $\le 1.86~(\times 10^{2})$ \\
Mrk\,1450   & $\le 10.97$ & $\le 14.09$ & $\le 8.63$ & $31.85\pm3.44$ &    $\le 11.22$ \\
NGC\,1140   & $\le 21.00$ & $\le 12.66$ & $\le 13.72$ & $3.28\pm0.30~(\times 10^{2})$ &    $\le 75.12$ \\
NGC\,1569   & $ -  $ & $29.45\pm8.33$ &    $\le 6.17~(\times 10^{2})$ & $1.61\pm0.22~(\times 10^{3})$ &    $\le 4.83~(\times 10^{2})$ \\
Pox\,186   & $\le 1.51$ & $\le 2.36$ & $\le 0.36$ & $\le 4.57$ & $\le 7.85$ \\
SBS\,0335-052   & $\le 2.22$ & $\le 6.33$ & $\le 5.01$ & $5.61\pm1.89$ &    $\le 11.93$ \\
SBS\,1159+545   & $\le 11.99$ & $\le 5.40$ & $\le 0.95$ & $\le 7.34$ & $\le 7.62$ \\
SBS\,1211+540   & $\le 3.13$ & $\le 2.75$ & $\le 2.01$ & $\le 10.29$ & $\le 35.40$ \\
SBS\,1249+493   & $\le 9.53$ & $\le 20.34$ & $\le 4.54$ & $\le 45.09$ & $\le 27.83$ \\
SBS\,1415+437   & $\le 4.67$ & $ -  $ &  $ -  $ &  $ -  $ &  $ -  $ \\
SBS\,1533+574   & $\le 2.66$ & $\le 7.76$ & $\le 9.59$ & $20.62\pm2.84$ &    $\le 11.13$ \\
Tol\,1214-277   & $\le 16.92$ & $\le 4.59$ & $\le 5.61$ & $\le 7.28$ & $13.18\pm3.27$ \\
UM\,448   & $55.15\pm25.71$ &    $\le 38.84$ & $\le 10.29$ & $4.89\pm0.49~(\times 10^{2})$ &    $\le 1.28~(\times 10^{2})$ \\
UM\,461   & $\le 4.73$ & $\le 8.43$ & $\le 3.83$ & $11.87\pm2.84$ &    $\le 16.98$ \\
    \hline \hline
    \vspace{-5pt}
  \end{tabular}
  \label{table:irsfluxes}
\end{table*}
\end{center}

\begin{center}
\begin{table*}[!ht]\small
  \caption{High-resolution MIR line fluxes in the compact galaxies of the DGS.\newline
Notes. Fluxes are in 10$^{-18}$\,W\,m$^{-2}$. 
Upper limits are given at 2$\sigma$. 
The symbol $^*$ indicates point-like sources reduced 
with the optimal extraction method. 
The other sources are extracted from the full apertures. 
We note that the observation of Pox\,186 (AOR key 9007360) 
is not included because it is not centered on the source. 
} 
\begin{tabular}{l c c c c c c}
    \hline\hline
     \vspace{-8pt}\\
    \multicolumn{1}{l}{Source Name} & 
    \multicolumn{5}{c}{Main Observed Spectral Lines} &
    \multicolumn{1}{c}{$HR/LR$} \\ \cline{2-6}
    \vspace{-8pt}\\
    \multicolumn{1}{l}{} &
    \multicolumn{1}{c}{\siv10.5\,\mum} &
    \multicolumn{1}{c}{\neii12.8\,\mum} &
    \multicolumn{1}{c}{\neiii15.6\,\mum} &
    \multicolumn{1}{c}{\siii18.7\,\mum} &
    \multicolumn{1}{c}{\siii33.5\,\mum} &
    \multicolumn{1}{c}{} \\
    \hline
    \vspace{-8pt}\\
Haro\,3   & $4.08\pm0.17~(\times 10^{2})$ &    $3.52\pm0.13~(\times 10^{2})$ &    $9.84\pm0.74~(\times 10^{2})$ &    $5.03\pm0.39~(\times 10^{2})$ &    $8.51\pm0.24~(\times 10^{2})$ &    $1.08$ \\
Haro\,11$^*$   & $4.94\pm0.11~(\times 10^{2})$ &    $3.27\pm0.09~(\times 10^{2})$ &    $1.12\pm0.05~(\times 10^{3})$ &    $5.31\pm0.29~(\times 10^{2})$ &    $8.17\pm0.70~(\times 10^{2})$ &    $1.04$ \\
He\,2-10   & $3.27\pm0.13~(\times 10^{2})$ &    $3.80\pm0.13~(\times 10^{3})$ &    $1.56\pm0.05~(\times 10^{3})$ &    $2.67\pm0.19~(\times 10^{3})$ &    $3.17\pm0.10~(\times 10^{3})$ &    $0.88$ \\
HS\,0822+3542$^*$   & $6.43\pm0.45$ &    $\le 0.24$ & $2.63\pm0.21$ &    $1.32\pm0.22$ &    $ -  $ &  $0.92$ \\
HS\,1442+4250$^*$   & $8.21\pm0.30$ &    $\le 0.51$ & $3.83\pm0.68$ &    $1.69\pm0.59$ &    $3.10\pm1.03$ & $1.19$ \\
II\,Zw\,40   & $2.00\pm0.10~(\times 10^{3})$ &    $73.52\pm7.93$ &    $1.41\pm0.09~(\times 10^{3})$ &    $5.21\pm0.22~(\times 10^{2})$ &    $7.82\pm0.21~(\times 10^{2})$ &    $0.97$ \\
I\,Zw\,18   & $5.69\pm1.42$ &    $\le 2.91$ & $7.89\pm1.02$ &    $5.04\pm1.29$ &    $3.94\pm0.91$ &    $1.17$ \\
Mrk\,153   & $24.01\pm5.47$ &    $3.91\pm0.77$ &    $31.22\pm1.72$ &    $14.19\pm2.32$ &    $28.89\pm6.97$ &    $0.99$ \\
Mrk\,209$^*$   & $85.71\pm2.88$ &    $2.45\pm0.31$ &    $54.49\pm1.74$ &    $24.34\pm1.49$ &    $49.69\pm5.70$ &    $1.07$ \\
Mrk\,930   & $1.39\pm0.05~(\times 10^{2})$ &    $39.39\pm2.44$ &    $2.12\pm0.06~(\times 10^{2})$ &    $94.43\pm4.98$ &    $1.28\pm0.11~(\times 10^{2})$ &    $1.12$ \\
Mrk\,1450$^*$   & $72.84\pm1.22$ &    $9.18\pm0.72$ &    $93.06\pm3.32$ &    $43.88\pm1.54$ &    $70.17\pm17.05$ &    $0.98$ \\
NGC\,1140   & $1.80\pm0.10~(\times 10^{2})$ &    $1.83\pm0.09~(\times 10^{2})$ &    $6.28\pm0.40~(\times 10^{2})$ &    $3.31\pm0.17~(\times 10^{2})$ &    $4.34\pm0.14~(\times 10^{2})$ &    $1.16$ \\
NGC\,1569   & $2.47\pm0.03~(\times 10^{3})$ &    $3.05\pm0.27~(\times 10^{2})$ &    $3.24\pm0.13~(\times 10^{3})$ &    $1.31\pm0.05~(\times 10^{3})$ &    $1.85\pm0.06~(\times 10^{3})$ &    $0.96$ \\
SBS\,0335-052$^*$   & $14.75\pm0.37$ &    $0.72\pm0.09$ & $12.37\pm0.74$ &    $3.49\pm1.05$ &    $4.70\pm1.58$ & $0.71$ \\
SBS\,1159+545$^*$   & $\le 5.10$ & $\le 2.19$ & $\le 2.97$ & $\le 4.23$ & $\le 12.83$ & $n.a.$ \\
SBS\,1415+437$^*$   & $12.02\pm0.92$ &    $2.31\pm0.29$ &    $11.33\pm0.99$ &    $7.58\pm1.20$ &    $14.04\pm3.42$ &    $0.81$ \\
Tol\,1214-277   & $8.86\pm0.47$ &    $\le 1.56$ & $5.57\pm0.55$ &    $\le 2.99$ & $\le 11.06$ & $0.89$ \\
UM\,448   & $2.32\pm0.04~(\times 10^{2})$ &    $3.14\pm0.12~(\times 10^{2})$ &    $6.38\pm0.25~(\times 10^{2})$ &    $3.21\pm0.12~(\times 10^{2})$ &    $4.92\pm0.21~(\times 10^{2})$ &    $0.92$ \\
UM\,461$^*$   & $46.71\pm0.90$ &    $1.58\pm0.39$ &    $29.11\pm1.24$ &    $8.71\pm0.80$ &    $12.64\pm4.60$ & $0.86$ \\
NGC\,625   & $5.72\pm0.07~(\times 10^{2})$ &    $1.66\pm0.05~(\times 10^{2})$ &    $7.46\pm0.26~(\times 10^{2})$ &    $4.40\pm0.15~(\times 10^{2})$ &    $7.78\pm0.39~(\times 10^{2})$ &    $n.a.$ \\
UM\,311   & $36.80\pm2.91$ &    $43.77\pm13.69$ &    $1.03\pm0.05~(\times 10^{2})$ &    $76.28\pm5.09$ &    $59.99\pm14.60$ &    $n.a.$ \\
VII\,Zw\,403   & $14.47\pm1.42$ &    $5.09\pm0.34$ &    $23.05\pm2.60$ &    $18.49\pm2.88$ &    $\le 13.60$ & $n.a.$ \\
    \hline 
  \end{tabular}
\begin{tabular}{l c c c c c}
    \hline
     \vspace{-8pt}\\
    \multicolumn{1}{l}{Source Name} & 
    \multicolumn{5}{c}{Other Observed Spectral Lines} \\ \cline{2-6}
    \vspace{-8pt}\\
    \multicolumn{1}{l}{} &
    \multicolumn{1}{c}{[Ar\,{\sc ii}]~6.99\,\mum} &
    \multicolumn{1}{c}{[Ne\,{\sc v}]~14.3\,\mum} &
    \multicolumn{1}{c}{[O\,{\sc iv}]~25.9\,\mum} &
    \multicolumn{1}{c}{[Si\,{\sc ii}]~34.8\,\mum} &
    \multicolumn{1}{c}{[Ne\,{\sc iii}]~36.0\,\mum} \\
    \hline
    \vspace{-8pt}\\
Haro\,3   & $ -  $ &  $\le 14.50$ & $17.93\pm5.00$ &    $3.99\pm0.24~(\times 10^{2})$ &    $1.17\pm0.29~(\times 10^{2})$ \\
Haro\,11$^*$   & $ -  $ &  $\le 5.29$ & $43.90\pm7.84$ &    $5.58\pm0.45~(\times 10^{2})$ &    $\le 1.92~(\times 10^{2})$ \\
He\,2-10   & $ -  $ &  $\le 20.84$ & $79.15\pm25.14$ &    $1.93\pm0.04~(\times 10^{3})$ &    $\le 2.92~(\times 10^{2})$ \\
HS\,0822+3542$^*$   & $ -  $ &  $\le 1.02$ & $ -  $ &  $ -  $ &  $ -  $ \\
HS\,1442+4250$^*$   & $ -  $ &  $\le 0.43$ & $2.37\pm0.74$ &    $3.40\pm1.39$ &    $\le 7.32$ \\
II\,Zw\,40   & $ -  $ &  $\le 7.46$ & $79.36\pm11.53$ &    $3.68\pm0.35~(\times 10^{2})$ &    $1.74\pm0.39~(\times 10^{2})$ \\
I\,Zw\,18   & $ -  $ &  $\le 4.02$ & $\le 4.08$ & $\le 4.37$ & $\le 10.87$ \\
Mrk\,153   & $ -  $ &  $\le 3.45$ & $5.10\pm1.92$ &    $23.78\pm6.17$ &    $\le 15.73$ \\
Mrk\,209$^*$   & $ -  $ &  $\le 1.87$ & $13.13\pm2.81$ &    $32.37\pm4.07$ &    $16.53\pm7.43$ \\
Mrk\,930   & $ -  $ &  $\le 3.29$ & $12.02\pm2.65$ &    $1.27\pm0.12~(\times 10^{2})$ &    $34.56\pm13.98$ \\
Mrk\,1450$^*$   & $ -  $ &  $\le 1.04$ & $8.31\pm0.88$ &    $30.72\pm5.42$ &    $\le 15.28$ \\
NGC\,1140   & $ -  $ &  $\le 9.94$ & $19.84\pm3.91$ &    $2.71\pm0.15~(\times 10^{2})$ &    $33.81\pm16.15$ \\
NGC\,1569   & $ -  $ &  $\le 8.18$ & $3.19\pm0.15~(\times 10^{2})$ &    $1.15\pm0.03~(\times 10^{3})$ &    $\le 5.77~(\times 10^{2})$ \\
SBS\,0335-052$^*$   & $ -  $ &  $\le 1.63$ & $5.22\pm1.15$ &    $\le 10.55$ & $\le 17.39$ \\
SBS\,1159+545$^*$   & $ -  $ &  $\le 3.64$ & $\le 2.35$ & $\le 13.85$ & $\le 18.92$ \\
SBS\,1415+437$^*$   & $ -  $ &  $\le 1.17$ & $6.51\pm2.02$ &    $17.77\pm3.91$ &    $\le 7.60$ \\
Tol\,1214-277$^*$   & $ -  $ &  $\le 0.75$ & $5.43\pm2.08$ &    $\le 8.89$ & $ -  $ \\ 
UM\,448   & $ -  $ &  $\le 5.64$ & $30.85\pm6.35$ &    $4.92\pm0.21~(\times 10^{2})$ &    $\le 77.59$ \\
UM\,461$^*$   & $ -  $ &  $\le 0.72$ & $2.77\pm0.90$ &    $8.76\pm2.71$ &    $\le 20.95$ \\
NGC\,625   & $ -  $ &  $\le 31.40$ & $\le 3.89$ & $1.85\pm0.12~(\times 10^{2})$ &    $68.04\pm22.33$ \\
UM\,311   & $ -  $ &  $\le 2.55$ & $6.39\pm2.62$ &    $56.23\pm5.82$ &    $\le 17.00$ \\
VII\,Zw\,403   & $ -  $ &  $\le 5.13$ & $\le 1.78$ & $\le 12.35$ & $\le 16.00$ \\
    \hline \hline
    \vspace{-5pt}
  \end{tabular}
  \label{table:irsfluxes_hires}
\end{table*}
\end{center}

\clearpage
\section{IRAS and PACS photometry of the compact sample}
\label{app:append-d}
\begin{table}[!ht]\small
   \caption{\ltir, IRAS 60/100, and PACS 70/100 band ratios. \newline
Notes. $n.a.$: not observed in the PACS 70\,\mum or 100\,\mum band. 
\ltir are SED-integrated values between 3-1\,100\,\mum, 
in units of solar luminosity for the $compact$ sample and 
W\,m$^{-2}$\,sr$^{-1}$ (peak surface brightness) for the $extended$ sample. 
The synthetic IRAS fluxes, indicated in brackets, are derived from 
the SED modeling of \cite{remy-2015}. 
}
\begin{tabular}{l c c c}
    \hline\hline
     \vspace{-8pt}\\
    \multicolumn{1}{l}{Source name} & 
    \multicolumn{1}{c}{60/100} & 
    \multicolumn{1}{c}{70/100} &
    \multicolumn{1}{c}{\ltir} \\
    \hline\hline
    \multicolumn{4}{c}{$Compact$ sample} \\
    \hline
Haro\,11   			&    $1.21$ &    $1.26$ &    $1.98 \times 10^{11}$ \\
Haro\,2   			&    $0.88$ &    $0.93$ &    $6.38 \times 10^{9}$ \\
Haro\,3   			&    $0.73$ &    $0.92$ &    $5.43 \times 10^{9}$ \\
He\,2-10   			&    $0.89$ &    $1.02$ &    $5.49 \times 10^{9}$ \\
HS\,0017+1055   	&    $[1.15]$ &    $1.40$ &    $9.70 \times 10^{8}$ \\
HS\,0052+2536   	&    $[0.72]$ &    $1.06$ &    $1.62 \times 10^{10}$ \\
HS\,0822+3542   	&    $[0.67]$ &    $1.11$ &    $1.05 \times 10^{7}$ \\
HS\,1222+3741   	&    $[0.88]$ &    $0.69$ &    $2.40 \times 10^{9}$ \\
HS\,1236+3937   	&  $ -  $ &  $0.83$ &    $ -  $ \\
HS\,1304+3529   	&    $[0.80]$ &    $0.80$ &    $1.69 \times 10^{9}$ \\
HS\,1319+3224   	&    $[0.82]$ &    $0.98$ &    $2.37 \times 10^{8}$ \\
HS\,1330+3651   	&    $[0.63]$ &    $0.83$ &    $1.37 \times 10^{9}$ \\
HS\,1442+4250   	&  $ -  $ &  $1.68$ &    $ -  $ \\
HS\,2352+2733   	&  $ -  $ &  $2.52$ &    $ -  $ \\
II\,Zw\,40   		&    $0.99$ &    $1.11$ &    $2.89 \times 10^{9}$ \\
I\,Zw\,18   			&    $[1.74]$ &    $2.46$ &    $3.34 \times 10^{7}$ \\
Mrk\,1089   		&    $0.67$ &    $0.94$ &    $3.68 \times 10^{10}$ \\
Mrk\,1450   		&    $0.49$ &    $1.22$ &    $3.03 \times 10^{8}$ \\
Mrk\,153   		&    $0.59$ &    $0.93$ &    $1.05 \times 10^{9}$ \\
Mrk\,209   		&    $0.97$ &    $0.95$ &    $2.98 \times 10^{7}$ \\
Mrk\,930   		&    $0.59$ &    $0.81$ &    $1.95 \times 10^{10}$ \\
NGC\,1140   		&    $0.68$ &    $0.95$ &    $3.73 \times 10^{9}$ \\
NGC\,1569   		&    $0.93$ &    $1.05$ &    $1.27 \times 10^{9}$ \\
NGC\,1705   		&    $0.48$ &    $0.80$ &    $6.29 \times 10^{7}$ \\
NGC\,2366   		&    $0.75$ &    $0.83$ &    $1.54 \times 10^{8}$ \\
NGC\,4214   		&    $0.62$ &    $0.76$ &    $5.34 \times 10^{8}$ \\
NGC\,4861   		&    $0.76$ &    $1.02$ &    $3.51 \times 10^{8}$ \\
NGC\,5253   		&    $1.00$ &    $1.01$ &    $1.63 \times 10^{9}$ \\
NGC\,625   		&    $0.66$ &    $0.61$ &    $2.85 \times 10^{8}$ \\
Pox\,186   		&    $[0.73]$ &    $0.74$ &    $5.64 \times 10^{7}$ \\
SBS\,0335-052   	&    $[2.76]$ &    $2.30$ &    $1.76 \times 10^{9}$ \\
SBS\,1159+545   	&    $[1.03]$ &    $1.01$ &    $2.00 \times 10^{8}$ \\
SBS\,1211+540   	&    $[1.28]$ &    $1.91$ &    $2.86 \times 10^{7}$ \\
SBS\,1249+493   	&    $[0.75]$ &    $0.93$ &    $9.34 \times 10^{8}$ \\
SBS\,1415+437   	&    $[1.00]$ &    $1.15$ &    $6.54 \times 10^{7}$ \\
SBS\,1533+574   	&    $0.63$ &    $0.75$ &    $2.04 \times 10^{9}$ \\
Tol\,1214-277   		&    $[0.83]$ &    $0.96$ &    $8.84 \times 10^{8}$ \\
UGC\,4483   		&    $[0.96]$ &    $0.43$ &    $2.60 \times 10^{6}$ \\
UM\,133   			&    $[0.73]$ &    $0.97$ &    $1.13 \times 10^{8}$ \\
UM\,311   			&    $[0.44]$ &    $0.57$ &    $5.61 \times 10^{9}$ \\
UM\,448   			&    $0.93$ &    $n.a.$ &  $9.00 \times 10^{10}$ \\
UM\,461   			&    $[0.93]$ &    $1.48$ &    $7.24 \times 10^{7}$ \\
VII\,Zw\,403   		&    $0.43$ &    $0.84$ &    $1.99 \times 10^{7}$ \\
    \hline\hline
    \multicolumn{4}{c}{$Extended$ sample} \\
    \hline
IC\,10   			&    $[0.93]$ &    $0.68$ &    $7.30 \times 10^{-5}$ \\
LMC-30Dor   		&    $[1.47]$ &    $n.a.$ &  $1.11 \times 10^{-3}$ \\
LMC-N11A   		&    $[0.90]$ &    $n.a.$ &  $7.92 \times 10^{-5}$ \\
LMC-N11B   		&    $[1.16]$ &    $n.a.$ &  $1.67 \times 10^{-4}$ \\
LMC-N11C   		&    $[1.09]$ &    $n.a.$ &  $8.99 \times 10^{-5}$ \\
LMC-N11I   		&    $[0.90]$ &    $n.a.$ &  $3.01 \times 10^{-5}$ \\
LMC-N158   		&    $[1.08]$ &    $n.a.$ &  $2.41 \times 10^{-4}$ \\
LMC-N159   		&    $[1.18]$ &    $n.a.$ &  $4.79 \times 10^{-4}$ \\
LMC-N160   		&    $[1.34]$ &    $n.a.$ &  $9.34 \times 10^{-4}$ \\
NGC\,4449   		&    $[1.06]$ &    $0.65$ &    $2.65 \times 10^{-5}$ \\
NGC\,6822   		&    $[0.87]$ &    $0.86$ &    $2.78 \times 10^{-5}$ \\
SMC-N66   		&    $[1.15]$ &    $n.a.$ &  $4.13 \times 10^{-5}$ \\
    \hline \hline
    \vspace{-5pt}
  \end{tabular}
  \label{table:bandratios}
\end{table}

\clearpage
\section{Line broadening and rotation in the PACS maps}
\label{app:append-e}
We find that the \oiiil and \oila lines, which are bright with smallest instrumental 
FWHM, are the most often broadened. 
The line widths are typically $\sim$100\,\kms larger than the instrumental 
profile in the brightest sources. For the fainter lines, it is more difficult 
to determine confidently whether or not broadening is present. 
Broadening of the \ciil line can only be accurately calculated for intrinsic line 
widths larger than 150\,\kms because of its large instrumental FWHM of 240\,\kms. 
When observed, \niiil is similarly broadened as are \oiiil and \oila. 
Such broadening indicates the possible presence of strong turbulent motions, 
or that multiple components contribute to the emission of those lines. 
The fact that we find similar intrinsic line widths for all tracers, despite 
their different origins, indicates that either they are spatially mixed 
within our beam (which is certainly the case for our compact sources), 
or that they are affected similarly by the source of this broadening 
(internal kinematics). 
Several galaxies also display clear signs of rotation in all observed spectral lines. 

\begin{center}
\begin{table*}[!ht]\small
  \caption{Broadening of the FIR lines (in \kms).} 
  \hfill{}
\begin{tabular}{l c c c c c c c c c}
    \hline\hline
     \vspace{-8pt}\\
    \multicolumn{1}{l}{} & 
    \multicolumn{6}{c}{Broadening} &
    \multicolumn{1}{c}{} & 
    \multicolumn{2}{c}{Rotation} \\ \cline{2-7} \cline{9-10}
    \multicolumn{1}{l}{} & 
    \multicolumn{1}{c}{\niii$_{57}$} & 
    \multicolumn{1}{c}{\oi$_{63}$} &
    \multicolumn{1}{c}{\oiii$_{88}$} &
    \multicolumn{1}{c}{\nii$_{122}$} &
    \multicolumn{1}{c}{\oi$_{145}$} &
    \multicolumn{1}{c}{\cii$_{157}$} & 
    \multicolumn{1}{c}{} & 
    \multicolumn{1}{c}{Axis} & 
    \multicolumn{1}{c}{$V_{rotation}$} \\ 
    \hline
    \vspace{-8pt}\\
	{PACS FWHM}		& 105	&  85		& 125	& 290	& 260	& 240	&& 		& \\
    \hline
    \vspace{-8pt}\\
	{Haro\,2}			& n. a.	& 90		& 140 	& n. a. 	& n. a.	& - 		&& -		& - \\
	{Haro\,3}			& 100	& 90		& 110 	& -	 	& -		& - 		&& NEE-SWW	& $\pm$30\,\kms \\
	{Haro\,11}			& 160	& 130	& 200 	& 200 	& 100	& 160	&& N-S		& $\pm$50\,\kms \\
	{He\,2-10}			& 160	& 100	& 170 	& -	 	& -		& - 		&& NE-SW	& $\pm$20\,\kms \\
	{HS\,1330+3651}	& n.a.	& -		& 150	& n.a.	& n.a.	& -		&& -		& - \\
	{II\,Zw\,40}		& n. a.	& 80		& 110 	& -	 	& -		& - 		&& N-S		& $\pm$30\,\kms \\
	{Mrk\,1089}		& n. a.	& 80		& 130 	& -	 	& -		& - 		&& NNE-SSW	& $\pm$80\,\kms \\
	{Mrk\,153}			& n. a.	& 80		& 120 	& n. a. 	& n. a.	& - 		&& NE-SW	& $\pm$30\,\kms \\
	{Mrk\,930}			& n. a.	& 120	& 120 	& -	 	& -		& - 		&& N-S		& $\pm$30\,\kms \\
	{NGC\,1140}		& n. a.	& 100	& 100 	& -	 	& -		& - 		&& NE-SW	& $\pm$40\,\kms \\
	{NGC\,1569}		& n. a.	& 80		& 100 	& n. a. 	& n. a.	& - 		&& -		& - \\ 
	{NGC\,1705}		& n. a.	& -		& 100 	& n. a. 	& n. a.	& - 		&& -		& - \\
	{NGC\,2366}		& n. a.	& 80		& 70	 	& n. a. 	& n. a.	& -	 	&& -		& - \\
	{NGC\,4214}		& n. a.	& 80		& 80	 	& -	 	& -		& - 		&& -		& - \\ 
	{SBS\,1533+574}	& n. a.	& -		& 80	 	& n. a. 	& n. a.	& - 		&& -		& - \\
	{UM\,448}			& n. a.	& 150	& 150 	& -	 	& 80		& 150 	&& NNE-SSW	& $\pm$60\,\kms \\ 
	{IC\,10}			& n. a.	& 90		& 70	 	& -	 	& -		& - 		&& -		& - \\
	{LMC}			& 150	& 150	& 150 	& 100	& 100	& - 		&& -		& - \\
	{SMC}			& n. a.	& 150	& 150 	& n. a. 	& n. a.	& - 		&& -		& - \\
	{NGC\,4449}		& n. a.	& 90		& 90	 	& -	 	& -		& - 		&& -		& - \\
	{NGC\,5253}		& n. a.	& 60		& 90	 	& -	 	& -		& - 		&& NE-SW	& $\pm$30\,\kms \\
	{NGC\,6822}		& n. a.	& 100	& 80	 	& -	 	& -		& - 		&& -		& - \\
    \hline \hline
    \vspace{-5pt}\\
  \end{tabular}
  \hfill{}
  \newline
The broadening (or intrinsic line width) is given as the observed FWHM subtracted by the instrumental FWHM. 
Uncertainties are on the order of 20\%. 
We indicate the broadening of the spaxel where the emission peaks. 
This value can be higher elsewhere because of rotation or instrumental effects (uncentered point-source). 
  \label{table:broad}
\end{table*}
\end{center}
}

\end{document}